\documentclass[preprint,aps,prd,showpacs,superscriptaddress,nofootinbib,tightenlines,11pt]{revtex4}
\usepackage{amsfonts}
\usepackage{multirow}
\usepackage{mathrsfs}
\usepackage{graphicx}
\usepackage{epstopdf}
\usepackage{amsmath}
\usepackage{amssymb}
\usepackage{bm}
\usepackage{bbm}
\usepackage{color}
\usepackage{array}
\usepackage{diagbox}
\usepackage{float}
\usepackage{braket}
\usepackage{upgreek}
\usepackage{subcaption}
\usepackage{diagbox}
\usepackage{slashed}
\usepackage[citecolor=blue]{hyperref}

\def\OMIT#1{}

\def\hlinew#1{%
  \noalign{\ifnum0=`}\fi\hrule \@height #1 \futurelet
   \reserved@a\@xhline}
\usepackage{array}
\newcommand{\PreserveBackslash}[1]{\let\temp=\\#1\let\\=\temp}
\newcolumntype{C}[1]{>{\PreserveBackslash\centering}p{#1}}
\newcolumntype{R}[1]{>{\PreserveBackslash\raggedleft}p{#1}}
\newcolumntype{L}[1]{>{\PreserveBackslash\raggedright}p{#1}}

\newcommand{\beq}{\begin{equation}}
\newcommand{\eeq}{\end{equation}}
\newcommand{\bqa}{\begin{eqnarray}}
\newcommand{\eqa}{\end{eqnarray}}
\newcommand{\bseq}{\begin{subequations}}
\newcommand{\eseq}{\end{subequations}}

\newcommand\fverb{\setbox\fverbbox=\hbox\bgroup\verb}
\newcommand\fverbdo{\egroup\medskip\noindent%
			\fbox{\unhbox\fverbbox}\ }
\newcommand\fverbit{\egroup\item[\fbox{\unhbox\fverbbox}]}
\newbox\fverbbox

\makeatletter

\newcommand{\Rmnum}[1]{\expandafter\@slowromancap\romannumeral #1@}

\usepackage{tikz}
\usetikzlibrary {shapes.misc}
\usetikzlibrary{arrows.meta}
\usetikzlibrary{calc}
\usetikzlibrary{
  decorations.pathmorphing,
  decorations.pathreplacing,
  decorations.markings
}
\usetikzlibrary{patterns}

\tikzset{
  every picture/.style={semithick, line cap=round},
  scalar/.style={dashed},
  fermion/.default=0.5,
  fermion/.style={postaction={decorate, decoration={
    markings,
    mark=at position #1 with {\arrow{Stealth[angle=30:7pt,inset=1.5pt]}},
    transform={xshift={3.5pt*cos(15)}}
  }}},
  antifermion/.default=0.5,
  antifermion/.style={postaction={decorate, decoration={
    markings,
    mark=at position #1 with {\arrowreversed{Stealth[angle=30:7pt,inset=1.5pt]}},
    transform={xshift={-3.5pt*cos(15)}}
  }}},
  gluon/.default=3pt,
  gluon/.style={decorate, decoration={
    coil,
    amplitude=0.5*#1,
    aspect=1,
    segment length=#1
  }},
  rgluon/.default=3pt,
  rgluon/.style={decorate, decoration={
    coil,
    amplitude=-0.5*#1,
    aspect=-1,
    segment length=#1
  }},
  gluonpre/.default=0pt,
  gluonpre/.style={decorate, decoration={
    coil,
    amplitude=1.5pt,
    aspect=1,
    segment length=3pt,
    pre length=#1
  }},
  rgluonpre/.default=0pt,
  rgluonpre/.style={decorate, decoration={
    coil,
    amplitude=-1.5pt,
    aspect=-1,
    segment length=3pt,
    pre length=#1
  }},
  crossmark/.style={cross out, draw=red, inner sep=2pt},
  counter/.style={path picture={
    \draw (path picture bounding box.south east) --
      (path picture bounding box.north west)
      (path picture bounding box.south west) --
      (path picture bounding box.north east);
  }},
  cnode/.default=8pt,
  cnode/.style={inner sep=0pt, minimum size=#1, circle}
}

\makeatother

\usepackage{babel}
\begin{document}
\title{\mbox{}\Large
Final-state rescattering mechanism of charmed baryon decays}

\author{Cai-Ping Jia\footnote{jiacp20@lzu.edu.cn, corresponding author}}
\affiliation{MOE Frontiers Science Center for Rare Isotopes, and School of Nuclear Science and Technology, Lanzhou University, Lanzhou 730000, China\vspace{0.2cm}}

\author{Hua-Yu Jiang\footnote{jianghuayu@htu.edu.cn, corresponding author}}
\affiliation{Institute of Particle and Nuclear Physics, Henan Normal University, Xinxiang 453007, China\vspace{0.2cm}}

\author{Jian-Peng Wang\footnote{wangjp20@lzu.edu.cn, corresponding author}}
\affiliation{MOE Frontiers Science Center for Rare Isotopes, and School of Nuclear Science and Technology, Lanzhou University, Lanzhou 730000, China\vspace{0.2cm}}

\author{Fu-Sheng Yu\footnote{yufsh@lzu.edu.cn, corresponding author}}
\affiliation{MOE Frontiers Science Center for Rare Isotopes, and School of Nuclear Science and Technology, Lanzhou University, Lanzhou 730000, China\vspace{0.2cm}}

\date{\today}
\begin{abstract}
The dynamical studies on the non-leptonic weak decays of charmed baryons are always challenging, due to the large non-perturbative contributions at the charm scale. In this work, we develop the final-state rescattering mechanism to study the two-body non-leptonic decays of charmed baryons. The final-state interaction is a physical picture of long-distance effects. Instead of using the Cutkosky rule to calculate the hadronic triangle diagrams which can only provide the imaginary part of decay amplitudes, we point out that the loop integral is more appropriate, as both the real parts and the imaginary parts of amplitudes can be calculated completely. In this way, it can be obtained for the non-trivial strong phases which are essential to calculate CP violations. With the physical picture of long-distance effects and the reasonable method of calculations, it is amazingly achieved that all the nine existing experimental data of branching fractions for the $\Lambda_c^+$ decays into an octet light baryon and a vector meson can be explained by only one parameter of the model. Besides, the decay asymmetries and CP violations are not sensitive to the model parameter, since the dependence on the parameter is mainly cancelled in the ratios, so that the theoretical uncertainties on these observables are lowered down.

\end{abstract}

\maketitle

\tableofcontents

\section{Introduction}
In the non-leptonic weak decays of heavy flavour hadrons, the effects of long-distance or soft QCD dynamics are important for generating strong phases, which are essential components of the direct $CP$ violation mechanism. 
$CP$ asymmetry is sensitive to strong phases and is a necessary condition for understanding the matter-antimatter asymmetry in our universe \cite{Sakharov:1967dj}.
However, there is no appropriate QCD-based approach for calculating the long-distance contributions due to the large soft-overlap effects of initial and final states. 
The QCD factorization (QCDF) \cite{Beneke:1999br,Beneke:2000ry,Beneke:2001ev,Beneke:2003zv} and perturbation QCD (PQCD) \cite{Keum:2000wi,Keum:2000ph,Lu:2000em} approach are mainly employed to calculate the contributions with small soft-overlap effects of hadron wave functions.
Therefore, it is crucial to build up an appropriate framework for estimating the long-distance non-perturbative effects in heavy flavour hadron decays. 
In this work, we take the two-body non-leptonic decays of charmed baryons as the starting point to study the long-distance non-perturbative dynamics and improve the calculation method that can reasonably describe the effects of the final-state interactions.

A lot of experimental measurements on the charmed baryon decays have been performed during the past decade \cite{LHCb:2017xtf,LHCb:2017hwf,LHCb:2017yqf,LHCb:2019nxp,LHCb:2019ldj,LHCb:2020zkk,LHCb:2022sck,LHCb:2023crj,Belle:2015wxn,Belle:2017tfw,Belle:2018gcs,Belle:2018lws,Belle:2019bgi,Belle:2020xku,Belle:2021mvw,Belle:2021crz,Belle:2021btl,Belle:2021avh,Belle:2021vyq,Belle:2022raw,Belle-II:2022ggx,Belle:2022uod,Belle:2022bsi,Belle:2022cbs,Belle:2022yaq,Belle:2022pwd,Belle:2022ywa,Belle:2023ngs,Belle-II:2024jql,BESIII:2015ysy,BESIII:2015bjk,BESIII:2016ozn,BESIII:2016yrc,BESIII:2017fim,BESIII:2017rfd,BESIII:2018cvs,BESIII:2018ciw,BESIII:2018mug,BESIII:2018cdl,BESIII:2019odb,BESIII:2020cpu,BESIII:2020kzc,BESIII:2022bkj,BESIII:2022wxj,BESIII:2022qaf,BESIII:2022izy,BESIII:2022ysa,BESIII:2022vrr,BESIII:2022tnm,BESIII:2022udq,BESIII:2022xne,BESIII:2022cmg,BESIII:2022aok,BESIII:2023jem,BESIII:2023rky,BESIII:2023vfi,BESIII:2023ooh,BESIII:2023jxv,BESIII:2023wrw,BESIII:2023iwu,BESIII:2023dvx,BESIII:2023uvs,BESIII:2023sdr,BESIII:2023pia,BESIII:2024xgl,BESIII:2024sfz,BESIII:2024mbf}.
The BESIII and Belle (II) experiments have measured the branching ratios and decay asymmetry parameters for a large amount of non-leptonic decay channels of charmed baryons. 
Besides, the most precise measurement on the $CP$ violation in charmed baryon decays up to date is given by LHCb as $A_{CP}^{dir}(\Lambda_c^+\to pK^+K^-)-A_{CP}^{dir}(\Lambda_c^+\to p\pi^+\pi^-)=(0.30\pm0.91\pm0.61)\%$\cite{LHCb:2017hwf} . 
Decay asymmetry parameters and $CP$ violation are of particular interest because they are sensitive to the strong phases in the decay and provide stronger constraints for theoretical analysis. 
More experimental results will be forthcoming with even larger data samples collected by BESIII, Belle II and LHCb in the near future. 
Therefore, the corresponding theoretical studies are urgently needed to understand the dynamics of charmed baryon decays and to accurately calculate the relevant observables.

The study of charmed baryon decays faces a lot of theoretical challenges, due to insufficient understanding of complicated dynamics in baryon system and non-perturbative dynamics at the charm scale.
From 1990s to now, there are many theoretical attempts to study the non-leptonic decays of charmed baryons \cite{Savage:1991wu,Sheikholeslami:1991ab,Verma:1995dk,Sharma:1996sc,Chen:2002jr,Lu:2016ogy,Wang:2017gxe,Geng:2017esc,Geng:2017mxn,Cheng:2018hwl,Zou:2019kzq,Jia:2019zxi,Hsiao:2019yur,Geng:2019xbo,Cen:2019ims,Meng:2020euv,Geng:2020zgr,Geng:2020tlx,Ke:2020uks,Li:2021qod,Xu:2021mkg,Hsiao:2021nsc,Wang:2022tcm,Cheng:2022kea,Zhong:2022exp,Liu:2023pyk,Xing:2023bzh,Liu:2023dvg,Xing:2023dni,Geng:2023pkr,Shi:2024plf,Zhong:2024zme,Zhong:2024qqs,Geng:2024sgq,Wang:2024ztg,Wang:2024nxb,Grossman:2018ptn,Wang:2019dls,He:2024pxh,Sun:2024mmk,Xing:2024nvg,Korner:1978ec,Kohara:1991ug,Cheng:1991sn,Korner:1992wi,Xu:1992vc,Xu:1992sw,Zenczykowski:1993jm,Cheng:1993gf,Uppal:1994pt,Chau:1995gk,Ivanov:1997ra,Sharma:1998rd,Gutsche:2018utw,Niu:2020gjw,Li:2024rqb,Zhang:2024jby,Lyu:2024qgc,Feng:2020jvp}. 
The most widely used analysis method is based on the flavor $SU(3)$ symmetry method, combined with an extensive global fit to the experimental data,
which does not provide enough information of the decay dynamics.
Compared with the case of charmed mesons, there are many more non-perturbative parameters that need to be fitted in the charmed baryon decays, which makes it currently mainly suitable for analyzing branching ratios.
While several studies have been devoted to fitting $CP$ violation in $SU(3)$ flavour symmetry \cite{Grossman:2018ptn,Wang:2019dls,He:2024pxh,Sun:2024mmk,Xing:2024nvg}, the analysis of asymmetry parameters and $CP$ violation in charmed baryon decays remains a major challenge.
In addition, there are some analyses of charmed baryon decays based on dynamical models such as pole models, current algebra, and quark model \cite{Korner:1978ec,Kohara:1991ug,Cheng:1991sn,Korner:1992wi,Xu:1992vc,Xu:1992sw,Zenczykowski:1993jm,Cheng:1993gf,Uppal:1994pt,Chau:1995gk,Ivanov:1997ra,Sharma:1998rd,Gutsche:2018utw,Niu:2020gjw}. However, these dynamical models usually have significant theoretical uncertainties, or are difficult to calculate the $CP$ asymmetries of charmed baryon decays.
Totally speaking, it is urgent to develop theoretical methods for describing the dynamics of charmed baryon decays and calculating its branching ratios, asymmetry parameters and $CP$ violation.

Final-state interaction is a physical picture for the long-distance contributions of the weak decays of heavy-flavor hadrons. 
The observed $CP$ violation of charmed meson decays in $D^0\to K^+K^-$ and $\pi^+\pi^-$ indicates that the penguin diagrams are at the same order as the tree diagrams apart from the CKM matrix elements \cite{LHCb:2019hro}. 
The enhancement of penguin diagrams might be induced by the long-distance contributions \cite{Li:2012cfa,Li:2019hho}.
The final-state-interaction effect can easily explain the similar size of penguin and tree diagrams, since they stem from the same hadronic triangle diagrams. 
For example, the final-state interactions have been used to explain the $CP$ violation in $D$ meson decays\cite{Bediaga:2022sxw,Pich:2023kim}, providing the results at the same order with the experimental observations.
The rescattering mechanism, modeled as single-particle exchanges (one loop triangle diagrams), has been widely applied to calculate the final state interactions in heavy hadron non-leptonic decays, such as the systematic analysis on $B$ meson two-body decays in \cite{Cheng:2004ru} and the successful prediction on the discovery channels of doubly charmed baryon $\Xi_{cc}^{++}$ in \cite{Yu:2017zst}.
In the previous studies, the computation of the triangle diagrams in rescattering mechanism is based on the optical theorem and Cutkosky cutting rules, where only the imaginary parts of triangle amplitudes are involved and therefore difficult to describe the strong phases.  
    
In this work, we improve the calculation method for the effects of final state interactions in the rescattering mechanism by utilizing the complete analytical expression of loop integrals and thus obtain the magnitude and strong phase of triangle diagrams.
The weak transition vertices are obtained from the weak effective Hamiltonian using the naive factorization approach for the short-distance contributions, while the hadron strong scatterings are calculated using the hadronic effective Lagrangian for the long-distance contributions. 
We adopt the Pauli-Villars scheme with the momentum cutoff to regularize the divergence in the loop integrals, with a model dependent parameter corresponding to the cutoff. It can be demonstrated that the modification of the propagator in the Pauli-Villars scheme is consistent with the traditional approach of adding a monopole form factor to the propagator. 

In our improved calculations, the explicit dependence of the branching fractions on the model parameters is significantly reduced compared to conventional methods.
As the $p\rho^0$ and $p\phi$ dominate the decays of $\Lambda_c^+\to p\pi^+\pi^-$ and $pK^+K^-$, we study $\Lambda_c^+\to\mathcal{B}_8V$ decays in this work for CP violation, with $\mathcal{B}_8$ and $V$ as octet baryons and vector mesons respectively.
It is amazingly achieved that the results of branching ratios with only one model parameter, are consistent with all the nine existing experimental data of the $\Lambda_c^+\to\mathcal{B}_8V$ decays.
Besides, we predict the $CP$ violation of $\Lambda_c^+\to\mathcal{B}_8V$ decays for the first time, benefited from the non-trivial strong phases.
It is also systematically studied for the decay asymmetry parameters in the angular distributions of final states.
Since the decay asymmetry parameters and the $CP$ asymmetry are defined as the ratios, the dependence of these observables on the model parameter and the input hadronic strong couplings is mostly cancelled in the ratios.
The model dependence in the improved theoretical approach is therefore significantly reduced and is expected to provide a reasonable description of long-distance final state interactions.

We stress that the framework building up in this work can be naturally extended to the study of non-leptonic decays of charmed meson and bottom hadrons.
For bottom hadron decays, although the factorizable contributions in perturbation theory may be important in magnitude, the effect of final state interactions is crucial for generating strong phases and thus for predicting $CP$ violation.

In section II, we introduce the theoretical framework, including the topological diagrams for the classification of weak transitions, the naive factorization approach for calculating the short-distance contributions and the rescattering mechanism for estimating the long-distance final state interactions.
In the third section, a discussion of the analytical expression of the three-point master integral will be involved, including the relation between the method of the complete loop integral and the Cutkosky cutting rule, proving the equivalence between the Pauli-Villars scheme with cutoff and the monopole form factors in the loop integral.
We present the determination of input parameters and the obtained numerical results (the branching ratios, $CP$ violation and decay asymmetry parameters for $\Lambda_c^+\to\mathcal{B}_8V$ decays) in section IV, where several channels, e.g. $\Lambda_c^+\to p\rho^0$ and $\Lambda_c^+\to p\phi$ , are used to demonstrate the dependence of various observables on model and input parameters.
A summary is given in the last section.
Essential details regarding the amplitudes for all of the $\Lambda_c^+\to\mathcal{B}_8V$ channels are provided in Appendix A, while Appendix B contains expressions for the hadronic effective Lagrangian and the heavy to light form factors.

\section{Theoretical Framework }

In this section, we firstly introduce the topological diagrams which are widely used to describe the heavy hadron non-leptonic decays\cite{Chau:1982da,Chau:1986jb,Biyajima:1987qm,Chau:1988kb,Chau:1995gk,Qin:2022nof}. Then we demonstrate the general framework for calculating the short-distance and long-distance dynamics in each kind of topological diagram.

\subsection{Topological diagrams and non-perturbative contribution}

\begin{figure}[tb]
    \centering
    \begin{minipage}{0.32\linewidth}
    \centering
        \includegraphics[scale=0.5]{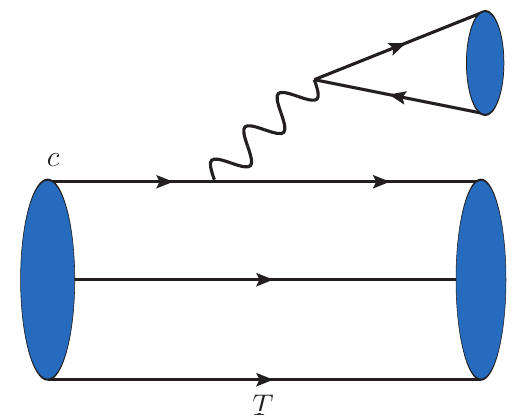}
    \end{minipage}
    \begin{minipage}{0.32\linewidth}
    \centering
        \includegraphics[scale=0.5]{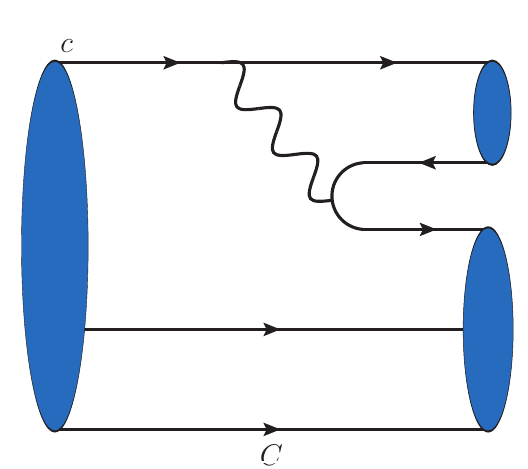}
    \end{minipage}
    \begin{minipage}{0.32\linewidth}
    \centering
        \includegraphics[scale=0.5]{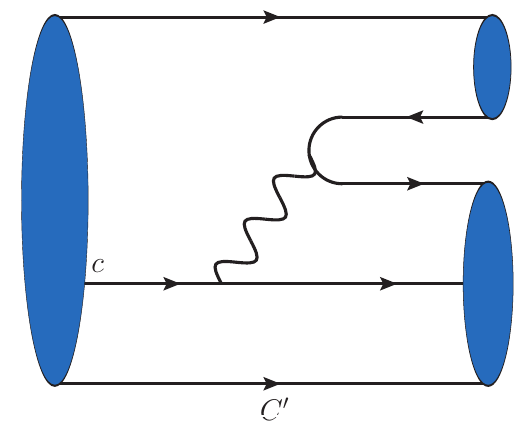}
    \end{minipage}
    
    \qquad
    
    \begin{minipage}{0.32\linewidth}
    \centering
        \includegraphics[scale=0.5]{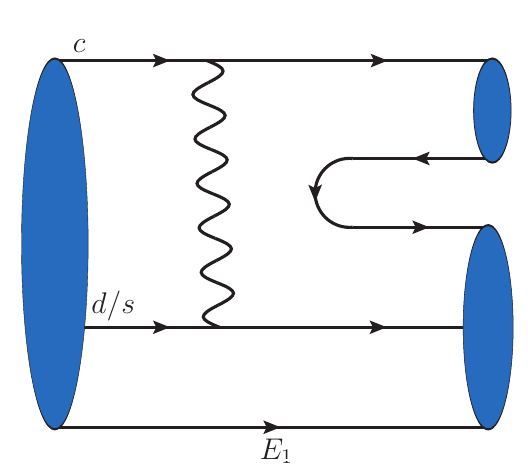}
    \end{minipage}
    \begin{minipage}{0.32\linewidth}
    \centering
        \includegraphics[scale=0.5]{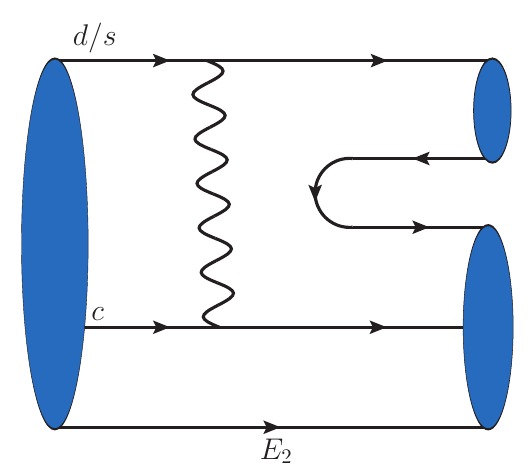}
    \end{minipage}
    \begin{minipage}{0.32\linewidth}
    \centering
        \includegraphics[scale=0.5]{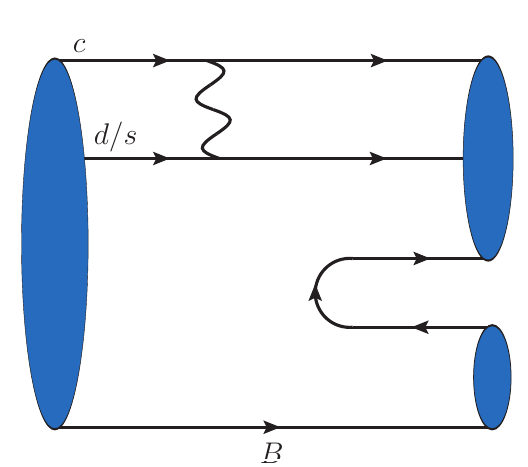}
    \end{minipage}
  
    \caption{Tree level topological diagrams for two-body nonleptonic decays of singly charmed baryon.}\label{topo fig}
\end{figure}

Topology diagrams contain perturbative and non-perturbative information in heavy baryon non-leptonic decays and provide an intuitive representation of the strong interaction dynamics in the weak decay.
According to the diagram viewpoint, the tree-level topological diagrams for two-body non-leptonic weak decays of charmed baryons are listed in Fig.\ref{topo fig}.
These diagrams are sorted by their different weak decay topologies\cite{Chau:1995gk}. $T$ stands for the color-allowed diagram with the external $W$-emission, $C$ and $C^\prime$ are both color-suppressed diagrams with internal $W$-emission, the difference between them is that the quark which is from the weak transition of charm quark flows into the final-state meson or baryon.
Similarly, there exist three different types of $W$-exchange diagrams which can be distinguished with the flow of quarks that generated from weak vertex and are respectively denoted as $E_1$, $E_2$ and $B$.


Generally, the topology diagrams involve factorization and non-factorization contributions. The $T$ topology is dominated by factorizable contributions according to the color transparency mechanism and can be well estimated within the naive factorization approach, i.e., it can be expressed as the convolution of baryon weak transition form factors and meson's decay constants. For the internal $W$-emission diagram $C$, the short-distance contributions can also be calculated from the naive factorization approach, however which are suppressed by the small effective Wilson coefficients of four-fermion operators at charm scale\cite{Fu-Sheng:2011fji,Li:2012cfa,Qin:2013tje}. Some studies have shown that the non-factorizable long-distance contributions of $C$ are much more important than the short-distance contributions in heavy baryon decays\cite{Yu:2017zst,Han:2021azw,Han:2021gkl}. Moreover, according to the power counting rules from the soft-collinear effective theory, the ratios among the topological diagrams have the following power counting relation, $\frac{|C|}{|T|}\sim\frac{|C^\prime|}{|C|}\sim\frac{|E_1|}{|C|}\sim\frac{|E_2|}{|C|}\sim\frac{|B|}{|C|}\sim\mathcal{O}\Big(\frac{\Lambda_{QCD}^h}{m_Q}\Big)$ \cite{Leibovich:2003tw,Mantry:2003uz}, this factor is about $\mathcal{O}\left(1\right)$ at $m_c=1.3GeV$. This means that the topology diagrams except for $T$ are non-factorization contributions dominated and must be well-estimated in charm system.


To calculate the contributions of different topological diagrams in charmed baryon decays, we face challenges due to their non-perturbative nature. Direct computation of all the topological diagrams is impractical. 
Instead, we employ different theoretical methods to calculate the short and long-distance contributions of these topology diagrams. 
Typically, we use a naive factorization approach to estimate factorizable contributions. For non-factorizable contributions, such as final state interaction effects mediated by loop mechanisms, we incorporate these effects into our model. 
This approach allows us to refine theoretical predictions by adjusting parameters and assumptions, ensuring better agreement with experimental data. 


\subsection{Short-distance contributions with factorization hypothesis}
In this part, we introduce the framework for estimating the short-distance factorizable contributions of the $T$ and $C$ diagram, which is the foundation for the calculation of the long-distance non-factorizable contributions based on the hypothesis of final-state rescattering mechanism.

Firstly, we present the weak effective Hamiltonian which is generally used to analyse the non-leptonic decays of heavy hadrons,
\begin{equation}
\mathcal{H}_{eff}=\frac{G_{F}}{\sqrt{2}}\sum_{q^{\prime}=d,s}V_{cq^{\prime}}^{*}V_{uq}\left[C_{1}(\mu)O_{1}(\mu)+C_{2}(\mu)O_{2}(\mu)\right]+h.c. \,,
\end{equation}
where $C_{1,2}(\mu)$ are Wilson coefficients including the short-distance QCD dynamics, $O_{1,2}(\mu)$ are the tree level four-fermion operators,
\begin{equation}
O_{1}=(\bar{u}_{\alpha}q_{\beta})_{V-A}(\bar{q}_{\beta}^{\prime}c_{\alpha})_{V-A}\,, \quad O_{2}=(\bar{u}_{\alpha}q_{\alpha})_{V-A}(\bar{q}_{\beta}^{\prime}c_{\beta})_{V-A} \,,
\end{equation}
then, the hadronic matrix element of charmed baryon decay can be derived
directly,
\begin{equation}
\left\langle \mathcal{B}_{8}M\left|\mathcal{H}_{eff}\right|\mathcal{B}_{c}\right\rangle =\frac{G_{F}}{\sqrt{2}}V_{cq^{\prime}}^{*}V_{uq}\sum_{i=1,2}C_{i}\left\langle \mathcal{B}_{8}M\left|O_{i}\right|\mathcal{B}_{c}\right\rangle \,,
\end{equation}
here, we neglect the contributions of penguin operators due to the large suppression of CKM matrix elements in charmed hadron decays.

A further step is to deal with the hadronic matrix element of four-fermion operators $\left\langle \mathcal{B}_{8}M\left|O_{i}\right|\mathcal{B}_{c}\right\rangle $. The well known approach is the naive factorization governed by the color transparency mechanism, the hadronic matrix element is factorized as the convolution of two parts
\begin{equation}
    \begin{aligned}
      \bra{\mathcal{B}_{8}M}O_{i}\ket{\mathcal{B}_{c}}\to \bra{M}\mathcal{J}^{\mu}_{uq}\ket{0}\otimes \bra{\mathcal{B}_{8}}\mathcal{J}_{\mu,cq^{\prime}}\ket{\mathcal{B}_{c}} \,,
    \end{aligned}
\end{equation}
where $\mathcal{J}^{\mu}_{uq}$ and $\mathcal{J}_{\mu,cq^{\prime}}$ represent the $V-A$ current with the quark flavor $(uq)$ and $(q^\prime c)$ respectively. The proof for this factorization can be extended from the studies of bottom hadron decays, as discussed in references \cite{Beneke:1999br,Beneke:2000ry,Beneke:2001ev,Beneke:2003zv}. 
Based on the factorization formula, the $T$ and $C$ topology are given by
\bseq
\begin{align} 
& \left\langle \mathcal{B}_{8}M\left|\mathcal{H}_{eff}\right|\mathcal{B}_{c}\right\rangle _{SD}^{T}=\frac{G_{F}}{\sqrt{2}}V_{cq^{\prime}}^{*}V_{uq}a_{1}(\mu)\left\langle M\left|\bar{u}\gamma^{\mu}(1-\gamma_{5})q\right|0\right\rangle \left\langle \mathcal{B}_{8}\left|\bar{q}^{\prime}\gamma_{\mu}(1-\gamma_{5})c\right|\mathcal{B}_{c}\right\rangle \,,\label{eq:NF_T} \\
& \left\langle \mathcal{B}_{8}M\left|\mathcal{H}_{eff}\right|\mathcal{B}_{c}\right\rangle _{SD}^{C}=\frac{G_{F}}{\sqrt{2}}V_{cq^{\prime}}^{*}V_{uq}a_{2}(\mu)\left\langle M\left|\bar{u}\gamma^{\mu}(1-\gamma_{5})q\right|0\right\rangle \left\langle \mathcal{B}_{8}\left|\bar{q}^{\prime}\gamma_{\mu}(1-\gamma_{5})c\right|\mathcal{B}_{c}\right\rangle \,, \label{eq:NF_C}
\end{align}
\eseq
where $a_{1,2}(\mu)$ are the effective Wilson coefficients
defined through $a_{1}(\mu)=C_{1}(\mu)+C_{2}(\mu)/3$,\\
$a_{2}(\mu)=C_{2}(\mu)+C_{1}(\mu)/3$ with $C_{1}(\mu)=1.21,C_{2}(\mu)=-0.42$ evolved from $M_{W}$
to the charm scale $\mu=m_{c}$\cite{Li:2012cfa}. 
In Eq.\eqref{eq:NF_T} and \eqref{eq:NF_C}, the first matrix element $\bra{M}\mathcal{J}^{\mu}_{uq}\ket{0}$
is parameterized as the meson's decay constants \cite{Cheng:1996cs}
\bseq
\begin{align}
 & \left\langle P(p)\left|\bar{u}\gamma^{\mu}(1-\gamma_{5})q\right|0\right\rangle =if_{P}p^{\mu} \,,\\
 & \left\langle V(p)\left|\bar{u}\gamma^{\mu}(1-\gamma_{5})q\right|0\right\rangle =m_{V}f_{V}\varepsilon^{\ast\mu} \,,
\end{align}
\eseq
where $f_{P,V}$ are the decay constant of pseudoscalar
and vector mesons respectively, and $\varepsilon^{\mu}$ is the polarization vector. 
The second one $\bra{\mathcal{B}_{8}}\mathcal{J}_{\mu,cq^{\prime}}\ket{\mathcal{B}_{c}} $ is formulated as the heavy-to-light transition form factors 
\begin{align}
 \big\langle \mathcal{B}_{8}(p_{f},s_{f})\left|\bar{q}^{\prime}\gamma_{\mu}(1-\gamma_{5})c\right|\mathcal{B}_{c}(p_{i},s_{i} & ) \big\rangle = 
 \bar{u}(p_{f},s_{f})\left[\gamma_{\mu}f_{1}(q^{2})+i\sigma_{\mu\nu}\frac{q^{\nu}}{m_{i}}f_{2}(q^{2})+\frac{q^{\mu}}{m_{i}}f_{3}(q^{2})\right]u(p_{i},s_{i}) \nonumber \\
 & -\bar{u}(p_{f},s_{f})\left[\gamma_{\mu}g_{1}(q^{2})+i\sigma_{\mu\nu}\frac{q^{\nu}}{m_{i}}g_{2}(q^{2})+\frac{q^{\mu}}{m_{i}}g_{3}(q^{2})\right]\text{\ensuremath{\gamma_{5}}}u(p_{i},s_{i}) \,,
\end{align}
with momentum transfer $q=p_{i}-p_{f}$, $f_{i}$ and $g_{i}$ are heavy-to-light transition form factors which can be calculated through non-perturbative approaches. Conventionally, the amplitude of two-body non-leptonic
charmed baryon decays is given in terms of partial wave scheme
\begin{equation} \label{eq:NF_BcBP}
\mathcal{A}(\mathcal{B}_{c}\to\mathcal{B}_{8}P)=i\bar{u}_{f}(A+B\gamma_{5})u_{i} \,,
\end{equation}
specifically, the invariant amplitudes $A$ and $B$ are expressed as the following factorization form by using the form factor $f_{i}$ and $g_{i}$ 
\begin{equation}
A=\lambda f_{P}(m_{i}-m_{f})f_{1}(m^{2}) \,, \quad B=\lambda f_{P}(m_{i}+m_{f})g_{1}(m^{2}) \,.
\end{equation} 
For vector meson, the amplitudes are
\begin{align} \label{eq:NF_BcBV}
\mathcal{A}(\mathcal{B}_{c}\to\mathcal{B}_{8}V)=\bar{u}_{f}\Big(A_{1}\gamma_{\mu}\gamma_{5}+A_{2}\frac{p_{f\mu}}{m_{i}}\gamma_{5}+B_{1}\gamma_{\mu}+B_{2}\frac{p_{f\mu}}{m_{i}}\Big)\varepsilon^{\ast \mu}u_{i} \,,
\end{align}
where
\begin{align}
 & A_{1}=-\lambda f_{V}m\Big(g_{1}(m^{2})+g_{2}(m^{2})\frac{m_{i}-m_{f}}{m_{i}}\Big),\quad A_{2}=-2\lambda f_{V}mg_{2}(m^{2})\,,\\
 & B_{1}=\lambda f_{V}m\Big(f_{1}(m^{2})-f_{2}(m^{2})\frac{m_{i}+m_{f}}{m_{i}}\Big),\quad B_{2}=2\lambda f_{V}mf_{2}(m^{2})\,,
\end{align}
with $\lambda=\frac{G_{F}}{\sqrt{2}}V^{*}_{cq^{\prime}}V_{uq}a_{1,2}(\mu)$ for simplicity, $m$ is the mass of final meson. 
As we mentioned before, the factorization formula in Eq.\eqref{eq:NF_BcBP} and \eqref{eq:NF_BcBV} are also essential ingredients for calculating the long-distance final-state interaction effects. And the decay width has the form with partial-wave amplitudes,
\begin{equation}
\Gamma(\mathcal{B}_{c}\to\mathcal{B}_{8}V)=\frac{p_{c}(E_{f}+m_{f})}{4\pi m_{i}} \Big[ 2 \big( \left|S\right|^{2}+\left|P_{2}^{2}\right| \big)+\frac{E_{V}^{2}}{m_{V}^{2}} \big( \left|S+D\right|^{2}+\left|P_{1}\right|^{2} \big) \Big] \,,
\end{equation}
with the relations as following,
\begin{align}
S & =-A_{1} \,, \nonumber\\
P_{1} & =-\frac{p_{c}}{E_{V}}\left(\frac{m_{i}+m_{f}}{E_{f}+m_{f}}B_{1}+B_{2}\right) \,,\nonumber\\
P_{2} & =\frac{p_{c}}{E_{f}+m_{f}}B_{1} \,,\nonumber\\
D & =-\frac{p_{c}^{2}}{E_{V}(E_{f}+m_{f})}(A_{1}-A_{2}) \,.
\end{align}

\subsection{Long-distance contributions with final-state rescattering mechanism}
In this section, we concentrate on the calculation of the long-distance contributions for the topology $C,C^\prime,E_1,E_2$ and $B$.
The long-distance contributions are considerably important in charmed
baryon decays, but it is notoriously difficult to study final-state interaction
effects (FSIs) in a systematical way as it is non-perturbative in nature. 

The final-state rescatterings provide a natural physical picture for the long-distance contributions of the heavy hadron decays. 
Wolfenstein and Sukuzi proposed the formalism of final-state interaction at the hadronic level based on $CPT$ invariance and unitarity~\cite{Wolfenstein:1990ks,Suzuki:1999uc}. Cheng, Chua and Soni performed a comprehensive study on $B$-meson two-body decays considering FSI effects~\cite{Cheng:2004ru}. 
Using a time evolution picture~\cite{Cheng:2020ipp}, the short-distance interactions happen rapidly and violently at the beginning of weak decays, while the long-distance interactions happen at a much later time. The full amplitude is expressed as~\cite{Cheng:2020ipp}
\begin{align}
\mathcal{A}=\mathcal{S}^{1/2}\mathcal{A}_0,
\end{align}
where $\mathcal{A}_0$ is free from any strong phase, and the $S$-matrix $\mathcal{S}^{1/2}$ is a time-evolution operator $U(\infty, 0)$. In the picture of final-state rescattering, the above formula could be expressed as $\mathcal{A}_i=\sum_j \mathcal{S}^{1/2}_{ij} \mathcal{A}_{0j}$. 
At a short time after the weak interaction, the quarks and gluons are good degrees of freedom, so that $\mathcal{A}_{0j}$ corresponds to the factorizable contributions given in the above subsection whose strong phase is vanishing. Then the states $j$ are rescattered into the final states $i$ via $\mathcal{S}^{1/2}_{ij}$ which usually contribute the non-zero strong phases.
The derivation and detailed explanations are referred to~\cite{Cheng:2020ipp,Chua:2007cm,Suzuki:1999uc}. 
It has been manifested that the final-state rescatterings are very important in the CP violation of three-body $B$ meson decays\cite{Suzuki:1999uc,Dedonder:2010fg,Bediaga:2013ela,Cheng:2013dua} and also in the CPV of charmed meson decays\cite{Buccella:2019kpn,Bediaga:2022sxw,Pich:2023kim}. 

\begin{figure}[h]
    \centering
    \begin{minipage}{0.24\linewidth}
    \centering
        \includegraphics[scale=0.4]{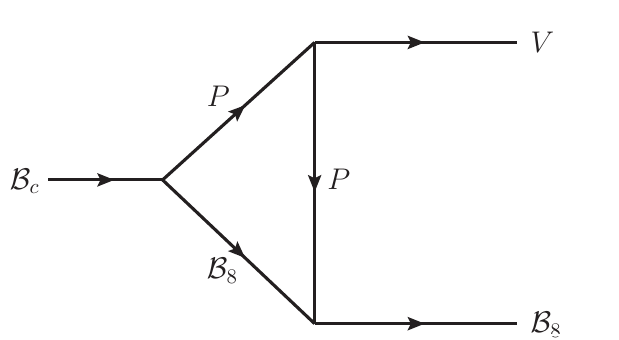}
    \end{minipage}
    \begin{minipage}{0.24\linewidth}
    \centering
        \includegraphics[scale=0.4]{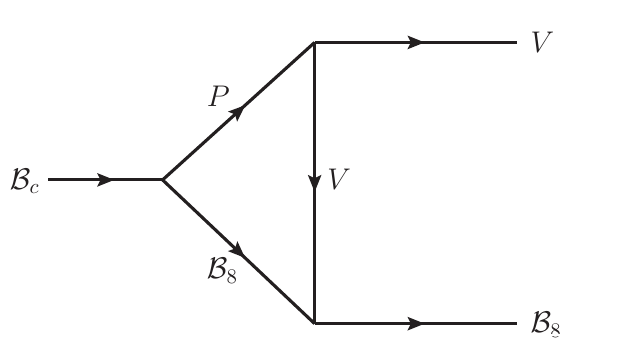}
    \end{minipage}
    \begin{minipage}{0.24\linewidth}
    \centering
        \includegraphics[scale=0.4]{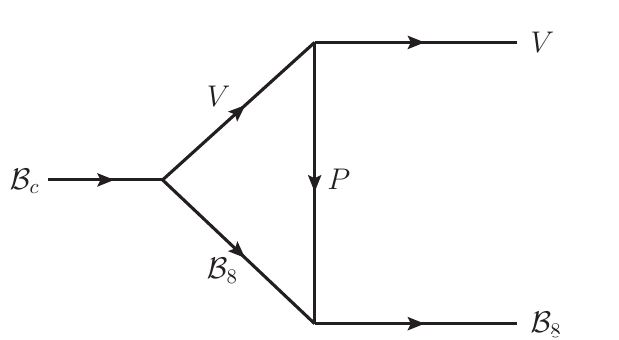}
    \end{minipage}    
    \begin{minipage}{0.24\linewidth}
    \centering
        \includegraphics[scale=0.4]{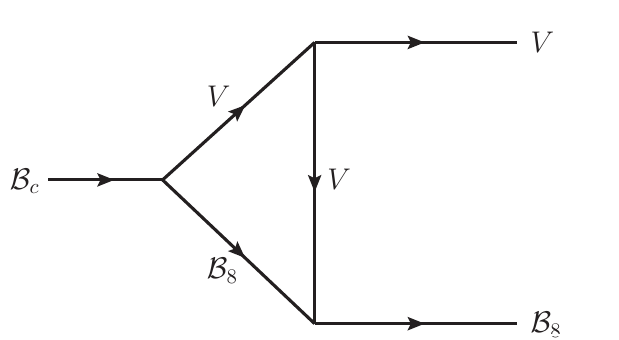}
    \end{minipage}    

    \begin{minipage}{0.24\linewidth}
    \centering
        \includegraphics[scale=0.4]{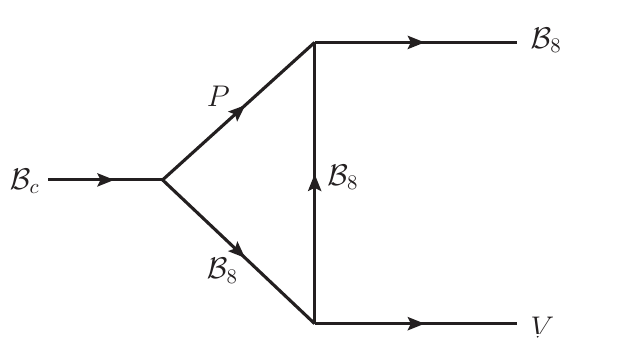}
    \end{minipage}
    \begin{minipage}{0.24\linewidth}
    \centering
        \includegraphics[scale=0.4]{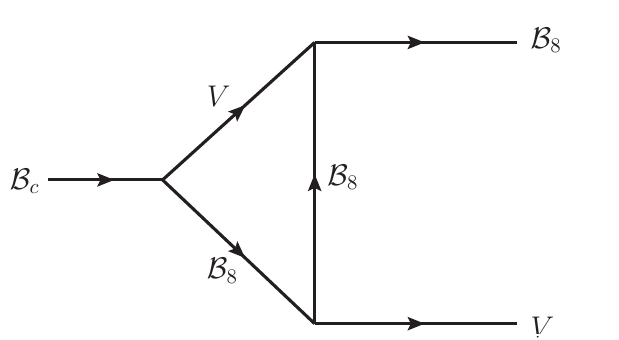}
    \end{minipage}  
    \begin{minipage}{0.24\linewidth}
    \centering
        \includegraphics[scale=0.4]{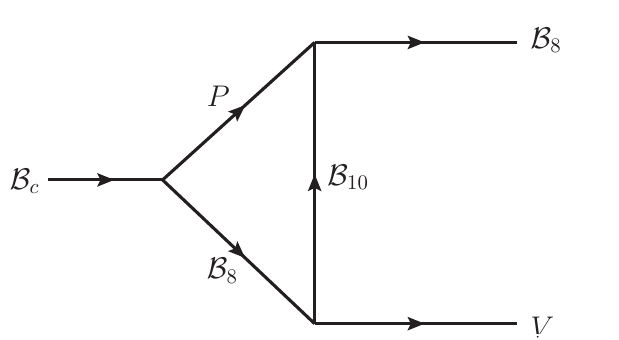}
    \end{minipage}
    \begin{minipage}{0.24\linewidth}
    \centering
        \includegraphics[scale=0.4]{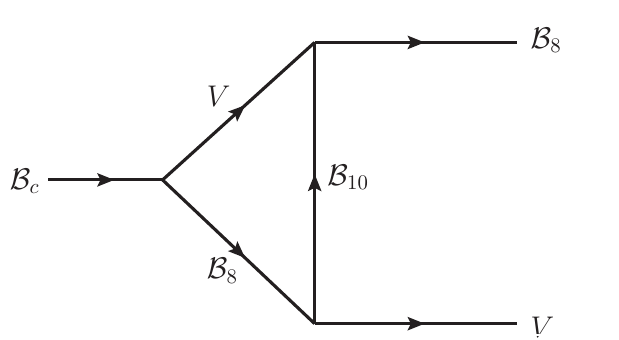}
    \end{minipage}
  
    \caption{The long-distance re-scattering contributions to $\Lambda_c^+\to\mathcal{B}_8V$ at hadron level.}
    \label{fig:triangle-diagram}
\end{figure}

Following the previous studies in \cite{Cheng:2004ru,Cheng:2010ry}, we can gain some control on re-scattering
effects by studying them in a phenomenological way, that is the FSIs can
be treated approximately as the one-particle-exchange processes at the hadron level. 
Under this mechanism, the charmed baryon two-body decays are described by triangle or bubble diagrams, the weak transitions are followed by the hadronic particle exchange/annihilation process.
The triangle diagrams for the description of FSIs in charmed baryon decays are listed in Fig.\ref{fig:triangle-diagram}.
 It has been argued that the two-body scatterings dominate the FSIs, while the three-body and four-body intermediate states are negligible or relatively small \cite{Magalhaes:2011sh,Garrote:2022uub} and we assume that two-body to n-body scatterings are negligible due to the limited phase space. We only consider the $2\to2$ scattering processes in present calculation.
%

In order to calculate the diagrams in Fig.\ref{fig:triangle-diagram}, we need to combine the formula of the weak transition vertex (described by the weak effective Hamiltonian and naive factorization) and the hadronic scattering amplitudes (governed by the effective hadronic Lagrangian which are collected in the App.\ref{App.B}).
To avoid double counting, the two-body weak transition amplitudes of charmed baryon only involve the short-distance factorizable contributions which we have introduced in last section.
For estimating the hadronic scattering amplitudes, we need the following Feynman rules derived from the matrix element of the effective hadronic Lagrangian with definite initial and final states,
\begin{align}\label{eq:strong vertex1}
&\langle V(k,\lambda_k)P(p_{2})|i\mathcal{L}_{VPP}|P(p_{1})\rangle=-ig_{VVP}\varepsilon^{*\mu}(k,\lambda_k)(p_{1}+p_{2})_{\mu}\,,\\
&\langle\mathcal{B}_{2}(p_{2})P(q)|i\mathcal{L}_{PBB}|\mathcal{B}_{1}(p_{1})\rangle=g_{BBP}\bar{u}(p_{2})i\gamma_{5}u(p_{1})\,,\\
& \langle\mathcal{B}_{2}(p_{2})V(q,\lambda_q)|i\mathcal{L}_{VBB}|\mathcal{B}_{1}(p_{1})\rangle=\bar{u}(p_{2})\left[f_{1}\gamma_{\nu}+f_{2}\frac{i}{m_{1}+m_{2}}\sigma_{\mu\nu}q^{\mu}\right]\varepsilon^{*\nu}(q,\lambda_q)u(p_{1})\,,  \\
&\langle V(p_{3},\lambda_{3})V(k,\lambda_{k})|i\mathcal{L}_{VVP}|P(p_{1})\rangle=-i\frac{g_{VVP}}{f_{p}}\epsilon^{\mu\nu\alpha\beta}p_{3\,\mu}\varepsilon^{*}_{\nu}(\lambda_{3},p_{3})k_{\alpha}\varepsilon^{*}_{\beta}(k,\lambda_{k})\,,\\
&\langle B(p_4,\lambda_{4})|i\mathcal{L}_{VBD}|D(k,\lambda_{k})V(p_{1},\lambda_{1})\rangle =-i\frac{g_{\rho N\Delta}}{m_{\rho}}\bar{u}(p_{4},\lambda_{4})\gamma^{5}\gamma^{\nu}u^{\mu}(k,\lambda_{k}) \nonumber\\
&\hspace{6.2cm} \times \left[p_{1\,\mu}\varepsilon_{\nu}(p_1,\lambda_1)-p_{1\,\nu}\varepsilon_{\mu}(p_1,\lambda_1)\right]\,, \\
& \langle B(p_{4,\lambda_{4}})|i\mathcal{L}_{VBD}|D(k,\lambda_{k})P(p_{1},\lambda_{1})\rangle
=\frac{g_{\pi N\Delta}}{m_{\pi}} \bar{u}(p_{4},\lambda_{4})p_{1\,\mu}u^{\mu}(k,\lambda_{k})\,,
\end{align}
\begin{align}
\label{eq:strong vertex2}
& \langle V(p_{3},\lambda_{3})V(k,\lambda_{k})|i\mathcal{L}_{VVV}|V(p_{1},\lambda_{1})\rangle
= -\frac{ig_{VVV}}{\sqrt{2}}\varepsilon_{\mu}(p_{1},\lambda_{1})\varepsilon^{\mu*}(p_{3},\lambda_{3})\varepsilon^{*}_{\nu}(k,\lambda_{k})\left(p^{\nu}_{3}+p^{\nu}_{1}\right)\nonumber\\
        & \hspace{6.2cm} -\frac{ig_{VVV}}{\sqrt{2}}\varepsilon^{*}_{\mu}(k,\lambda_{k})\varepsilon^{\mu}(p_{1},\lambda_{1})\varepsilon^{*}_{\nu}(p_{3},\lambda_{3})\left(-p^{\nu}_{1}-p^{\nu}_{k}\right)\nonumber\\
        & \hspace{6.2cm} -\frac{ig_{VVV}}{\sqrt{2}}\varepsilon^{*}_{\mu}(p_{3},\lambda_{3})\varepsilon^{*\mu}(k,\lambda_{k})\varepsilon_{\nu}(p_{1},\lambda_{1})\left(p^{\nu}_{k}-p^{\nu}_{3}\right) .
\end{align}

Then, we arrive at the analytical expression of the eight hardonic loop diagrams by applying the factorization formula of the weak transition amplitudes and the above Feynman rules of the hadronic couplings. For simplicity, we unify the symbols of momentum and spin polarization through following rule $\Lambda^{+}_{c}(p_{i},s_{i})\to \mathcal{B}_{8}(p_{3},s_{3})V(p_{4},\varepsilon_{4})$ with re-scattering particle $\mathcal{B}_{8}(p_{1},s_{1})P(p_{2})/V(p_{2},\varepsilon_{2})$ and exchanged one $P(k),V(k,\varepsilon_{k}),B/D(k,s_{k})$.
Under this convention, we obtain the amplitudes of the eight triangle diagrams with the loop integral variable $k$,
\begin{align}
\mathcal{\mathcal{M}}[P,\mathcal{B}_8;P] = & -i\int\frac{d^{4}k}{(2\pi)^{4}}g_{BBP}\cdot g_{VPP}\bar{u}(p_{4},s_{4})\gamma_{5}(\slashed{p}_2+m_{2})(A+B\gamma_{5})u(p,s) \nonumber \\
 &\times\varepsilon^{*}_{\mu}(p_{3},\lambda_{3})(p_{1}+k)^{\mu}
 \frac{\mathcal{F}}{(p_{1}^{2}-m_{1}^{2}+i\epsilon)(p_{2}^{2}-m_{2}^{2}+i\epsilon)(k^{2}-m_{k}^{2}+i\epsilon)}\,,
\\
\mathcal{\mathcal{M}}[P,\mathcal{B}_8;V] = & -\int\frac{d^{4}k}{(2\pi)^{4}}\frac{4g_{VVP}}{f_{P}}\bar{u}(p_{4},s_{4})\left[f_{1BBV}\gamma_{\delta}-\frac{if_{2BBV}}{m_{2}+m_{4}}\sigma_{\sigma\delta}k^{\sigma}\right] \nonumber \\
&\times(A+B\gamma_{5})u(p,s)\epsilon_{\mu\nu\alpha\delta}p_{3}^{\mu}\varepsilon^{*\nu}(p_3,\lambda_3)k^{\alpha}(\slashed{p}_2+m_{2}) \nonumber \\
&\times\frac{\mathcal{F}}{(p_{1}^{2}-m_{1}^{2}+i\epsilon)(p_{2}^{2}-m_{2}^{2}+i\epsilon)(k^{2}-m_{k}^{2}+i\epsilon)}\,,
\\
\mathcal{\mathcal{M}}[V,\mathcal{B}_8;P]= & i\int\frac{d^{4}k}{(2\pi)^{4}}\frac{4g_{VVP}}{f_{p}}g_{BBP}\bar{u}(p_{4},s_{4})\gamma_{5}(\slashed{p}_2+m_{2}) \nonumber \\
&\times\left(A_{1}\gamma_{\delta}\gamma_{5}+A_{2}\frac{p_{2\delta}}{m_{i}}\gamma_{5}+B_{1}\gamma_{\delta}+B_{2}\frac{p_{2\delta}}{m_{i}}\right)\epsilon_{\mu\nu\alpha\delta}p_{3}^{\mu}\varepsilon^{*\nu}(p_3,\lambda_3)p_{1}^{\alpha}u(p,s) \nonumber \\
&\times\frac{\mathcal{F}}{(p_{1}^{2}-m_{1}^{2}+i\epsilon)(p_{2}^{2}-m_{2}^{2}+i\epsilon)(k^{2}-m_{k}^{2}+i\epsilon)}\,,
\end{align}
\begin{align}
\mathcal{\mathcal{M}}[V,\mathcal{B}_8;V]= & -i\int\frac{d^{4}k}{(2\pi)^{4}}\bar{u}(p_{4},s_{4})(f_{1BBV}\gamma_{\delta}-\frac{if_{2BBV}}{m_{2}+m_{4}}\sigma_{\sigma\delta}k^{\sigma})(\slashed{p}_2+m_{2}) \nonumber \\
&\times\left(A_{1}\gamma^{\alpha}\gamma_{5}+A_{2}\frac{p_{2}^{\alpha}}{m_{i}}\gamma_{5}+B_{1}\gamma^{\alpha}+B_{2}\frac{p_{2}^{\alpha}}{m_{i}}\right)u(p,s) \nonumber \\
 &\times g_{VVV}\left[\left(-g^{\delta\mu}+\frac{k^{\delta}k^{\mu}}{m_k^{2}}\right)\left(-g_{\mu\alpha}+\frac{p_{1\mu}p_{1\alpha}}{m_{1}^{2}}\right)\varepsilon_{\nu}^{*}(p_{3},\lambda_3)(k+p_{1})^{\nu}\right. \nonumber \\
 &\left.+\varepsilon_{\mu}^{*}(p_{3},\lambda_3)\left(-g^{\delta\mu}+\frac{k^{\delta}k^{\mu}}{m_k^{2}}\right)\left(-g_{\nu\alpha}+\frac{p_{1\nu}p_{1\alpha}}{m_{1}^{2}}\right)(p_{3}-k)^{\nu}\right. \nonumber \\
 & \left.+\left(-g_{\mu\alpha}+\frac{p_{1\mu}p_{1\alpha}}{m_{1}^{2}}\right)\varepsilon^{\mu*}(p_{3},\lambda_3)\left(-g^{\delta\nu}+\frac{k^{\delta}k^{\nu}}{m_k^{2}}\right)(-p_{1}-p_{3})_{\nu}\right] \nonumber \\
 &\times\frac{\mathcal{F}}{(p_{1}^{2}-m_{1}^{2}+i\epsilon)(p_{2}^{2}-m_{2}^{2}+i\epsilon)(k^{2}-m_{k}^{2}+i\epsilon)}\,, \\
\mathcal{\mathcal{M}}[P,\mathcal{B}_8;\mathcal{B}_8]= & -i\int\frac{d^{4}k}{(2\pi)^{4}}g_{BBP}\bar{u}(p_{4},s_{4})\gamma_{5}(\slashed{k}+m_{k})\left[f_{1BBV}\gamma_{\nu}+\frac{if_{2BBV}}{m_{2}+m_{k}}\sigma_{\mu\nu}p_{3}^{\mu}\right] \nonumber \\
&\times\varepsilon^{\nu*}(p_{3},\lambda_3)(\slashed{p}_2+m_{2})(A+B\gamma_{5})u(p,s) \nonumber \\
&\times\frac{\mathcal{F}}{(p_{1}^{2}-m_{1}^{2}+i\epsilon)(p_{2}^{2}-m_{2}^{2}+i\epsilon)(k^{2}-m_{k}^{2}+i\epsilon)}\,,
\\
\mathcal{\mathcal{M}}[V,\mathcal{B}_8;\mathcal{B}_8]= & i\int\frac{d^{4}p_{1}}{(2\pi)^{4}}\bar{u}(p_{4},s_{4})\left[f_{11BBV}\gamma_{\delta}-\frac{if_{12BBV}}{m_{4}+m_{k}}\sigma_{\sigma\delta}p_{1}^{\sigma}\right](\slashed{k}+m_{k}) \nonumber\\
&\times\left[f_{21BBV}\gamma_{\nu}+\frac{if_{22BBV}}{m_{2}+m_{k}}\sigma_{\mu\nu}p_{3}^{\mu}\right]\left(-g^{\delta\alpha}+\frac{p_{1}^{\delta}p_{1}^{\alpha}}{m_{1}^{2}}\right) \nonumber\\
&\times\varepsilon^{*\nu}(p_{3},\lambda_3)(\slashed{p}_2+m_{2})
(A_{1}\gamma_{\alpha}\gamma_{5}+A_{2}\frac{p_{2\alpha}}{m_{i}}\gamma_{5}+B_{1}\gamma_{\alpha}+B_{2}\frac{p_{2\alpha}}{m_{i}})u(p,s) \nonumber\\
 &\times\frac{\mathcal{F}}{(p_{1}^{2}-m_{1}^{2}+i\epsilon)(p_{2}^{2}-m_{2}^{2}+i\epsilon)(k^{2}-m_{k}^{2}+i\epsilon)} \,,
\end{align}
\begin{align}
 \mathcal{M}[P,\mathcal{B}_{8},\mathcal{B}_{10}]
= & i\frac{g_{\rho\Delta N}g_{\pi\Delta N}}{m_{\rho}m_{\pi}}\int\frac{d^{4}k}{(2\pi)^{4}}\bar{u}(p_{4},\lambda_{4})p_{1,\mu}P^{\mu\alpha}\gamma^{5}\gamma^{\beta}(\not{p}_{2}+m_{2}) \nonumber\\
&\times\left[p_{3\alpha}\varepsilon^{*}_{\beta}(p_3,\lambda_{3})-p_{3\beta}\varepsilon^{*}_{\alpha}(p_3,\lambda_{3})\right]\left[A+B\gamma_{5}\right]u(p_{i},\lambda_{i}) \nonumber\\
&\times\frac{\mathcal{F}}{(p_{1}^{2}-m_{1}^{2}+i\epsilon)(p_{2}^{2}-m_{2}^{2}+i\epsilon)(k^{2}-m_{k}^{2}+i\epsilon)}\,,
\\
 \mathcal{M}[V,\mathcal{B}_{8},\mathcal{B}_{10}]
 = & \frac{g_{\rho\Delta N1}g_{\rho\Delta N2}}{m_{\rho1}m_{\rho2}}\int\frac{d^{4}k}{(2\pi)^{4}}\bar{u}(p_{4},\lambda_{4})\gamma^{5}\gamma^{\nu}P^{\mu\beta}\gamma^{5}\gamma^{\sigma}(\not{p}_{2}+m_{2}) \nonumber \\
&\times\left[A_{1}\gamma_{\alpha}\gamma_{5}+A_{2}\frac{p_{2\alpha}}{m_{i}}\gamma_{5}+B_{1}\gamma_{\alpha}+B_{2}\frac{p_{2\alpha}}{m_{i}}\right](-g_{\alpha\nu}p_{1\mu}+g_{\alpha\mu}p_{1\nu}) \nonumber \\
&\times(p_{3\beta}\varepsilon^{*}_{\sigma}(p_3,\lambda_{3})-p_{3\sigma}\varepsilon^{*}_{\beta}(p_3,\lambda_{3})) \nonumber \\
&\times\frac{\mathcal{F}}{(p_{1}^{2}-m_{1}^{2}+i\epsilon)(p_{2}^{2}-m_{2}^{2}+i\epsilon)(k^{2}-m_{k}^{2}+i\epsilon)}\,,
\end{align}
where $P^{\mu\alpha}$ denotes $\sum_{s}u_{\mu}(p,s)\bar{u}_{\alpha}(p,s)$ and the monopole form factor
\begin{equation}
    \mathcal{F}=\left(\frac{\Lambda_{1}^{2}-m_{1}^{2}}{\Lambda_{1}^{2}-p_{1}^{2}}\right)\left(\frac{\Lambda_{2}^{2}-m_{2}^{2}}{\Lambda_{2}^{2}-p_{2}^{2}}\right)\,,
\end{equation}
is to regulate the possible divergence in the loop integrals.
For completing the calculation, we need the following spin summation formula 
\begin{align}
\sum_{s}u(p,s)\bar{u}(p,s) & =\slashed{p}+m\,,\\
\sum_{s}u_{\mu}(p,s)\bar{u}_{\nu}(p,s) & =(\slashed{p}+m) \left\{ -g_{\mu\nu}+\frac{\gamma_{\mu}\gamma_{\nu}}{3}+\frac{2p_{\mu}p_{\nu}}{3m^{2}}-\frac{p_{\mu}\gamma_{\nu}-p_{\nu}\gamma_{\mu}}{3m} \right\}\,, 
\end{align}
for $\frac{1}{2}$ and $\frac{3}{2}$ spinor respectively and the polarization summation of massive vector meson
\begin{align}
\sum_{\lambda_{1}}\varepsilon^{*\rho}(p_{1},\lambda_{1})\varepsilon^{\nu}(p_{1},\lambda_{1})&=-g^{\rho\nu}+\frac{p_1^{\rho}p_1^{\nu}}{m^{2}_{1}} .
\end{align}

In principle, the sum over all allowed intermediate two-body states is required by the unitarity and completeness of theory.
However, as aforementioned, the contributions of multi-body states and much higher excited states in the intermediate state are suppressed due to the constraints of phase space. 

In addition, it is also difficult to calculate the contributions from all orders. 
Therefore, in practical applications, we focus on the most important contributions of the intermediate processes and include as many computable contributions as possible in order to more accurately describe the impact of the intermediate states.

\begin{figure}[tb]
  \centering
  \includegraphics[width=0.5\textwidth]{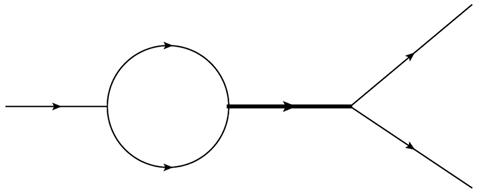} \\
  \caption{Final state interaction effects with $s$-channel.}\label{fig:FSI-S}
\end{figure}

It should be noted that the final-state interactions not only involve the above-mentioned $t/u$ channel single-particle exchange contributions, but also have the $s$ channel contributions from resonance states which is formed as the bubble diagram in Fig.\ref{fig:FSI-S}.
For $b$-hadron decays, the contributions from $s$ channel can be neglected safely, the reason is that the mass of $b$-hadron is above $5$GeV, and the light hadrons and charmed hadrons produced in this decay do not have such high excited states. In charm decays, there are several light hadron excited states at the energy region of charmed hadron. However, the excited states have large width in general, and the width effects of high excited states in propagator can not be neglected,   
\begin{equation}
    S\sim\frac{1}{p^2-m^2+im\Gamma}\,,
\end{equation}
hence, the contribution of $s$-channel for charmed hadron decay will be suppressed by width effects of resonance states. 

In essence, incorporating all conceivable intermediate states into final state interaction calculations proves unfeasible. 
The reliability of strong coupling inputs diminishes notably for higher excited states, resulting in increased uncertainties. 
Even if these states contribute, excessive uncertainty undermines the validity of our calculations. 
Therefore, our focus remains on $S$-wave states exclusively, currently disregarding $s$-channel contributions. 
This means our analysis includes only pseudoscalar and vector mesons, along with octet and decuplet baryons in triangle diagrams.

\section{Analytical discussions}

\subsection{Cutkosky cutting rule and Feynman loop integral}
In the traditional approach \cite{Cheng:2004ru}, the amplitude of the triangle diagram is usually calculated by utilizing optical theorem and Cutkosky cutting rule to obtain the imaginary part. The real part of the triangle diagram can be derived from the dispersion relation in principle, but due to the large ambiguity, the amplitude in the whole region cannot be reliably described, so this effect is ignored in many works. Therefore, the traditional method is not thorough in calculating the triangle diagram and therefore loses the information

\begin{figure}[tb]
  \centering
  \includegraphics[width=0.45\textwidth]{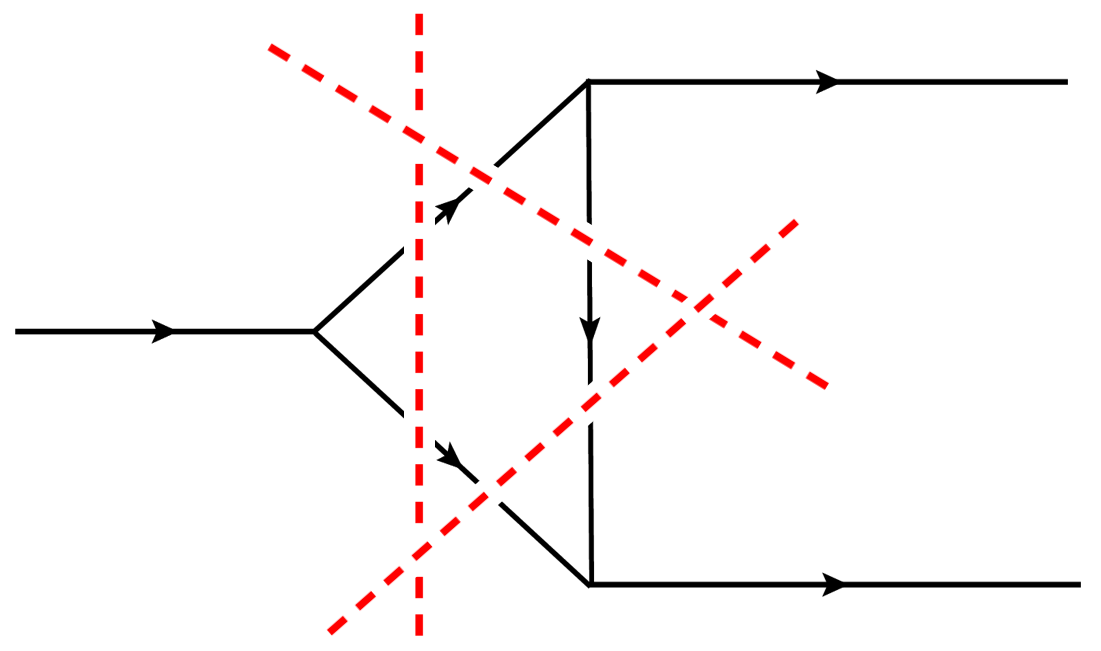}
  \caption{The Cutkosky cutting rules of three-point master integrals.}\label{fig:cutting rule:all}
\end{figure}

In order to obtain the complete amplitude of the triangle diagram, the most natural approach is to calculate the loop integrals. 
In this work, we adopt the the Passarino-Veltman (PV) method to reduce the tensor integrals and the method of integral by parts (IBP) to obtain the expressions with master integrals. 
Then we obtain the numerical results for the master integrals by means of Looptools.
In the following, we present the proof that the imaginary part of a general three-point master integral (displayed in Fig.\ref{fig:cutting rule:all}) which is directly calculated from the analytical expression is consistent with the one by summing over all of cutting ways which is required by Cutkosky cutting rule.
For simplicity, we adopt the convention $p^2=(p_3+p_4)^2=m_i^2,~p_3^2=p_4^2=m_f^2,~m_1^2=m_2^2=m_k^2=m_1^2$. The imaginary part of the the three-point master integral is obtained from the directly computed analytical expression by integrating out the loop-momentum $k$,
\begin{align}\label{eq:prove:1}
  &\text{Im}(-i)\int\frac{d^4 k}{(2\pi)^4}\frac{1}{k^2-m_1^2+i\varepsilon}
  \frac{1}{(k-p_1)^2-m_1^2+i\varepsilon}\frac{1}{(k-p_1-p_2)^2-m_1^2+i\varepsilon}
  \nonumber \\
  =&-\frac{1}{16\pi^2}\,\text{Im}\int_0^1 d\,x\int_0^{1-x} \frac{d\,y}
  {m_i^2\,x^2+m_f^2\,y^2+m_i^2\,x\,y-m_i^2\,x-m_f^2\,y+m_1^2-i\varepsilon} \nonumber \\
  =&-\frac{1}{16\pi}\int_0^1 d\,x\int_0^{1-x}d\,y\,
  \delta(m_i^2\,x^2+m_f^2\,y^2+m_i^2\,x\,y-m_i^2\,x-m_f^2\,y+m_1^2) \nonumber \\
  =&\left\{ 
  \begin{array}{l}
  \frac{1}{16\pi m_i\sqrt{(m_i^2-4 m_f^2)}}\ln \biggl{[} \frac{m_i^2-2 m_f^2-
  \sqrt{(m_i^2-4 m_f^2)(m_i^2-4 m_1^2)}}{m_i^2-2 m_f^2+
  \sqrt{(m_i^2-4 m_f^2)(m_i^2-4 m_1^2)}}\left( \frac{m_i m_f+
  \sqrt{(m_i^2-4 m_f^2)(m_f^2-4 m_1^2)}}{m_i m_f-
  \sqrt{(m_i^2-4 m_f^2)(m_f^2-4 m_1^2)}} \right)^2 \biggr{]}\\
  \hspace{10.4cm} \forall\, m_1 \in (0, \frac{m_f}{2} ) \,,\\
  \frac{1}{16\pi m_i\sqrt{m_i^2-4 m_f^2}}\ln\frac{m_i^2-2 m_f^2-
  \sqrt{(m_i^2-4 m_f^2)(m_i^2-4 m_1^2)}}{m_i^2-2 m_f^2+
  \sqrt{(m_i^2-4 m_f^2)(m_i^2-4 m_1^2)}}
  \hspace{2.64cm} \forall\, m_1\in  (\frac{m_f}{2},\frac{m_i}{2})
  \,, \\ \vspace{1mm} \\   0
  \hspace{10.25cm} \forall\, m_1\in  (\frac{m_i}{2},\infty)  \,,
  \end{array} 
  \right.
\end{align}
and the one calculated from the Cutkosky cutting rule is,
\begin{align}\label{eq:prove:2}
  &\text{Im}(-i)\int\frac{d^4 k}{(2\pi)^4}\frac{1}{k^2-m_1^2+i\varepsilon}
  \frac{1}{(k-p_1)^2-m_1^2+i\varepsilon}\frac{1}{(k-p_1-p_2)^2-m_1^2+i\varepsilon}
  \nonumber \\
  =&P\int\frac{d^4 k}{8\pi^2}\biggl{[} \frac{\theta[k_0]\delta(k^2-m_1^2)
  \theta[p_{10}+p_{20}-k_0]\delta((p_1+p_2-k)^2-m_1^2)}{(k-p_1)^2-m_1^2+i\varepsilon} \nonumber \\
  &+\frac{\theta[k_0]\delta(k^2-m_1^2)\theta[p_{10}-k_0]\delta((p_1-k)^2-m_1^2)}
  {(k-p_1-p_2)^2-m_1^2+i\varepsilon} \nonumber \\
  &+\frac{\theta[k_0-p_{10}]\delta((k-p_1)^2-m_1^2)\theta[p_{10}+p_{20}-k_0]
  \delta((p_1+p_2-k)^2-m_1^2)}{k^2-m_1^2+i\varepsilon} \biggr{]} \nonumber \\
  =&\frac{\theta[m_i-2 m_1]}{16\pi m_i\sqrt{m_i^2-4 m_f^2}}
  \ln\left|\frac{m_i^2-2m_f^2-\sqrt{(m_i^2-4 m_f^2)(m_i^2-4 m_1^2)}}{m_i^2-2m_f^2+
  \sqrt{(m_i^2-4 m_f^2)(m_i^2-4 m_1^2)}}\right| \nonumber \\
  &+\frac{\theta[m_f-2m_1]}{8\pi m_i\sqrt{m_i^2-4 m_f^2}}
  \ln\left|\frac{m_i m_f+\sqrt{(m_i^2-4 m_f^2)(m_f^2-4 m_1^2)}}
  {m_i m_f-\sqrt{(m_i^2-4 m_f^2)(m_f^2-4 m_1^2)}}\right|\,.
\end{align}
based on the properties of the Theta function, it is clear that the results in first line of Eq.\eqref{eq:prove:2} satisfy the range $m_1\in  (0,\frac{m_i}{2})$, it includes both the region $m_1\in  (0,\frac{m_f}{2})$ and the  region $m_1\in  (\frac{m_f}{2},\frac{m_i}{2})$, and the results in second line only satisfy the range $m_1\in  (0,\frac{m_f}{2})$. Then, it is not difficult to transform the results in Eq.\eqref{eq:prove:2} into the results in Eq.\eqref{eq:prove:1}. Hence, we can demonstrate the results in Eq.\eqref{eq:prove:1} and Eq.\eqref{eq:prove:2} are consistent with each other.  
The corresponding proof can be extended to the case of any other one-loop master integrals and even higher-loop diagrams~\cite{Zhou:2004gm,Andreas:1995ap}.

\subsection{Pauli-Villars regularization and model form factors}
In our calculation, the hadronic couplings are essential inputs and usually are estimated with the initial and final hadron states on-shell.
In the triangle diagram, the hadronic couplings need to be modified with the effects that the internal particles have off-shell momentum. This is the first reason to introduce the form factor parameterized as follows,
\begin{align}\label{eq:monopole FF}
\mathcal{F}=\left(\frac{m^2\pm\Lambda^2}{p^2\pm\Lambda^2}\right)^n,\quad \Lambda=m+\eta\Lambda_{QCD}\,,
\end{align}
where the signs $``+/-"$ are corresponding to the s-channel and t/u-channel respectively, $m$ and $p$ are the mass and momentum of propagator and $\Lambda_{QCD}=330$MeV is the nonperturbative scale in charm quark decay. 
It has been argued that the monopole behavior of the form factor with $n=1$ is much more consistent with the results that expected by the QCD sum rule~\cite{Gortchakov:1995im}.

The second reason is that we need to introduce an appropriate regularization scheme to regulate the inevitable divergence appearing in the master integrals of the triangle diagram amplitude. The introduced form factor in Eq.\eqref{eq:monopole FF} is consistent with the Pauli-Villars regularization scheme.
We present some details for deriving the form factor from the Pauli-Villars scheme in the following. The idea is to introduce a heavy particle that coupled to the propagator, that is the propagator can be rewritten as,
\begin{align} \label{eq:PV_reg}
\frac{1}{k^{2}-m^{2}+i\epsilon} 
  \Rightarrow ~& \frac{1}{k^{2}-m^{2}+i\epsilon}+\frac{a_{1}}{k^{2}-\Lambda_{1}^{2}+i\epsilon}\nonumber \\
  \Rightarrow ~& \frac{k^{2}-\Lambda_{1}^{2}+a_{1}(k^{2}-m^{2})}{(k^{2}-m^{2})(k^{2}-\Lambda_{1}^{2})}\,,
\end{align}
where $\Lambda_1\gg m^2$ is the cutoff scale, $a_1$ is an undetermined coefficient. When we take the value of $a_1$ is equal to $-1$, the superficial degree of divergence of the integral is obviously reduced and the expression in Eq.\eqref{eq:PV_reg} can be simplified as,
\begin{equation}
\frac{1}{k^{2}-m^{2}+i\epsilon}\to\frac{m^{2}-\Lambda_{1}^{2}}{(k^{2}-m^{2})(k^{2}-\Lambda_{1}^{2})}=\frac{1}{k^{2}-m^{2}}\frac{m^{2}-\Lambda_{1}^{2}}{k^{2}-\Lambda_{1}^{2}}\,,
\end{equation}
where the factor $(m^{2}-\Lambda_{1}^{2})/(k^{2}-\Lambda_{1}^{2})$ is nothing but the monopole form factor we introduced in Eq.\eqref{eq:monopole FF}. 
If we encounter integrals with higher divergences, we can address this by introducing couplings with heavier particles. However, we won't delve into this approach here. The cutoff scale $\Lambda$ is chosen to follow the same parameterized form as in the conventional method \cite{Cheng:2004ru}, 
\begin{equation}\label{eq:FF:Lambda}
\Lambda=m+\eta\Lambda_{QCD}\,,
\end{equation}
where $m$ represents the mass of the propagator, the only parameter is $\eta$, which can be determined from experimental data.

\section{Numerical results and discussion}
In this section, we present the determination of the various input parameters and the numerical results for the branching ratios, decay asymmetry parameters and $CP$ violations of the $\Lambda_c \to \mathcal{B}_8 V$ channels.
\subsection{Inputs}
We take the hadron masses and the decay constants of pseudoscalar mesons from the Particle Data Group \cite{ParticleDataGroup:2022pth}. 
The decay constants of vector mesons are obtained from the comprehensive analysis in \cite{Ball:2006eu}. 
The values for all of the decay constants are listed in Table.\ref{tab:decay_constant}.
\begin{table}[tb]
\renewcommand{\arraystretch}{1.5}
\addtolength{\arraycolsep}{3pt}
	\centering
	\caption{ The decay constants of pseudoscalar \cite{ParticleDataGroup:2022pth} and vector \cite{Ball:2006eu} mesons.} \label{tab:decay_constant}
	\begin{tabular}{c|c|c|c|c|c|c} 
		\hline
		\hline
Decay constant &  $f_\pi$ & $f_K$  &  $f_\rho$  & $f_\omega$  & $f_\phi$ &  $f_{K^*}$    \\
\hline
Value [MeV] &  $ 130.2\pm0.12 $ & $ 155.7\pm0.3 $ &  $ 216\pm3  $ & $  187\pm5 $ & $ 215\pm5 $ & $220\pm5$ \\ 
		\hline
		\hline 
	\end{tabular} 
\end{table}

In the calculation of short-distance contributions, the weak transition form factors of $\Lambda^{+}_{c}$ to octet baryons are necessary inputs and have been investigated in the several QCD methods, such as Lattice QCD (LQCD) \cite{Detmold:2016pkz,Meinel:2016dqj,Meinel:2017ggx,Bahtiyar:2021voz}, light-front quark model (LFQM) \cite{Zhao:2018zcb,Li:2021qod}, light-cone sum rules (LCSR) \cite{Liu:2009sn}, covariant quark model \cite{Gutsche:2015rrt} and relativistic quark model \cite{Faustov:2018dkn}.
Considering that the transition form factors from LQCD have higher accuracy, we use the calculation results with higher-order fit in \cite{Meinel:2016dqj,Meinel:2017ggx}, see Table.\ref{tab:weak_form_factors}.
\begin{table}[tb]
\renewcommand{\arraystretch}{1.5}
\addtolength{\arraycolsep}{6pt}
	\centering
	\caption{The transition form factors of $\Lambda^{+}_{c}$ to octet baryons from the LQCD method \cite{Meinel:2016dqj,Meinel:2017ggx} and evolved to the physical region via BCL parameterization with high-order fit.}\label{tab:weak_form_factors}
\begin{tabular}{ccccc} 
		\hline
		\hline
Decay   \ \ \ \ \ \ \ \ \ & $f_{1}(0)$  &\ \ \ \ \ \ \ \ \  $f_{2}(0)$ \ \ \ \ \ \ \ \ \ & $g_{1}(0)$ &\ \ \ \ \ \ \ \ \  $g_{2}(0)$ \ \ \ \ \ \ \ \ \ \\ 
\hline
$ \Lambda^{+}_{c}\to \Lambda$ \cite{Meinel:2016dqj} &  $ 0.655\pm0.022 $ & $ -0.307\pm0.042 $ & $0.578\pm0.015$ & $0.002\pm0.030$ \\
$ \Lambda^{+}_{c}\to N$ \cite{Meinel:2017ggx} & $ 0.662\pm0.039 $ & $ -0.322\pm0.050 $ & $0.599\pm0.031$ & $-0.002\pm0.053$ \\
\hline
\hline
	\end{tabular} 
\end{table}

In addition, the strong coupling constants are crucial non-perturbative parameters in our calculation. 
For the strong couplings between octet baryons and light mesons, we use their values obtained with the method of LCSRs under the $SU(3)$ flavor symmetry \cite{Aliev:2006xr,Aliev:2009ei}. 
The couplings among decuplet baryons, octet baryons and light mesons are obtained from the experiment data about $NN$ potential \cite{Janssen:1996kx}, that is $g^{2}_{\Delta N\pi}/(4\pi)=0.36$ and $g^{2}_{\Delta N\rho}/(4\pi)=20.45$, the rest of this type of couplings are derived from $SU(3)$ flavor symmetry.
All of the three kinds of meson couplings are determined based on the exact $SU(3)$ flavor symmetry and the value of the couplings $g_{\rho\pi\pi}$, $g_{\rho\rho\rho}$, $g_{\omega\rho\pi}$. 
The coupling constant $g_{\rho\pi\pi} = 6.05 \pm 0.02$ quantifies how the $\rho$ meson interacts with two pions, extracted from experimental data \cite{Cheng:2004ru}. 
The hidden local symmetry theory relates the $\rho\rho\rho$ coupling constant, $g_{\rho\rho\rho}$, to the $\rho$ meson mass ($m_\rho$) and the pion physical decay constant ($F_\pi = 93$MeV), given by $g_{\rho\rho\rho} = \frac{m\rho}{2 F_\pi}$ \cite{Oset:2010tof}. 
Additionally, the coupling constant $g_{\omega \rho \pi}$ can be expressed as $g_{\omega \rho \pi} = \frac{3}{16 \pi^2} g_{\rho\rho\rho}^2$, describing the interaction between the $\omega$ meson, $\rho$ meson, and a pion \cite{Bramon:1994pq}.

\subsection{Branching ratios}
By summing the squares of all helicity amplitudes labeled by the spins of the initial and final states, and averaging over the initial state particle spins, we obtain the formula for the decay width
\begin{equation}\label{eq:width:helicity}
\Gamma(\mathcal{B}_{c}\to\mathcal{B}_{8}V)=\frac{p_{c}}{8\pi m_{i}^{2}}\enskip\frac{1}{2}\enskip\sum_{\lambda_i\lambda_f\lambda_V}\left| H_{\lambda_f,\,\lambda_V}^{\lambda_i} (\mathcal{B}_{c}\to\mathcal{B}_{8}V)\right|^{2} \,,
\end{equation}
where $p_c=\frac{1}{2m_i}\sqrt{\left[m_i^2-(m_f^2+m_V^2)\right]\left[m_i^2-(m_f^2-m_V^2)\right]}$, $m_i$ and $m_f$ are the masses of the initial and final state baryons,  $m_V$ is the mass of vector mesons. $\lambda_i,\,\lambda_f$ and $\lambda_V$ label the helicity of the initial, final state baryon and the vector meson respectively. 
Then dividing by the total width of $\Lambda_c$ we can get the branching ratio for each channel.

With the inputs discussed in last section, we obtain the branching fractions of $\Lambda_c \to \mathcal{B}_8 V$ as shown in Table.\ref{tab:BR}, where the decay modes are classified into three categories (Cabbibo favoured (CF), single Cabbibo suppressed (SCS) and double Cabbibo suppressed (DCS)) according to CKM matrix element.
\begin{table}[tb]
\renewcommand{\arraystretch}{1.5}
\addtolength{\arraycolsep}{3pt}
	\centering
	\caption{The branching ratio of $\Lambda_c^+\to\mathcal{B}_8V$ processes with $\eta=0.6\pm0.1$.}\label{tab:BR}
	\begin{tabular}{cccccc} 
		\hline\hline
Decay modes   &  Topology  & $\mathcal{B}R_{\text{SD}}(\%)$  &   $\mathcal{B}R_{\text{LD}}(\%)$  &  $\mathcal{B}R_{\text{tot}}(\%)$ &  $\mathcal{B}R_{\text{exp}}(\%)$ \\ \hline
$ \Lambda^{+}_{c}\to \Lambda^0\rho^{+}$ & $ T,C',E_2,B  $ &  $ 6.12 $& $ 2.30^{+1.18}_{-1.94} $& $6.26^{+2.44}_{-1.39}$& $4.06\pm0.52$ \\	
$ \Lambda^{+}_{c}\to \Sigma^+\rho^{0}$ & $ C',E_2,B $ &  $-$ & $ - $ & $0.77^{+1.38}_{-0.53} $& $<1.7 $ \\
$ \Lambda^{+}_{c}\to \Sigma^+\omega$ & $ C',E_2,B $ &  $ - $ & $ - $ & $2.06^{+0.40}_{-1.78}$& $1.7\pm0.21$ \\
$ \Lambda^{+}_{c}\to \Sigma^+\phi$ & $ E_1 $ &  $ - $ & $ - $ & $0.33^{+0.08}_{-0.29}$& $0.39\pm0.06$ \\
$\Lambda^{+}_{c}\to p\bar{K}^{*0}$ & $ C,E_1  $ &  $ 3.26\times10^{-3} $& $ 3.76^{+1.37}_{-3.43} $& $3.70^{+1.29}_{-3.39}$&$1.96\pm0.27$ \\
$\Lambda^{+}_{c}\to \Xi^0 K^{*+}$ & $ E_2,B  $ &  $ - $ & $ - $ & $1.94^{+0.40}_{-1.68}$&  $ - $\\\hline\hline
Decay modes   &  Topology  & $\mathcal{B}R_{\text{SD}}(\times10^{-3})$  &   $\mathcal{B}R_{\text{LD}}(\times10^{-3})$  &  $\mathcal{B}R_{\text{tot}}(\times10^{-3})$  &  $\mathcal{B}R_{\text{exp}}(\times10^{-3})$\\ \hline
$ \Lambda^{+}_{c}\to \Lambda^0 K^{*+}$ & $ T,C',E_2,B$ & $ 2.92 $& $ 2.78^{+1.28}_{-1.02} $& $4.71^{+0.48}_{-0.20}$& $ - $\\
$ \Lambda^{+}_{c}\to \Sigma^0 K^{*+}$ & $ C',E_2,B $ & $ - $ & $ - $ & $1.60^{+0.89}_{-0.62}$&  $ - $\\
$ \Lambda^{+}_{c}\to \Sigma^+ K^{*0}$ & $  C',E_1 $ & $ - $ & $ - $ & $2.10^{+1.37}_{-0.86}$&  $3.5\pm1.0$\\
$ \Lambda^{+}_{c}\to p\phi$ & $ C $ &  $ 1.78\times10^{-3} $ & $ 1.44^{+1.14}_{-0.66} $& $1.37^{+1.13}_{-0.65}$&  $1.06\pm0.14$\\
$\Lambda^{+}_{c}\to p\omega$ & $ C,C',E_1,E_2,B  $ &  $ 1.48\times10^{-3} $ & $ 1.28^{+0.46}_{-0.37} $& $1.26^{+0.45}_{-0.37}$& $0.83\pm0.11$\\
$\Lambda^{+}_{c}\to p\rho^0$ & $ C,C',E_1,E_2,B $ &  $ 1.81\times10^{-3} $ & $ 2.79^{+1.89}_{-1.29} $& $2.72^{+1.87}_{-1.27}$&  $ 1.52\pm0.44 $\\
$ \Lambda^{+}_{c}\to n\rho^+$ & $ T,C',E_2,B  $ &  $ 7.14 $& $ 8.50^{+9.57}_{-4.72} $& $26.39^{+13.71}_{-8.86}$& $ - $
 \\\hline\hline 
Decay modes   &  Topology  & $\mathcal{B}R_{\text{SD}}(\times10^{-4})$  &   $\mathcal{B}R_{\text{LD}}(\times10^{-4})$  &  $\mathcal{B}R_{\text{tot}}(\times10^{-4})$  &  $\mathcal{B}R_{\text{exp}}$\\ \hline
$ \Lambda^{+}_{c}\to p K^{*0}$ & $ C,C'  $ &  $ 9.28\times10^{-4} $ & $ 0.53^{+3.67}_{-0.38} $& $0.55^{+3.71}_{-0.39}$& $ - $\\	
$ \Lambda^{+}_{c}\to n K^{*+}$ & $ T,C' $ &  $ 3.66 $ & $ 0.44^{+1.64}_{-0.30} $& $5.08^{+1.95}_{-0.66}$&  $ - $\\\hline\hline
	\end{tabular} 
\end{table}

Based on the results in Table \ref{tab:BR}, it is evident that the relative magnitudes of the factorizable contributions from the $T$ and $C$ diagrams are consistent with the relative sizes of their corresponding effective Wilson coefficients, in accordance with theoretical expectations. 
Moreover, for decay processes with the same decay mode, i.e., involving the same CKM matrix element, the decay branching ratios of processes that include $T$ diagram and those that do not are of the same order of magnitude. 
This underscores the significant role of non-factorizable contributions in charmed baryon decays, which cannot be ignored. 
Additionally, for several processes with identical topological diagrams, the variations in their branching ratios are primarily due to the CKM matrix elements, further quantitatively confirming the universality of the topological diagram approach.

Using the process $\Lambda_c^+ \to \Lambda^0 \rho^+$ as an example, where short-distance contributions dominate, the decay width comprises both short and long-distance contributions, with the former being predominant. Consequently, the decay branching ratio exhibits a weak dependence on $\eta$. However, as $\eta$ increases, the long-distance contribution also gradually increases, eventually becoming comparable to the short-distance contribution. At this point, the dependence of the decay branching ratio on the $\eta$ parameter becomes stronger. The accompanying graph illustrates the dependence of the decay branching ratio on $\eta$ for this process.

\begin{figure}[tb]
  \centering
  \includegraphics[width=0.4\textwidth]{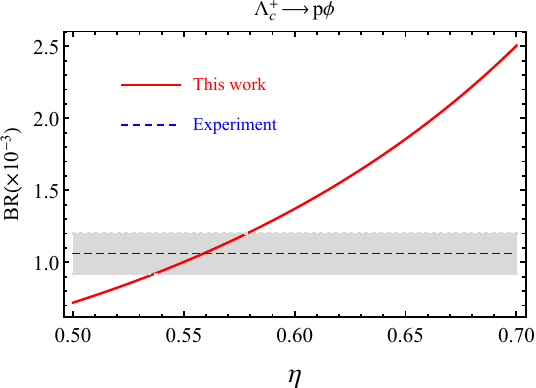}
  \includegraphics[width=0.4\textwidth]{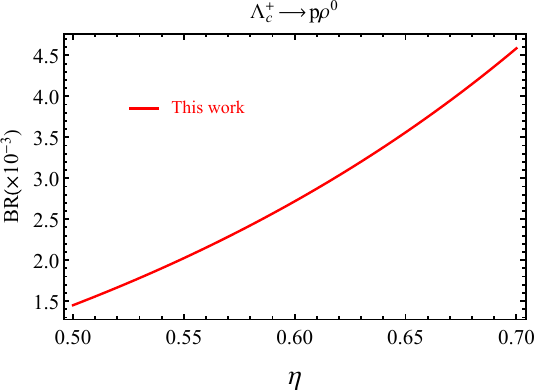}
  \caption{The dependence of branching ratio on $\eta$ for decay mode $\Lambda_c^+\to p\phi$ and $p\rho^0$.}\label{fig:LambRho:BR:eta}
\end{figure}
At $\eta=0.6$, the contribution from long-distance effects rises to a point where it matches the contribution from short-distance effects. This results in a notable shift in the $\eta$ dependence.
The $\eta$ parameter strongly influences the decay branching ratios across various processes. However, the sensitivity of the ratio of these branching ratios to $\eta$ diminishes considerably.

In addition to the model parameter $\eta$, the observables of interest are influenced by non-perturbative input parameters such as heavy-to-light form factors and strong coupling constants. These parameters are typically determined using non-perturbative methods or extracted from experimental data, but they can carry significant uncertainties. 
The heavy-to-light form factors are particularly crucial as they appear in both short-distance and long-distance contributions. Their impact on the decay branching ratios, such as in the $\Lambda_c^+ \to \Lambda^0 \rho^+$ process, is substantial. 
To illustrate, consider the dependence of the decay branching ratio on two significant $\Lambda_c \to \Lambda$ transition form factors, denoted as $f_1$ and $g_1$. Graphically depicting this dependence helps visualize how variations in these form factors affect the overall branching ratio of the decay process. This sensitivity underscores the importance of accurately determining these form factors to improve the precision of theoretical predictions and to interpret experimental data effectively.
\begin{figure}[tb]
  \centering
  \includegraphics[width=0.4\textwidth]{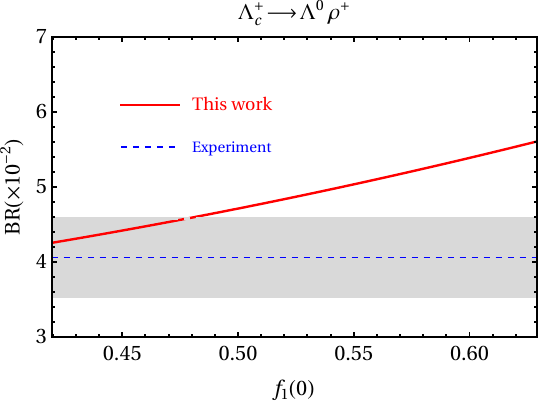}
  \includegraphics[width=0.4\textwidth]{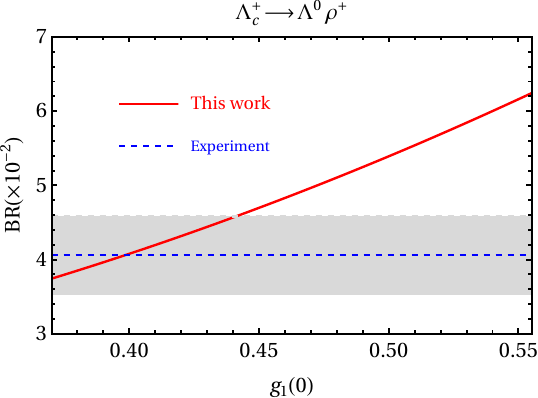}
  \caption{The dependence of branching ratio on heavy to light form factors for decay mode $\Lambda_c^+\to\Lambda^0\rho^+$.}\label{fig:LambRho:BR:FF}
\end{figure}
In Fig.\ref{fig:LambRho:BR:FF}, it is evident that the heavy-to-light form factor $g_1$ exerts a stronger influence on the decay branching ratio compared to $f_1$, especially when considering variations within a 20\% error margin. These form factors, derived from lattice QCD calculations, show discrepancies particularly in the low $q^2$ region where they tend to exceed experimental measurements. These discrepancies introduce significant uncertainties into the predicted decay branching ratio.
Therefore, improving the precision of these form factor determinations, especially to better align with experimental data across different momentum transfer ranges, is crucial for enhancing the reliability of theoretical predictions for processes like $\Lambda_c^+ \to \Lambda^0 \rho^+$. This alignment is essential for reducing uncertainties and improving the reliability of theoretical models in describing such decay processes accurately.

The strong coupling constant plays a crucial role in theoretical calculations, especially in processes involving long-distance effects. 
Variations in its values from different theoretical approaches, as highlighted in \cite{Aliev:2009ei}, can lead to significant uncertainties, including the possibility of differences in sign. These variations directly translate into uncertainties in the predicted decay branching ratios of particles.
In particular, the impact of the strong coupling constant is more pronounced for vertices involving vector particles compared to those involving pseudoscalar particles. 
This difference arises primarily due to the larger contributions from vector particle-related triangle diagrams in the decay processes.
To illustrate this, let's consider the example of the $\Lambda_c^+ \to \Sigma^+ \phi$ decay process, where we analyze the effects of the strong coupling constants $D_2$ and $F_2$ associated with the $\mathcal{B}_8\mathcal{B}_8 V$ vertex. 
When we adopt a $50\%$ uncertainty in these coupling constants, the branching ratio of this decay process can vary significantly—up to 6 to 7 times. 
This sensitivity underscores the importance of accurately determining these strong coupling constants to reduce uncertainties in theoretical predictions.
Thus, improving the precision of these non-perturbative parameters is crucial for refining theoretical models and ensuring they align closely with experimental observations, thereby enhancing our understanding of particle decay dynamics.

In certain decay channels, such as $\Lambda_c^+ \to \Sigma^+ \phi$ and $\Lambda_c^+ \to p \phi$, each involving a single topological diagram (denoted as $E_1$ and $C$ respectively), the computed amplitude magnitude reflects the size of these diagrams. This characteristic allows for extracting information about the topological diagram itself. After normalizing for the effects of CKM matrix elements specific to the decay processes, the relative sizes of the $E_1$ and $C$ diagrams are determined to lie between $0.5^{+0.2}_{-0.1}$ when the parameter $\eta$ varies from $0.5$ to $0.7$. This finding is consistent with analyses conducted using the SCET method. Additionally, the branching fractions of $\Lambda_c^+\to p\omega$ and $\Lambda_c^+\to p\rho^0$ determined from LHCb collaboration \cite{LHCb:2024djr} are consistent with our calculations, as shown in Table \ref{tab:BR}.

\subsection{Decay parameters}
\begin{figure}[tb]
  \centering
  \includegraphics[width=0.5\textwidth]{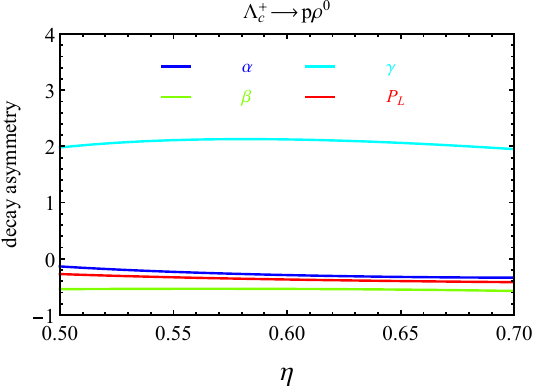}
  \caption{The dependence of decay asymmetry parameters on $\eta$ for decay mode $\Lambda_c^+\to p\rho^0$.}\label{fig:pRho:decay asy:eta}
\end{figure}

The asymmetry parameters in the decay process $\Lambda_c^+ \to \mathcal{B}_8 V$ quantify differences in the magnitudes squared of various helicity amplitudes. This approach helps reduce the influence of the $\eta$ parameter. Specifically, this decay involves four distinct non-zero helicity amplitudes. The BESIII experiment, as defined in the measurement methodology outlined by Hong et al. (2022)~\cite{Hong:2022prk}, defines several types of asymmetry parameters for measurement. These parameters play a crucial role in analyzing how decay events distribute across the final-state particles $\mathcal{B}_8$ and $V$, providing valuable insights into the angular correlations and dynamics of strong interactions within this decay process,
\begin{align}
\alpha  =\text{\ensuremath{\frac{\left|H_{1,\frac{1}{2}}\right|^{2}-\left|H_{-1,-\frac{1}{2}}\right|^{2}}{\left|H_{1,\frac{1}{2}}\right|^{2}+\left|H_{-1,-\frac{1}{2}}\right|^{2}}}},\quad
\beta =\text{\ensuremath{\frac{\left|H_{0,\frac{1}{2}}\right|^{2}-\left|H_{0,-\frac{1}{2}}\right|^{2}}{\left|H_{0,\frac{1}{2}}\right|^{2}+\left|H_{0,-\frac{1}{2}}\right|^{2}}}},\quad
\gamma =\text{\ensuremath{\frac{\left|H_{1,\frac{1}{2}}\right|^{2}+\left|H_{-1,-\frac{1}{2}}\right|^{2}}{\left|H_{0,\frac{1}{2}}\right|^{2}+\left|H_{0,-\frac{1}{2}}\right|^{2}}},}
\end{align}
\begin{equation}
P_{L} =\text{\ensuremath{\frac{\left|H_{1,\frac{1}{2}}\right|^{2}-\left|H_{-1,-\frac{1}{2}}\right|^{2}+\left|H_{0,\frac{1}{2}}\right|^{2}-\left|H_{0,-\frac{1}{2}}\right|^{2}}{\left|H_{1,\frac{1}{2}}\right|^{2}+\left|H_{-1,-\frac{1}{2}}\right|^{2}+\left|H_{0,\frac{1}{2}}\right|^{2}+\left|H_{0,-\frac{1}{2}}\right|^{2}}}} \,.
\end{equation}
these parameters are not all independent, they are related as follows,
\begin{equation}\label{eq:asy:relation}
P_{L}=\frac{\beta+\alpha\cdot\gamma}{1+\gamma} \,,
\end{equation}
which can be used to check whether the results are correct. These definitions are related to the helicity amplitudes labeled by the expectation value of final state helicity as mentioned in Eq.\eqref{eq:width:helicity}.
We calculated the decay asymmetry parameters with $\eta=0.6\pm0.1$ and presented in Table.\ref{Tab:asy}.
\begin{table}[tb]
\renewcommand{\arraystretch}{1.5}
\addtolength{\arraycolsep}{3pt}
	\centering
	\caption{The decay asymmetry parameters of $\Lambda_c^+\to\mathcal{B}_8V$ processes with $\eta=0.6\pm0.1$.}\label{Tab:asy}
	\begin{tabular}{ccccc} 
		\hline\hline
~~~Decay modes~~~   &  ~~~~~~~~~$\alpha$~~~~~~~~~  &  ~~~~~~~~~$\beta$~~~~~~~~~  &  ~~~~~~~~~$\gamma$~~~~~~~~~  &  ~~~~~~~~~$P_L$~~~~~~~~~  \\ \hline
$ \Lambda^{+}_{c}\to \Lambda^0\rho^{+}$ & $ -0.30^{+0.45}_{-0.40} $&  $ -0.67^{+0.06}_{-0.28} $& $ 0.30^{+0.20}_{-0.19} $& $-0.58^{+0.06}_{-0.28}$\\	
$ \Lambda^{+}_{c}\to \Sigma^+\rho^{0}$ & $ -0.82^{+0.04}_{-0.17}$&  $-0.54^{+0.08}_{-0.02}$& $ 0.74^{+0.95}_{-0.14} $& $-0.66^{+0.05}_{-0.16} $\\
$ \Lambda^{+}_{c}\to \Sigma^+\omega$  & $0.85^{+0.002}_{-0.07} $&  $ 0.58^{+0.12}_{-0.001} $& $ 3.27^{+1.17}_{-1.17}  $& $0.81^{+0.02}_{-0.07} $\\
$ \Lambda^{+}_{c}\to \Sigma^+\phi$ & $ -0.11^{+0.002}_{-0.03} $&  $ 0.47^{+0.11}_{-0.10} $& $ 2.12^{+0.08}_{-0.06} $& $ 0.08^{+0.04}_{-0.05} $\\
$\Lambda^{+}_{c}\to p\bar{K}^{*0}$ & $ 0.15^{+0.01}_{-0.15} $&  $ 0.73^{+0.21}_{-0.52} $& $ 3.15^{+1.35}_{-0.02} $& $0.29^{+0.19}_{-0.12}$\\
$\Lambda^{+}_{c}\to \Xi^0 K^{*+}$ & $ -0.12^{+0.06}_{-0.15}  $&  $  -0.03^{+0.03}_{-0.005}  $& $  1.56^{+0.14}_{-0.03}  $& $ -0.08^{+0.05}_{-0.10} $\\\hline\hline
$ \Lambda^{+}_{c}\to \Lambda^0 K^{*+}$ & $  -0.77^{+0.14}_{-0.08}  $& $  -0.39^{+0.25}_{-0.22}  $& $  1.54^{+0.21}_{-0.27}  $& $ -0.62^{+0.17}_{-0.13} $\\
$ \Lambda^{+}_{c}\to \Sigma^0 K^{*+}$ & $ -0.03^{+0.02}_{-0.01}  $& $  0.31^{+0.04}_{-0.07}  $& $  2.21^{+0.48}_{-0.33}  $& $ 0.08^{+0.04}_{-0.03} $\\
$ \Lambda^{+}_{c}\to \Sigma^+ K^{*0}$ & $ 0.07^{+0.06}_{-0.004}  $& $ 0.40^{+0.02}_{-0.03}  $& $  1.52^{+0.08}_{-0.07}  $& $ 0.20^{+0.02}_{-0.01}  $\\
$ \Lambda^{+}_{c}\to p\phi$ & $ -0.11^{+0.06}_{-0.004} $&  $ -0.16^{+0.12}_{-0.07} $& $ 5.98^{+1.07}_{-0.76} $& $-0.12^{+0.07}_{-0.01}$\\
$\Lambda^{+}_{c}\to p\omega$ & $ 0.31^{+0.14}_{-1.02} $&  $ 0.08^{+0.04}_{-0.05} $& $ 0.12^{+0.17}_{-0.06} $& $ 0.11^{+0.02}_{-0.11}$\\
$\Lambda^{+}_{c}\to p\rho^0$ & $ -0.29^{+0.15}_{-0.05} $&  $ -0.54^{+0.01}_{-0.04} $& $ 2.12^{+0.14}_{-0.17} $& $ -0.37^{+0.10}_{-0.05}$\\
$ \Lambda^{+}_{c}\to n\rho^+$ & $ -0.95^{+0.003}_{-0.004} $&  $ -0.61^{+0.14}_{-0.11} $& $ 0.36^{+0.01}_{-0.01} $& $ -0.70^{+0.10}_{-0.08}$\\\hline\hline 
$ \Lambda^{+}_{c}\to p K^{*0}$ & $ 0.45^{+0.07}_{-0.14} $&  $ -0.27^{+0.48}_{-0.16} $& $ 9.87^{+3.61}_{-3.64} $& $ 0.39^{+0.09}_{-0.18}$\\	
$ \Lambda^{+}_{c}\to n K^{*+}$ & $ -0.89^{+0.29}_{-0.05} $&  $ -0.83^{+0.37}_{-0.10} $& $ 1.04^{+0.34}_{-0.14} $& $ -0.86^{+0.32}_{-0.07}$\\\hline\hline
	\end{tabular} 
\end{table}
The decay asymmetry parameters are the ratios of different linear combinations of helicity amplitudes, and therefore to some extent, they mitigate the dependence on model parameters and input parameters. 
Whether the process is dominated by short-distance contributions or long-distance contributions, the decay asymmetry parameter has a weaker dependence on the $\eta$ parameter compared to the branching ratios. 
The curves in Fig.\ref{fig:pRho:decay asy:eta} show the behavior of the four asymmetry parameters as the $\eta$ parameter varies.
It can be clearly seen that the asymmetry parameters show a slight dependency on the $\eta$ parameter. Theoretical calculations of these parameters for $\Lambda_c^+ \to \Lambda^0 \rho^+$ closely match experimental measurements, indicating that the computed relative magnitudes and phases of helicity amplitudes are consistent with observations.

Similar to how the asymmetry parameters respond to changes in the $\eta$ parameter, their sensitivity to variations in the heavy-to-light form factors is notably reduced. The graph illustrates a minimal correlation between the asymmetry parameters and the $f_1$ and $g_1$ form factors during the $\Lambda_c\to\Lambda$ transition, aligning closely with theoretical predictions. 
Likewise, the overall impact of the strong coupling constants on these decay asymmetry parameters is also significantly diminished, indicating a weak dependence on these coupling constants.

\subsection{$CP$ asymmetries}
In this part, we present the numerical predictions on direct $CP$ violation in charmed baryon decays, which is a crucial aspect in understanding the asymmetry between matter and antimatter in the universe. 
The direct $CP$ asymmetry is defined as follows
\begin{align}\label{eq:direct_CPV}
    A^\text{dir}_{CP}(\Lambda_c^+\to p\rho^0) 
    & = \frac{\Gamma(\Lambda_c^+\to p\rho^0)-\bar{\Gamma}(\Lambda_c^+\to p\rho^0)}{\Gamma(\Lambda_c^+\to p\rho^0) + \bar{\Gamma}(\Lambda_c^+\to p\rho^0)} \nonumber\\
    & = \frac{2r\sin\Delta\delta\sin\Delta\phi}{1+r^2+2r\cos\Delta\delta\cos\Delta\phi} \,,   
\end{align}
where $r$ is the ratio of amplitudes, $\Delta\delta$ and $\Delta\phi$ are the strong and weak phase difference respectively. 

Apart from this, there exist $CP$ asymmetries induced by decay asymmetry parameters,
\begin{align} 
\alpha_{CP}  =\frac{\alpha+\bar{\alpha}}{\alpha-\bar{\alpha}},\qquad
\beta_{CP} & =\frac{\beta+\bar{\beta}}{\beta-\bar{\beta}},\qquad
\gamma_{CP}  =\frac{\gamma-\bar{\gamma}}{\gamma+\bar{\gamma}},\qquad
P_{L,CP}  =\frac{P_L-\bar{P}_L}{P_L+\bar{P}_L}.
\end{align}
With the parameter $\eta=0.6\pm0.1$ is adopted, the predictions of $CP$ violations for charmed baryon decays are presented in Table \ref{Tab:CPV}. 
\begin{table}[tb]
\renewcommand{\arraystretch}{1.5}
\addtolength{\arraycolsep}{5pt}
	\centering
	\caption{The $CP$ asymmetries($\times10^{-4}$) of $\Lambda_c^+\to\mathcal{B}_8V$ processes with  $\eta=0.6\pm0.1$.}\label{Tab:CPV}
	\begin{tabular}{cccccc} 
		\hline\hline
~~~Decay modes~~~   & ~~~~~$A^\text{dir}_{CP}$~~~~~ & ~~~~~$\alpha_{CP}$~~~~~  &  ~~~~~$\beta_{CP}$~~~~~  &  ~~~~~$\gamma_{CP}$~~~~~ & ~~~~~$P_{L,CP}$~~~~~ \\ \hline
$ \Lambda^{+}_{c}\to \Lambda^0 K^{*+}$ & $ 0.89^{+0.91}_{-0.61} $& $ -0.84^{+0.62}_{-0.97} $& $  -2.12^{+1.60}_{-8.07} $& $ 0.61^{+0.40}_{-0.31}$& $ -1.07^{+0.77}_{-1.43}$\\
$ \Lambda^{+}_{c}\to \Sigma^0 K^{*+}$ & $ 2.28^{+0.92}_{-0.85} $& $ 13.75^{+18.22}_{-2.24} $& $  2.55^{+2.02}_{-0.76} $& $ -1.00^{+0.18}_{-0.50}$& $ 0.69^{+2.28}_{-0.89}$\\
$ \Lambda^{+}_{c}\to \Sigma^+ K^{*0}$ & $ -1.99^{+0.80}_{-0.74} $& $ 5.51^{+0.06}_{-1.11} $& $  0.95^{+0.02}_{-0.20} $& $ 0.64^{+0.08}_{-0.05}$& $ 1.64^{+0.26}_{-0.18}$\\
$\Lambda^{+}_{c}\to p\omega$ & $ 4.55^{+0.36}_{-0.81} $& $ 19.61^{+8.95}_{-9.35} $& $  -14.80^{-0.30}_{-1.99} $& $ 8.32^{+0.28}_{-8.17}$& $ -2.16^{+0.01}_{-2.21}$\\
$\Lambda^{+}_{c}\to p\rho^0$ & $ 3.73^{+0.95}_{-1.16} $& $ 0.48^{+0.54}_{-0.98} $& $  2.88^{+0.09}_{-0.71} $& $ -1.23^{+0.90}_{-0.42}$& $ 1.77^{+0.20}_{-0.05}$\\
$ \Lambda^{+}_{c}\to n\rho^+$ & $ -1.45^{+0.29}_{-0.52} $& $ 0.01^{+0.32}_{-0.07} $& $  1.86^{+1.34}_{-1.00} $& $ -1.21^{+0.40}_{-0.01}$& $ 1.08^{+0.71}_{-0.59}$\\\hline\hline 
	\end{tabular} 
\end{table}

Due to the significantly suppressed CKM matrix elements associated with penguin operators, this study exclusively considers $CP$ violation induced by tree-level operators. Specifically, for singly Cabibbo suppressed processes, two types of CKM matrix elements are involved: $V_{cd} V_{ud}^*$ and $V_{cs} V_{us}^*$. 
In the rescattering mechanism framework, both types contribute to the same process, generating a non-zero weak phase difference. Moreover, the long-distance final state interactions provide strong phase information for the decay process. 
Thus, our theoretical approach naturally computes the magnitude of $CP$ violation and currently stands as the sole method capable of quantifying $CP$ violation in charmed baryon decays.
With the decay amplitude written as $\mathcal{A}=\lambda_d\mathcal{A}_d+\lambda_s\mathcal{A}_s$, $\lambda_q=V_{cq}V_{uq}^*$, $q=d,s$, the $CP$ violation can then be expressed as 
\begin{align}
A_{CP}^{\rm dir}\approx -2\frac{{\rm Im}(\lambda_d^*\lambda_s)}{|\lambda_d|^2} \frac{{\rm Im}(\mathcal{A}_d^*\mathcal{A}_s)}{|\mathcal{A}_d-\mathcal{A}_s|^2}.
\end{align} 
As displayed in Eq.\eqref{eq:direct_CPV}, the direct $CP$ violation is proportional to $\sin\Delta\phi$, where $\Delta\phi$ represents the weak phase difference, for tree-level operators in charmed baryon decays, $\displaystyle\Delta\phi=\arctan\left[\frac{\text{Im}(V_{cd})}{\text{Re}(V_{cd})}\right] \sim 6\times10^{-4}$. Consequently, the anticipated magnitude of $CP$ violation in charmed baryon decays is expected to align with this level, consistent with the theoretical predictions in Table \ref{Tab:CPV}.

\begin{figure}[!ht]
    \centering
    \begin{minipage}{0.24\linewidth}
    \centering
        \includegraphics[width=1.1\textwidth]{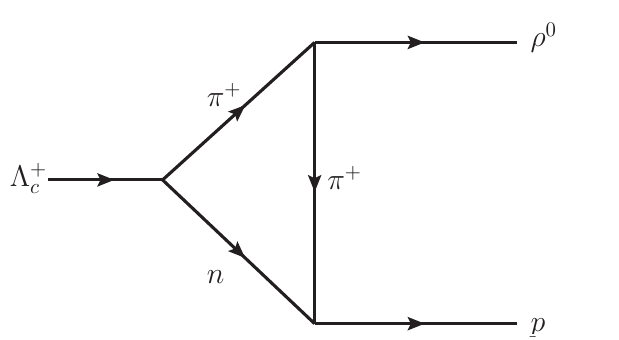}
    \end{minipage}
    \begin{minipage}{0.24\linewidth}
    \centering
        \includegraphics[width=1.1\textwidth]{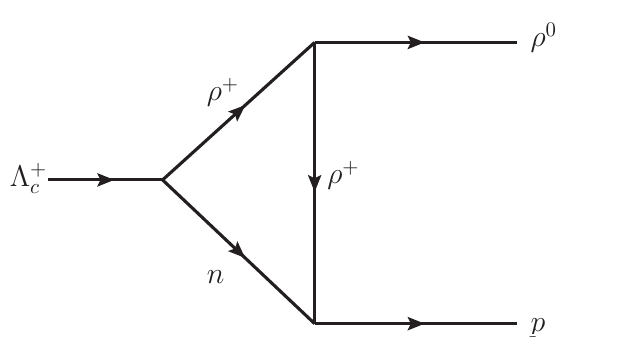}
    \end{minipage}    
    \begin{minipage}{0.24\linewidth}
    \centering
        \includegraphics[width=1.1\textwidth]{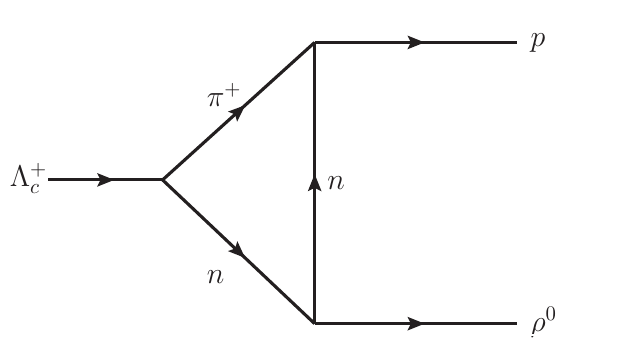}
    \end{minipage}    
        \qquad       
    \begin{minipage}{0.24\linewidth}
    \centering
        \includegraphics[width=1.1\textwidth]{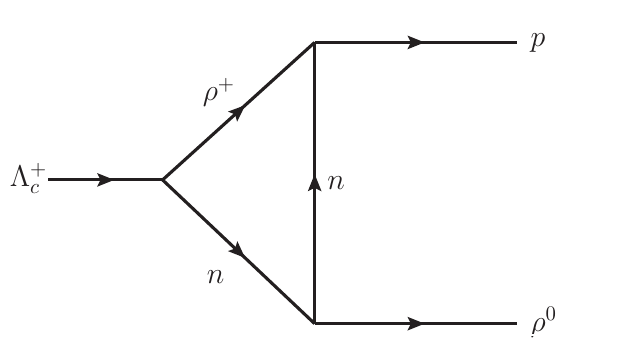}
    \end{minipage}       
    \begin{minipage}{0.24\linewidth}
    \centering
        \includegraphics[width=1.1\textwidth]{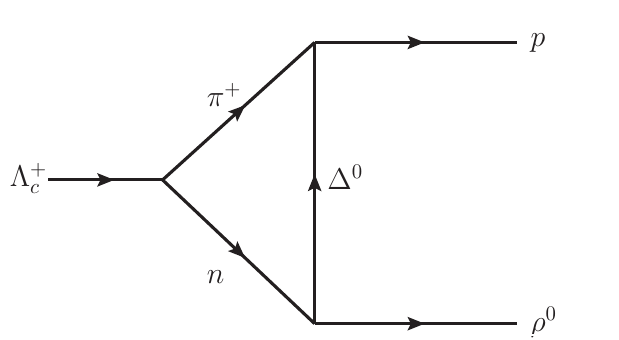}
    \end{minipage}
    \begin{minipage}{0.24\linewidth}
    \centering
        \includegraphics[width=1.1\textwidth]{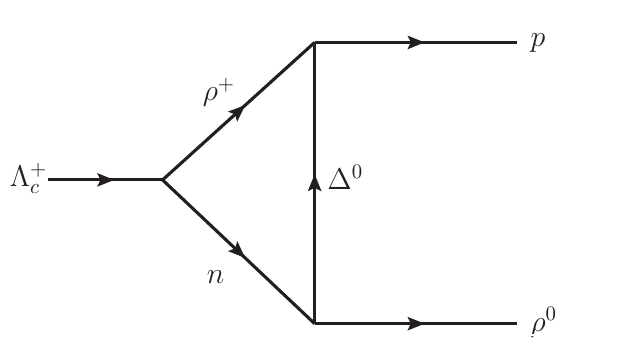}
    \end{minipage} 
    \caption{The triangle diagrams for $\Lambda_c^+\to p\rho^0$ with CKM matrix element $V_{cd}^*V_{ud}$.}
    \label{fig:prho:long1}
\end{figure}

\begin{figure}[tb]
    \flushleft
    \begin{minipage}{0.24\linewidth}
    \centering
        \includegraphics[width=1.1\textwidth]{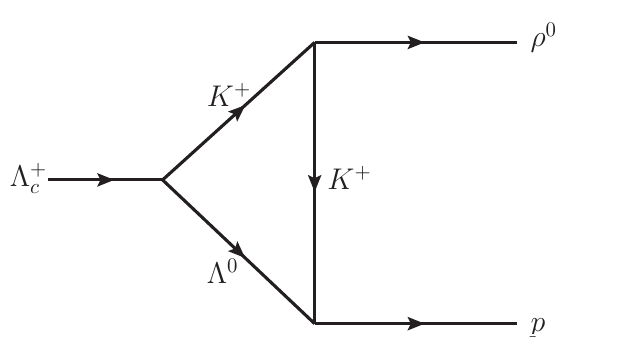}
    \end{minipage}
    \begin{minipage}{0.24\linewidth}
    \centering
        \includegraphics[width=1.1\textwidth]{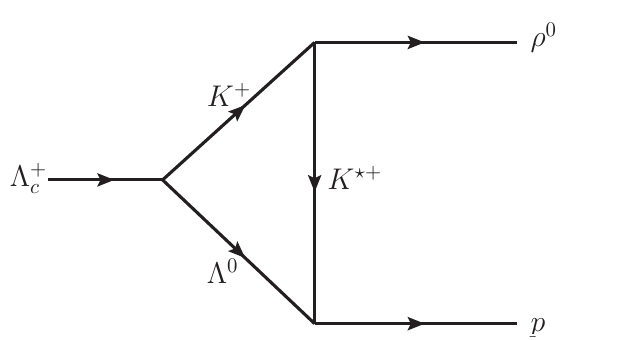}
    \end{minipage}    
    \begin{minipage}{0.24\linewidth}
    \centering
        \includegraphics[width=1.1\textwidth]{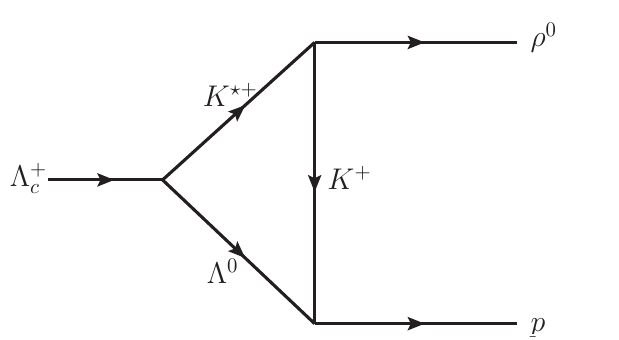}
    \end{minipage}           
    \begin{minipage}{0.24\linewidth}
    \centering
        \includegraphics[width=1.1\textwidth]{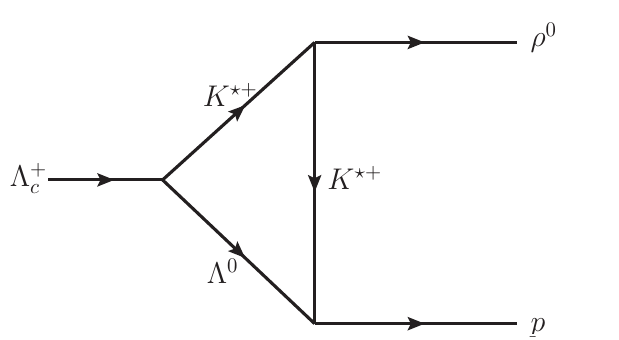}
    \end{minipage}  
    \qquad 
    \begin{minipage}{0.24\linewidth}
    \centering
        \includegraphics[width=1.1\textwidth]{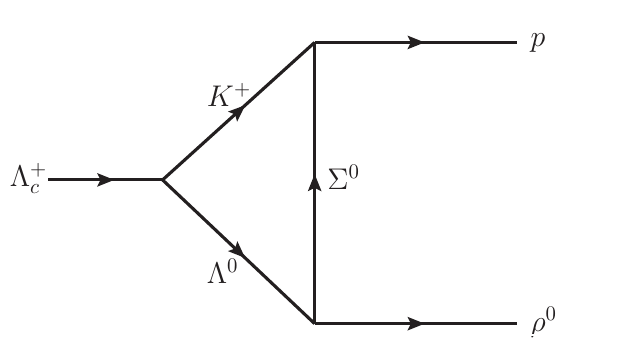}
    \end{minipage}
    \begin{minipage}{0.24\linewidth}
    \centering
        \includegraphics[width=1.1\textwidth]{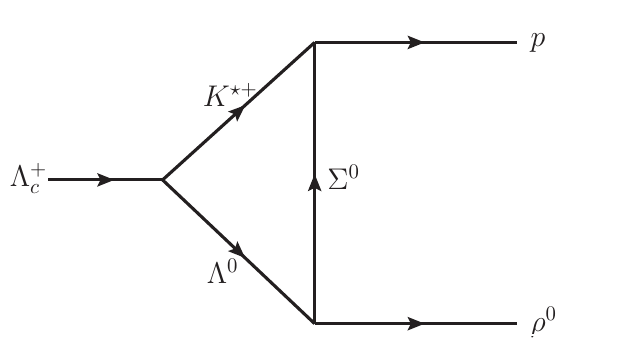}
    \end{minipage}       
    \begin{minipage}{0.24\linewidth}
    \centering
        \includegraphics[width=1.1\textwidth]{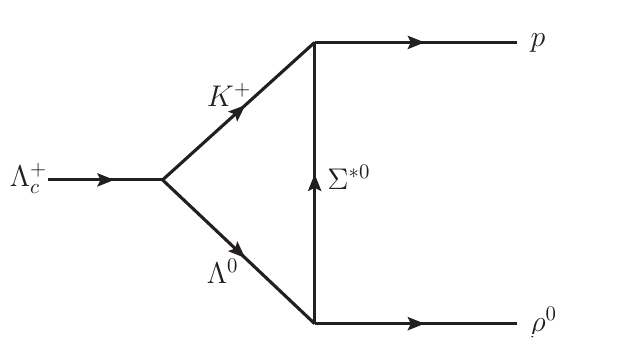}
    \end{minipage}       
    \begin{minipage}{0.24\linewidth}
    \centering
        \includegraphics[width=1.1\textwidth]{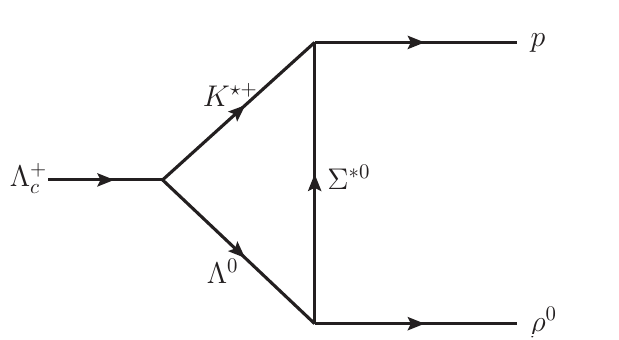}
    \end{minipage}
    \caption{The triangle diagrams for $\Lambda_c^+\to p\rho^0$ with CKM matrix element $V_{cs}^*V_{us}$.}
    \label{fig:prho:long2}
\end{figure}

For instance, the decay channel $\Lambda_c^+ \to p\rho^0$ involves $CP$ violation induced by tree-level operators mediated by CKM matrix elements $V_{cd}V_{ud}$ and $V_{cs} V_{us}$. These elements contribute to the triangle diagram amplitudes depicted in Fig.\ref{fig:prho:long1} and \ref{fig:prho:long2}, influencing the decay amplitude and introducing a weak phase difference $\Delta\phi$. 
This phase difference, combining with the strong phases introduced from the triangle integrals, is critical for understanding and predicting $CP$ violation in charmed baryon decays. 

\begin{figure}[tb]
  \centering
  \includegraphics[width=0.5\textwidth]{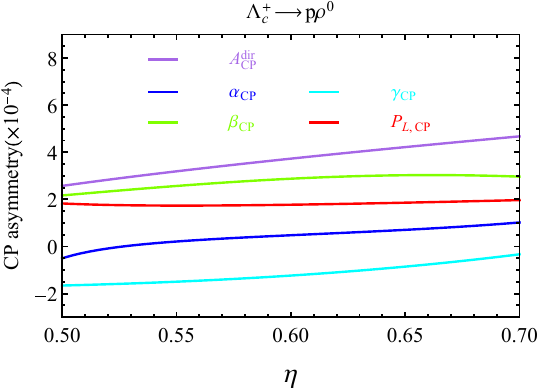}
  \caption{The dependence of $CP$ asymmetries on $\eta$ for decay mode $\Lambda_c^+\to p\rho^0$ .}\label{fig:pRho:CPV:eta}
\end{figure}

To demonstrate the reliability of the theoretical method in predicting the magnitude of $CP$ violation, it is crucial to analyze the dependence of $CP$ violation parameters on model parameters, such as $\eta$. In the decay $\Lambda_c^+ \to p\rho^0$, the variation of $CP$ violation parameters with respect to the $\eta$ parameter is illustrated in Fig.\ref{fig:pRho:CPV:eta}.
This analysis helps establish how sensitive $CP$ violation effects are to changes in the model parameter $\eta$.
The observation that both direct $CP$ violation and asymmetry parameter-induced $CP$ violation exhibit very weak dependence on model parameters. 
It suggests that the predictions for $CP$ violation observables are minimally affected by variations in the underlying theoretical assumptions or model parameters. 
Such findings are crucial for validating the theoretical approach and ensuring that the predicted magnitudes of $CP$ violation are trustworthy and reflective of the physical processes involved in these decays. 

\section{Summary\label{sec:summary}}
In this work, we seriously treated the final-state interaction effects in charmed baryon decay, due to its importance for understanding the dynamics of non-perturbative contributions at the charm scale. The calculations of the long-distance final state interactions are improved with a comprehensive treatment of hadronic-loop diagrams, rather than the Cutkosky rule which computed the imaginary parts of amplitudes barely. In the improved method, the complete real and imaginary parts of the amplitudes have been simultaneously calculated, so that it provides the strong phase to calculate the $CP$ asymmetries. By accurately incorporating long-distance final state interactions, our method allows for precise predictions of decay observables for charmed baryon decays. The results within this method successfully match experimental data of branching ratio across various decay channels such as $\Lambda_c^+\to\mathcal{B}_8V$. Furthermore, the decay asymmetries and $CP$ asymmetries, being defined as ratios, exhibit very weak dependence on model parameters.

Moreover, our approach holds potential beyond charmed baryon decays, potentially extending to the study of weak decays in other heavy-flavor hadrons. Future advancements will focus on refining form factor treatments to further enhance predictive accuracy and reduce remaining model dependencies. This work thus represents a significant step forward in theoretical modeling of heavy-flavor hadron decays and $CP$ violation phenomena.

\begin{acknowledgments}
The authors are grateful to Hai-Yang Cheng and Cai-Dian Lü for the useful discussions on the analysis of final-state interaction. Cai-Ping Jia is also grateful to Jia-Jie Han and Ming-Xiao Duan for the discussions of program.
This work is supported in part by Natural Science Foundation of China under
grant No. 12335003 and 12405114, and by the Fundamental Research Funds for the Central Universities under No. lzujbky-2024-oy02 and lzujbky-2021-it11. 
  \end{acknowledgments}

\appendix
\section{Amplitudes of each modes}
\subsection{Cabibbo-favored}
 \begin{align*}
\mathcal{A}(\Lambda_{c}^{+}\to\Lambda^{0}\rho^{+})= & \mathcal{T}_{SD}(\Lambda_{c}^{+}\to\Lambda^{0}\rho^{+})+\mathcal{M}(\pi^{+},\Lambda^{0};\omega)+\mathcal{M}(\rho^{+},\Lambda^{0};\eta_{8})+\mathcal{M}(\rho^{+},\Lambda^{0};\eta_{1})\\
 +&\mathcal{M}(\pi^{+},\Lambda^{0};\Sigma^{-})+\mathcal{M}(\rho^{+},\Lambda^{0};\Sigma^{-})+\mathcal{M}(\pi^{+},\Lambda^{0};\Sigma^{*-})+\mathcal{M}(\rho^{+},\Lambda^{0};\Sigma^{*-})
 \end{align*}
\begin{align*}
\mathcal{A}(\Lambda_{c}^{+}\to\Sigma^{+}\rho^{0})= & \mathcal{M}(\pi^{+},\Lambda^{0};\pi^{+})+\mathcal{M}(\rho^{+},\Lambda^{0};\rho^{+})+\mathcal{M}(\pi^{+},\Lambda^{0};\Sigma^{0})+\mathcal{M}(\rho^{+},\Lambda^{0};\Sigma^{0})\\
 +&\mathcal{M}(\pi^{+},\Lambda^{0};\Sigma^{*0})+\mathcal{M}(\rho^{+},\Lambda^{0};\Sigma^{*0})
 \end{align*}
 \begin{align*}
\mathcal{A}(\Lambda_{c}^{+}\to\Sigma^{0}\rho^{+})= &\mathcal{M}(\pi^{+},\Lambda^{0};\pi^{0})+\mathcal{M}(\rho^{+},\Lambda^{0};\rho^{0})\\
 +&\mathcal{M}(\pi^{+},\Lambda^{0};\Sigma^{-})+\mathcal{M}(\rho^{+},\Lambda^{0};\Sigma^{-})+\mathcal{M}(\pi^{+},\Lambda^{0};\Sigma^{*-})+\mathcal{M}(\rho^{+},\Lambda^{0};\Sigma^{*-})
 \end{align*}
  \begin{align*}
\mathcal{A}(\Lambda_{c}^{+}\to\Sigma^{+}\omega)= & \mathcal{M}(\pi^{+},\Lambda^{0};\rho^{+})+\mathcal{M}(\rho^{+},\Lambda^{0};\pi^{+})+\mathcal{M}(\pi^{+},\Lambda^{0};\Lambda^{0})+\mathcal{M}(\rho^{+},\Lambda^{0};\Lambda^{0})
\end{align*}
\begin{align*}
\mathcal{A}(\Lambda_{c}^{+}\to\Sigma^{+}\phi)= & \mathcal{M}(\pi^{+},\Lambda^{0};\Lambda^{0})+\mathcal{M}(\rho^{+},\Lambda^{0};\Lambda^{0})
\end{align*}
\begin{align*}
\mathcal{A}(\Lambda_{c}^{+}\to p\bar{K}^{*0})=&\mathcal{C}_{SD}(\Lambda_{c}^{+}\to p\bar{K}^{*0})+\mathcal{M}(\pi^{+},\Lambda^{0};K^{+})+\mathcal{M}(\pi^{+},\Lambda^{0};K^{*+})+\mathcal{M}(\rho^{+},\Lambda^{0};K^{+})\\
+&\mathcal{M}(\rho^{+},\Lambda^{0};K^{*+})
 +\mathcal{M}(\pi^{+},\Lambda^{0};n)+\mathcal{M}(\rho^{+},\Lambda^{0};n)
 \end{align*}
 \begin{align*}
\mathcal{A}(\Lambda_{c}^{+}\to\Xi^{0}K^{*+})= & \mathcal{M}(\pi^{+},\Lambda^{0};\bar{K}^{0})+\mathcal{M}(\pi^{+},\Lambda^{0};\bar{K}^{*0})+\mathcal{M}(\rho^{+},\Lambda^{0};\bar{K}^{0})+\mathcal{M}(\rho^{+},\Lambda^{0};\bar{K}^{*0})\\
+&\mathcal{M}(\pi^{+},\Lambda^{0};\Xi^{-})+\mathcal{M}(\rho^{+},\Lambda^{0};\Xi^{-})+\mathcal{M}(\pi^{+},\Lambda^{0};\Xi^{*-})+\mathcal{M}(\rho^{+},\Lambda^{0};\Xi^{*-})
\end{align*}

\subsection{Singly Cabibbo-suppressed}
\begin{align*}
\mathcal{A}(\Lambda_{c}^{+}\to\Lambda^{0}K^{*+})= & \mathcal{T}_{SD}(\Lambda_{c}^{+}\to\Lambda^{0}K^{*+})+\mathcal{M}(\pi^{+},n;\bar{K}^{0})+\mathcal{M}(\pi^{+},n;\bar{K}^{*0})+\mathcal{M}(\rho^{+},n;\bar{K}^{0})\\
 +&\mathcal{M}(\rho^{+},n;\bar{K}^{*0})+\mathcal{M}(\pi^{+},n;\Sigma^{-})+\mathcal{M}(\rho^{+},n;\Sigma^{-})+\mathcal{M}(\pi^{+},n;\Sigma^{*-})+\mathcal{M}(\rho^{+},n;\Sigma^{*-})\\
 +&\mathcal{M}(K^{+},\Lambda^{0};\eta_{8})+\mathcal{M}(K^{+},\Lambda^{0};\eta_{1})+\mathcal{M}(K^{+},\Lambda^{0};\omega)+\mathcal{M}(K^{+},\Lambda^{0};\phi)\\
 +&\mathcal{M}(K^{*+},\Lambda^{0};\eta_{8})+\mathcal{M}(K^{*+},\Lambda^{0};\eta_{1})+\mathcal{M}(K^{*+},\Lambda^{0};\omega)+\mathcal{M}(K^{*+},\Lambda^{0};\phi)\\
 +&\mathcal{M}(K^{+},\Lambda^{0};\Xi^{-})+\mathcal{M}(K^{*+},\Lambda^{0};\Xi^{-})+\mathcal{M}(K^{+},\Lambda^{0};\Xi^{*-})+\mathcal{M}(K^{*+},\Lambda^{0};\Xi^{*-})
\end{align*}
\begin{align*}
\mathcal{A}(\Lambda_{c}^{+}\to\Sigma^{0}K^{*+})= & \mathcal{M}(\pi^{+},n;\bar{K}^{0})+\mathcal{M}(\pi^{+},n;\bar{K}^{*0})+\mathcal{M}(\rho^{+},n;\bar{K}^{0})+\mathcal{M}(\rho^{+},n;\bar{K}^{*0})\\
 +&\mathcal{M}(\pi^{+},n;\Sigma^{-})+\mathcal{M}(\rho^{+},n;\Sigma^{-})+\mathcal{M}(\pi^{+},n;\Sigma^{*-})+\mathcal{M}(\rho^{+},n;\Sigma^{*-})\\
 +&\mathcal{M}(K^{+},\Lambda^{0};\pi^{0})+\mathcal{M}(K^{+},\Lambda^{0};\rho^{0})+\mathcal{M}(K^{*+},\Lambda^{0};\pi^{0})+\mathcal{M}(K^{*+},\Lambda^{0};\rho^{0})\\
 +&\mathcal{M}(K^{+},\Lambda^{0};\Xi^{-})+\mathcal{M}(K^{*+},\Lambda^{0};\Xi^{-})+\mathcal{M}(K^{+},\Lambda^{0};\Xi^{*-})+\mathcal{M}(K^{*+},\Lambda^{0};\Xi^{*-})
\end{align*}
\begin{align*}
\mathcal{A}(\Lambda_{c}^{+}\to\Sigma^{+}K^{*0})= & \mathcal{M}(\pi^{+},n;\Lambda^{0})+\mathcal{M}(\rho^{+},n;\Lambda^{0})+\mathcal{M}(\pi^{+},n;\Sigma^{0})+\mathcal{M}(\rho^{+},n;\Sigma^{0})+\mathcal{M}(\pi^{+},n;\Sigma^{*0})\\
 +&\mathcal{M}(\rho^{+},n;\Sigma^{*0})+\mathcal{M}(K^{+},\Lambda^{0};\pi^{+})+\mathcal{M}(K^{+},\Lambda^{0};\rho^{+})+\mathcal{M}(K^{*+},\Lambda^{0};\pi^{+})\\
 +&\mathcal{M}(K^{*+},\Lambda^{0};\rho^{+})+\mathcal{M}(K^{+},\Lambda^{0};\Xi^{0})+\mathcal{M}(K^{*+},\Lambda^{0};\Xi^{0})+\mathcal{M}(K^{+},\Lambda^{0};\Xi^{*0})\\
 +&\mathcal{M}(K^{*+},\Lambda^{0};\Xi^{*0})
\end{align*}
\begin{align*}
\mathcal{A}(\Lambda_{c}^{+}\to p\phi)= & \mathcal{C}_{SD}(\Lambda_{c}^{+}\to p\phi)+\mathcal{M}(K^{+},\Lambda^{0};K^{+})+\mathcal{M}(K^{+},\Lambda^{0};K^{*+})+\mathcal{M}(K^{*+},\Lambda^{0};K^{+})\\
 +&\mathcal{M}(K^{*+},\Lambda^{0};K^{*+})+\mathcal{M}(K^{+},\Lambda^{0};\Lambda^{0})+\mathcal{M}(K^{*+},\Lambda^{0};\Lambda^{0})
\end{align*}
\begin{align*}
\mathcal{A}(\Lambda_{c}^{+}\to p\omega)=&\mathcal{C}_{SD}(\Lambda_{c}^{+}\to p\omega)+\mathcal{M}(\pi^{+},n;\rho^{+})+\mathcal{M}(\rho^{+},n;\pi^{+})+\mathcal{M}(\pi^{+},n;n)+\mathcal{M}(\rho^{+},n;n)\\
 +&\mathcal{M}(K^{+},\Lambda^{0};K^{+})+\mathcal{M}(K^{+},\Lambda^{0};K^{*+})+\mathcal{M}(K^{*+},\Lambda^{0};K^{+})+\mathcal{M}(K^{*+},\Lambda^{0};K^{*+})\\
 +&\mathcal{M}(K^{+},\Lambda^{0};\Lambda^{0})+\mathcal{M}(K^{*+},\Lambda^{0};\Lambda^{0})
\end{align*}
\begin{align*}
\mathcal{A}(\Lambda_{c}^{+}\to p\rho^{0})= & \mathcal{C}_{SD}(\Lambda_{c}^{+}\to p\rho^{0})+\mathcal{M}(\pi^{+},n;\pi^{+})+\mathcal{M}(\rho^{+},n;\rho^{+})+\mathcal{M}(\pi^{+},n;n)+\mathcal{M}(\rho^{+},n;n)\\
 +&\mathcal{M}(\pi^{+},n;\Delta^{0})+\mathcal{M}(\rho^{+},n;\Delta^{0})+\mathcal{M}(K^{+},\Lambda^{0};K^{+})+\mathcal{M}(K^{+},\Lambda^{0};K^{*+})\\
 +&\mathcal{M}(K^{*+},\Lambda^{0};K^{+})+\mathcal{M}(K^{*+},\Lambda^{0};K^{*+})+\mathcal{M}(K^{+},\Lambda^{0};\Sigma^{0})+\mathcal{M}(K^{*+},\Lambda^{0};\Sigma^{0})\\
 +&\mathcal{M}(K^{+},\Lambda^{0};\Sigma^{*0})+\mathcal{M}(K^{*+},\Lambda^{0};\Sigma^{*0})
\end{align*}
\begin{align*}
\mathcal{A}(\Lambda_{c}^{+}\to n\rho^{+})= & \mathcal{T}_{SD}(\Lambda_{c}^{+}\to n\rho^{+})+\mathcal{M}(\pi^{+},n;\pi^{0})+\mathcal{M}(\rho^{+},n;\rho^{0})+\mathcal{M}(\pi^{+},n;\omega)\\
 +&\mathcal{M}(\rho^{+},n;\eta_{8})+\mathcal{M}(\rho^{+},n;\eta_{1})+\mathcal{M}(\pi^{+},n;\Delta^{-})+\mathcal{M}(\rho^{+},n;\Delta^{-})\\
 +&\mathcal{M}(K^{+},\Lambda^{0};K^{0})+\mathcal{M}(K^{+},\Lambda^{0};K^{*0})+\mathcal{M}(K^{*+},\Lambda^{0};K^{0})+\mathcal{M}(K^{*+},\Lambda^{0};K^{*0})\\
 +&\mathcal{M}(K^{+},\Lambda^{0};\Sigma^{-})+\mathcal{M}(K^{*+},\Lambda^{0};\Sigma^{-})+\mathcal{M}(K^{+},\Lambda^{0};\Sigma^{*-})+\mathcal{M}(K^{*+},\Lambda^{0};\Sigma^{*-})
\end{align*}

\subsection{Doubly Cabibbo-suppressed}
\begin{align*}
\mathcal{A}(\Lambda_{c}^{+}\to pK^{*0})= & \mathcal{C}_{SD}(\Lambda_{c}^{+}\to pK^{*0})+\mathcal{M}(K^{+},n;\pi^{+})+\mathcal{M}(K^{+},n;\rho^{+})+\mathcal{M}(K^{*+},n;\pi^{+})\\
 +&\mathcal{M}(K^{*+},n;\rho^{+}) +\mathcal{M}(K^{+},n;\Lambda^{0})+\mathcal{M}(K^{*+},n;\Lambda^{0})+\mathcal{M}(K^{+},n;\Sigma^{0})\\
 +&\mathcal{M}(K^{*+},n;\Sigma^{0}) +\mathcal{M}(K^{+},n;\Sigma^{*0})+\mathcal{M}(K^{*+},n;\Sigma^{*0})\\
 \end{align*}
 \begin{align*}
\mathcal{A}(\Lambda_{c}^{+}\to nK^{*+})= & \mathcal{T}_{SD}(\Lambda_{c}^{+}\to nK^{*+})+\mathcal{M}(K^{+},n;\pi^{0})+\mathcal{M}(K^{+},n;\eta_{8})+\mathcal{M}(K^{+},n;\eta_{1})\\
 +&\mathcal{M}(K^{+},n;\rho^{0})+\mathcal{M}(K^{+},n;\omega)+\mathcal{M}(K^{*+},n;\pi^{0})+\mathcal{M}(K^{*+},n;\eta_{8})\\
 +&\mathcal{M}(K^{*+},n;\eta_{1}) +\mathcal{M}(K^{*+},n;\rho^{0})+\mathcal{M}(K^{*+},n;\omega)+\mathcal{M}(K^{+},n;\Sigma^{-})\\
 +&\mathcal{M}(K^{*+},n;\Sigma^{-})+\mathcal{M}(K^{+},n;\Sigma^{*-})+\mathcal{M}(K^{*+},n;\Sigma^{*-})
\end{align*}

\section{Effective Lagrangian}\label{App.B}
The effective Lagrangian used in the re-scattering mechanism are \cite{Yu:2017zst,Fajfer:2003ag,Ronchen:2012eg,Aliev:2009ei,Aliev:2006xr}
\begin{equation}\label{effective Lag}
	\begin{aligned}
\mathcal{L}_{VPP}&=\frac{ig_{\rho\pi\pi}}{\sqrt{2}}Tr\left[V^{\mu}\left[P,\partial_{\mu}P\right]\right]\\
\mathcal{L}_{VVV}&=\frac{ig_{\rho\rho\rho}}{\sqrt{2}}Tr\left[\left(\partial_{\nu}V_{\mu}V^{\mu}-V_{\mu}\partial_{\nu}V^{\mu}\right)V^{\nu}\right]\\
\mathcal{L}_{VVP}&=\frac{4g_{VVP}}{f_{P}}\epsilon^{\mu\nu\alpha\beta}Tr\left[\partial_{\mu}V_{\nu}\partial_{\alpha}V_{\beta}P\right]\\
\mathcal{L}_{\mathcal{B}_8\mathcal{B}_8P}&=\sqrt{2}DTr\left[\bar{\mathcal{B}}_8i\gamma_{5}\left\{P,\mathcal{B}_8\right\}\right]+\sqrt{2}FTr\left[\bar{\mathcal{B}}_8i\gamma_{5}\left[P,\mathcal{B}_8\right]\right]\\
\mathcal{L}_{\mathcal{B}_8\mathcal{B}_8V}&=\sqrt{2}D^{\prime}Tr\left[\bar{\mathcal{B}}_8\gamma_{\mu}\left\{V^{\mu},\mathcal{B}_8\right\}\right]+\sqrt{2}F^{\prime}Tr\left[\bar{\mathcal{B}}_8\gamma_{\mu}\left[V^{\mu},\mathcal{B}_8\right]\right]\\
&-\sqrt{2}(D^{\prime}-F^{\prime})Tr(\bar{\mathcal{B}}_8\gamma_{\mu}\mathcal{B}_8)TrV^{\mu}\\
\mathcal{L}_{\mathcal{B}_{10}\mathcal{B}_8P}&=\frac{g_{\Delta N\pi}}{m_{\pi}}\epsilon_{ijk}(\bar{\mathcal{B}}_8)^{j}_{l}(\mathcal{B}_{10})^{mkl}_{\mu}\partial^{\mu}P^{i}_{m}\\
\mathcal{L}_{\mathcal{B}_{10}\mathcal{B}_8V}&=-i\frac{g_{\rho N\Delta}}{m_{\rho}}\left[(\bar{\mathcal{B}}_{10})^{\mu}\gamma^{5}\gamma^{\nu}\mathcal{B}_8+\bar{\mathcal{B}}_8\gamma^{5}\gamma^{\nu}(\mathcal{B}_{10})^{\mu}\right](\partial_{\mu}\rho_{\nu}-\partial_{\nu}\rho_{\mu})
	\end{aligned}
\end{equation}
where the corresponding $V,P,\mathcal{B}_8,\mathcal{B}_{10}$ represent the vector and pseudo-scalar meson octet, baryon octet and decuplet respectively under $SU(3)$ flavor group. 


\bibliography{references}

\begin{thebibliography}{184}
\expandafter\ifx\csname natexlab\endcsname\relax\def\natexlab#1{#1}\fi
\providecommand{\url}[1]{\texttt{#1}}
\providecommand{\href}[2]{#2}
\providecommand{\path}[1]{#1}
\providecommand{\DOIprefix}{doi:}
\providecommand{\ArXivprefix}{arXiv:}
\providecommand{\URLprefix}{URL: }
\providecommand{\Pubmedprefix}{pmid:}
\providecommand{\doi}[1]{\href{http://dx.doi.org/#1}{\path{#1}}}
\providecommand{\Pubmed}[1]{\href{pmid:#1}{\path{#1}}}
\providecommand{\bibinfo}[2]{#2}
\ifx\xfnm\relax \def\xfnm[#1]{\unskip,\space#1}\fi
\bibitem[{Sakharov(1967)}]{Sakharov:1967dj}
\bibinfo{author}{A.~D. Sakharov},
\newblock \bibinfo{title}{{Violation of CP Invariance, C asymmetry, and baryon
  asymmetry of the universe}},
\newblock \bibinfo{journal}{Pisma Zh. Eksp. Teor. Fiz.} \bibinfo{volume}{5}
  (\bibinfo{year}{1967}) \bibinfo{pages}{32--35}.
  \DOIprefix\doi{10.1070/PU1991v034n05ABEH002497}.
\bibitem[{Beneke et~al.(1999)Beneke, Buchalla, Neubert, and
  Sachrajda}]{Beneke:1999br}
\bibinfo{author}{M.~Beneke}, \bibinfo{author}{G.~Buchalla},
  \bibinfo{author}{M.~Neubert}, \bibinfo{author}{C.~T. Sachrajda},
\newblock \bibinfo{title}{{QCD factorization for B ---\ensuremath{>} pi pi
  decays: Strong phases and CP violation in the heavy quark limit}},
\newblock \bibinfo{journal}{Phys. Rev. Lett.} \bibinfo{volume}{83}
  (\bibinfo{year}{1999}) \bibinfo{pages}{1914--1917}.
  \DOIprefix\doi{10.1103/PhysRevLett.83.1914}.
  \href{http://arxiv.org/abs/hep-ph/9905312}{{\tt arXiv:hep-ph/9905312}}.
\bibitem[{Beneke et~al.(2000)Beneke, Buchalla, Neubert, and
  Sachrajda}]{Beneke:2000ry}
\bibinfo{author}{M.~Beneke}, \bibinfo{author}{G.~Buchalla},
  \bibinfo{author}{M.~Neubert}, \bibinfo{author}{C.~T. Sachrajda},
\newblock \bibinfo{title}{{QCD factorization for exclusive, nonleptonic B meson
  decays: General arguments and the case of heavy light final states}},
\newblock \bibinfo{journal}{Nucl. Phys. B} \bibinfo{volume}{591}
  (\bibinfo{year}{2000}) \bibinfo{pages}{313--418}.
  \DOIprefix\doi{10.1016/S0550-3213(00)00559-9}.
  \href{http://arxiv.org/abs/hep-ph/0006124}{{\tt arXiv:hep-ph/0006124}}.
\bibitem[{Beneke et~al.(2001)Beneke, Buchalla, Neubert, and
  Sachrajda}]{Beneke:2001ev}
\bibinfo{author}{M.~Beneke}, \bibinfo{author}{G.~Buchalla},
  \bibinfo{author}{M.~Neubert}, \bibinfo{author}{C.~T. Sachrajda},
\newblock \bibinfo{title}{{QCD factorization in B ---\ensuremath{>} pi K, pi pi
  decays and extraction of Wolfenstein parameters}},
\newblock \bibinfo{journal}{Nucl. Phys. B} \bibinfo{volume}{606}
  (\bibinfo{year}{2001}) \bibinfo{pages}{245--321}.
  \DOIprefix\doi{10.1016/S0550-3213(01)00251-6}.
  \href{http://arxiv.org/abs/hep-ph/0104110}{{\tt arXiv:hep-ph/0104110}}.
\bibitem[{Beneke and Neubert(2003)}]{Beneke:2003zv}
\bibinfo{author}{M.~Beneke}, \bibinfo{author}{M.~Neubert},
\newblock \bibinfo{title}{{QCD factorization for B ---\ensuremath{>} PP and B
  ---\ensuremath{>} PV decays}},
\newblock \bibinfo{journal}{Nucl. Phys. B} \bibinfo{volume}{675}
  (\bibinfo{year}{2003}) \bibinfo{pages}{333--415}.
  \DOIprefix\doi{10.1016/j.nuclphysb.2003.09.026}.
  \href{http://arxiv.org/abs/hep-ph/0308039}{{\tt arXiv:hep-ph/0308039}}.
\bibitem[{Keum et~al.(2001{\natexlab{a}})Keum, Li, and Sanda}]{Keum:2000wi}
\bibinfo{author}{Y.~Y. Keum}, \bibinfo{author}{H.-N. Li},
  \bibinfo{author}{A.~I. Sanda},
\newblock \bibinfo{title}{{Penguin enhancement and $B \to K \pi$ decays in
  perturbative QCD}},
\newblock \bibinfo{journal}{Phys. Rev. D} \bibinfo{volume}{63}
  (\bibinfo{year}{2001}{\natexlab{a}}) \bibinfo{pages}{054008}.
  \DOIprefix\doi{10.1103/PhysRevD.63.054008}.
  \href{http://arxiv.org/abs/hep-ph/0004173}{{\tt arXiv:hep-ph/0004173}}.
\bibitem[{Keum et~al.(2001{\natexlab{b}})Keum, Li, and Sanda}]{Keum:2000ph}
\bibinfo{author}{Y.-Y. Keum}, \bibinfo{author}{H.-n. Li},
  \bibinfo{author}{A.~I. Sanda},
\newblock \bibinfo{title}{{Fat penguins and imaginary penguins in perturbative
  QCD}},
\newblock \bibinfo{journal}{Phys. Lett. B} \bibinfo{volume}{504}
  (\bibinfo{year}{2001}{\natexlab{b}}) \bibinfo{pages}{6--14}.
  \DOIprefix\doi{10.1016/S0370-2693(01)00247-7}.
  \href{http://arxiv.org/abs/hep-ph/0004004}{{\tt arXiv:hep-ph/0004004}}.
\bibitem[{Lu et~al.(2001)Lu, Ukai, and Yang}]{Lu:2000em}
\bibinfo{author}{C.-D. Lu}, \bibinfo{author}{K.~Ukai}, \bibinfo{author}{M.-Z.
  Yang},
\newblock \bibinfo{title}{{Branching ratio and CP violation of B
  ---\ensuremath{>} pi pi decays in perturbative QCD approach}},
\newblock \bibinfo{journal}{Phys. Rev. D} \bibinfo{volume}{63}
  (\bibinfo{year}{2001}) \bibinfo{pages}{074009}.
  \DOIprefix\doi{10.1103/PhysRevD.63.074009}.
  \href{http://arxiv.org/abs/hep-ph/0004213}{{\tt arXiv:hep-ph/0004213}}.
\bibitem[{Aaij et~al.(2018{\natexlab{a}})}]{LHCb:2017xtf}
\bibinfo{author}{R.~Aaij}, et~al. (\bibinfo{collaboration}{LHCb}),
\newblock \bibinfo{title}{{Measurements of the branching fractions of
  $\Lambda_{c}^{+} \rightarrow p \pi^{-} \pi^{+}$, $\Lambda_{c}^{+} \rightarrow
  p K^{-} K^{+}$, and $\Lambda_{c}^{+} \rightarrow p \pi^{-} K^{+}$}},
\newblock \bibinfo{journal}{JHEP} \bibinfo{volume}{03}
  (\bibinfo{year}{2018}{\natexlab{a}}) \bibinfo{pages}{043}.
  \DOIprefix\doi{10.1007/JHEP03(2018)043}.
  \href{http://arxiv.org/abs/1711.01157}{{\tt arXiv:1711.01157}}.
\bibitem[{Aaij et~al.(2018{\natexlab{b}})}]{LHCb:2017hwf}
\bibinfo{author}{R.~Aaij}, et~al. (\bibinfo{collaboration}{LHCb}),
\newblock \bibinfo{title}{{A measurement of the $CP$ asymmetry difference in
  $\varLambda_{c}^{+} \to pK^{-}K^{+}$ and $p\pi^{-}\pi^{+}$ decays}},
\newblock \bibinfo{journal}{JHEP} \bibinfo{volume}{03}
  (\bibinfo{year}{2018}{\natexlab{b}}) \bibinfo{pages}{182}.
  \DOIprefix\doi{10.1007/JHEP03(2018)182}.
  \href{http://arxiv.org/abs/1712.07051}{{\tt arXiv:1712.07051}}.
\bibitem[{Aaij et~al.(2018{\natexlab{c}})}]{LHCb:2017yqf}
\bibinfo{author}{R.~Aaij}, et~al. (\bibinfo{collaboration}{LHCb}),
\newblock \bibinfo{title}{{Search for the rare decay $\Lambda_{c}^{+} \to
  p\mu^+\mu^-$}},
\newblock \bibinfo{journal}{Phys. Rev. D} \bibinfo{volume}{97}
  (\bibinfo{year}{2018}{\natexlab{c}}) \bibinfo{pages}{091101}.
  \DOIprefix\doi{10.1103/PhysRevD.97.091101}.
  \href{http://arxiv.org/abs/1712.07938}{{\tt arXiv:1712.07938}}.
\bibitem[{Aaij et~al.(2019{\natexlab{a}})}]{LHCb:2019nxp}
\bibinfo{author}{R.~Aaij}, et~al. (\bibinfo{collaboration}{LHCb}),
\newblock \bibinfo{title}{{Observation of the doubly Cabibbo-suppressed decay
  $\Xi_{c}^{+}\to p\phi$}},
\newblock \bibinfo{journal}{JHEP} \bibinfo{volume}{04}
  (\bibinfo{year}{2019}{\natexlab{a}}) \bibinfo{pages}{084}.
  \DOIprefix\doi{10.1007/JHEP04(2019)084}.
  \href{http://arxiv.org/abs/1901.06222}{{\tt arXiv:1901.06222}}.
\bibitem[{Aaij et~al.(2019{\natexlab{b}})}]{LHCb:2019ldj}
\bibinfo{author}{R.~Aaij}, et~al. (\bibinfo{collaboration}{LHCb}),
\newblock \bibinfo{title}{{Precision measurement of the $\Lambda_c^+$,
  $\Xi_c^+$ and $\Xi_c^0$ baryon lifetimes}},
\newblock \bibinfo{journal}{Phys. Rev. D} \bibinfo{volume}{100}
  (\bibinfo{year}{2019}{\natexlab{b}}) \bibinfo{pages}{032001}.
  \DOIprefix\doi{10.1103/PhysRevD.100.032001}.
  \href{http://arxiv.org/abs/1906.08350}{{\tt arXiv:1906.08350}}.
\bibitem[{Aaij et~al.(2020)}]{LHCb:2020zkk}
\bibinfo{author}{R.~Aaij}, et~al. (\bibinfo{collaboration}{LHCb}),
\newblock \bibinfo{title}{{Search for $CP$ violation in ${{{\varXi }} ^+_{c}}
  \rightarrow {p} {{K} ^-} {{\pi } ^+} $ decays using model-independent
  techniques}},
\newblock \bibinfo{journal}{Eur. Phys. J. C} \bibinfo{volume}{80}
  (\bibinfo{year}{2020}) \bibinfo{pages}{986}.
  \DOIprefix\doi{10.1140/epjc/s10052-020-8365-0}.
  \href{http://arxiv.org/abs/2006.03145}{{\tt arXiv:2006.03145}}.
\bibitem[{Aaij et~al.(2023{\natexlab{a}})}]{LHCb:2022sck}
\bibinfo{author}{R.~Aaij}, et~al. (\bibinfo{collaboration}{LHCb}),
\newblock \bibinfo{title}{{Amplitude analysis of the
  \ensuremath{\Lambda}c+\textrightarrow{}pK-\ensuremath{\pi}+ decay and
  \ensuremath{\Lambda}c+ baryon polarization measurement in semileptonic beauty
  hadron decays}},
\newblock \bibinfo{journal}{Phys. Rev. D} \bibinfo{volume}{108}
  (\bibinfo{year}{2023}{\natexlab{a}}) \bibinfo{pages}{012023}.
  \DOIprefix\doi{10.1103/PhysRevD.108.012023}.
  \href{http://arxiv.org/abs/2208.03262}{{\tt arXiv:2208.03262}}.
\bibitem[{Aaij et~al.(2023{\natexlab{b}})}]{LHCb:2023crj}
\bibinfo{author}{R.~Aaij}, et~al. (\bibinfo{collaboration}{LHCb}),
\newblock \bibinfo{title}{{$ {\Lambda}_c^{+} $ polarimetry using the dominant
  hadronic mode}},
\newblock \bibinfo{journal}{JHEP} \bibinfo{volume}{07}
  (\bibinfo{year}{2023}{\natexlab{b}}) \bibinfo{pages}{228}.
  \DOIprefix\doi{10.1007/JHEP07(2023)228}.
  \href{http://arxiv.org/abs/2301.07010}{{\tt arXiv:2301.07010}}.
\bibitem[{Yang et~al.(2016)}]{Belle:2015wxn}
\bibinfo{author}{S.~B. Yang}, et~al. (\bibinfo{collaboration}{Belle}),
\newblock \bibinfo{title}{{First Observation of Doubly Cabibbo-Suppressed Decay
  of a Charmed Baryon: $\Lambda^{+}_{c} \rightarrow p K^{+} \pi^{-}$}},
\newblock \bibinfo{journal}{Phys. Rev. Lett.} \bibinfo{volume}{117}
  (\bibinfo{year}{2016}) \bibinfo{pages}{011801}.
  \DOIprefix\doi{10.1103/PhysRevLett.117.011801}.
  \href{http://arxiv.org/abs/1512.07366}{{\tt arXiv:1512.07366}}.
\bibitem[{Pal et~al.(2017)}]{Belle:2017tfw}
\bibinfo{author}{B.~Pal}, et~al. (\bibinfo{collaboration}{Belle}),
\newblock \bibinfo{title}{{Search for $\Lambda_c^+\to\phi p \pi^0$ and
  branching fraction measurement of $\Lambda_c^+\to K^-\pi^+ p \pi^0$}},
\newblock \bibinfo{journal}{Phys. Rev. D} \bibinfo{volume}{96}
  (\bibinfo{year}{2017}) \bibinfo{pages}{051102}.
  \DOIprefix\doi{10.1103/PhysRevD.96.051102}.
  \href{http://arxiv.org/abs/1707.00089}{{\tt arXiv:1707.00089}}.
\bibitem[{Berger et~al.(2018)}]{Belle:2018gcs}
\bibinfo{author}{M.~Berger}, et~al. (\bibinfo{collaboration}{Belle}),
\newblock \bibinfo{title}{{Measurement of the Decays $\Lambda_c\to
  \Sigma\pi\pi$ at Belle}},
\newblock \bibinfo{journal}{Phys. Rev. D} \bibinfo{volume}{98}
  (\bibinfo{year}{2018}) \bibinfo{pages}{112006}.
  \DOIprefix\doi{10.1103/PhysRevD.98.112006}.
  \href{http://arxiv.org/abs/1802.03421}{{\tt arXiv:1802.03421}}.
\bibitem[{Sumihama et~al.(2019)}]{Belle:2018lws}
\bibinfo{author}{M.~Sumihama}, et~al. (\bibinfo{collaboration}{Belle}),
\newblock \bibinfo{title}{{Observation of $\Xi(1620)^0$ and evidence for
  $\Xi(1690)^0$ in $\Xi_c^+ \rightarrow \Xi^-\pi^+\pi^+$ decays}},
\newblock \bibinfo{journal}{Phys. Rev. Lett.} \bibinfo{volume}{122}
  (\bibinfo{year}{2019}) \bibinfo{pages}{072501}.
  \DOIprefix\doi{10.1103/PhysRevLett.122.072501}.
  \href{http://arxiv.org/abs/1810.06181}{{\tt arXiv:1810.06181}}.
\bibitem[{Li et~al.(2019)}]{Belle:2019bgi}
\bibinfo{author}{Y.~B. Li}, et~al. (\bibinfo{collaboration}{Belle}),
\newblock \bibinfo{title}{{First measurements of absolute branching fractions
  of the $\Xi_c^+$ baryon at Belle}},
\newblock \bibinfo{journal}{Phys. Rev. D} \bibinfo{volume}{100}
  (\bibinfo{year}{2019}) \bibinfo{pages}{031101}.
  \DOIprefix\doi{10.1103/PhysRevD.100.031101}.
  \href{http://arxiv.org/abs/1904.12093}{{\tt arXiv:1904.12093}}.
\bibitem[{Lee et~al.(2021)}]{Belle:2020xku}
\bibinfo{author}{J.~Y. Lee}, et~al. (\bibinfo{collaboration}{Belle}),
\newblock \bibinfo{title}{{Measurement of branching fractions of
  $\Lambda_{c}^{+} \rightarrow \eta\Lambda\pi^{+}$, $\eta \Sigma^{0} \pi^{+}$,
  $\Lambda(1670) \pi^{+}$, and $\eta \Sigma(1385)^{+}$}},
\newblock \bibinfo{journal}{Phys. Rev. D} \bibinfo{volume}{103}
  (\bibinfo{year}{2021}) \bibinfo{pages}{052005}.
  \DOIprefix\doi{10.1103/PhysRevD.103.052005}.
  \href{http://arxiv.org/abs/2008.11575}{{\tt arXiv:2008.11575}}.
\bibitem[{Li et~al.(2021{\natexlab{a}})}]{Belle:2021mvw}
\bibinfo{author}{S.~X. Li}, et~al. (\bibinfo{collaboration}{Belle}),
\newblock \bibinfo{title}{{Measurements of the branching fractions of
  $\Lambda_c^+ \to p \eta$ and $\Lambda_c^+ \to p \pi^0$ decays at Belle}},
\newblock \bibinfo{journal}{Phys. Rev. D} \bibinfo{volume}{103}
  (\bibinfo{year}{2021}{\natexlab{a}}) \bibinfo{pages}{072004}.
  \DOIprefix\doi{10.1103/PhysRevD.103.072004}.
  \href{http://arxiv.org/abs/2102.12226}{{\tt arXiv:2102.12226}}.
\bibitem[{Li et~al.(2021{\natexlab{b}})}]{Belle:2021crz}
\bibinfo{author}{Y.~B. Li}, et~al. (\bibinfo{collaboration}{Belle}),
\newblock \bibinfo{title}{{Measurements of the branching fractions of the
  semileptonic decays $\Xi_{c}^{0} \to \Xi^{-} \ell^{+} \nu_{\ell}$ and the
  asymmetry parameter of $\Xi_{c}^{0} \to \Xi^{-} \pi^{+}$}},
\newblock \bibinfo{journal}{Phys. Rev. Lett.} \bibinfo{volume}{127}
  (\bibinfo{year}{2021}{\natexlab{b}}) \bibinfo{pages}{121803}.
  \DOIprefix\doi{10.1103/PhysRevLett.127.121803}.
  \href{http://arxiv.org/abs/2103.06496}{{\tt arXiv:2103.06496}}.
\bibitem[{Li et~al.(2021{\natexlab{c}})}]{Belle:2021btl}
\bibinfo{author}{S.~X. Li}, et~al. (\bibinfo{collaboration}{Belle}),
\newblock \bibinfo{title}{{Measurement of the branching fraction of
  $\Lambda_c^+ \to p \omega$ decay at Belle}},
\newblock \bibinfo{journal}{Phys. Rev. D} \bibinfo{volume}{104}
  (\bibinfo{year}{2021}{\natexlab{c}}) \bibinfo{pages}{072008}.
  \DOIprefix\doi{10.1103/PhysRevD.104.072008}.
  \href{http://arxiv.org/abs/2108.11301}{{\tt arXiv:2108.11301}}.
\bibitem[{Li et~al.(2022{\natexlab{a}})}]{Belle:2021avh}
\bibinfo{author}{Y.~Li}, et~al. (\bibinfo{collaboration}{Belle}),
\newblock \bibinfo{title}{{Measurements of the branching fractions of $\Xi_c^0
  \to \Lambda K_S^0$, $\Xi_c^0 \to \Sigma^0 K_S^0$, and $\Xi_c^0 \to \Sigma^+
  K^-$ decays at Belle}},
\newblock \bibinfo{journal}{Phys. Rev. D} \bibinfo{volume}{105}
  (\bibinfo{year}{2022}{\natexlab{a}}) \bibinfo{pages}{L011102}.
  \DOIprefix\doi{10.1103/PhysRevD.105.L011102}.
  \href{http://arxiv.org/abs/2111.08981}{{\tt arXiv:2111.08981}}.
\bibitem[{Li et~al.(2022{\natexlab{b}})}]{Belle:2021vyq}
\bibinfo{author}{S.~X. Li}, et~al. (\bibinfo{collaboration}{Belle}),
\newblock \bibinfo{title}{{First Measurement of the $\Lambda_c^+ \to p \eta'$
  decay}},
\newblock \bibinfo{journal}{JHEP} \bibinfo{volume}{03}
  (\bibinfo{year}{2022}{\natexlab{b}}) \bibinfo{pages}{090}.
  \DOIprefix\doi{10.1007/JHEP03(2022)090}.
  \href{http://arxiv.org/abs/2112.14276}{{\tt arXiv:2112.14276}}.
\bibitem[{Li et~al.(2023)}]{Belle:2022raw}
\bibinfo{author}{Y.~Li}, et~al. (\bibinfo{collaboration}{Belle}),
\newblock \bibinfo{title}{{First search for the weak radiative decays
  \ensuremath{\Lambda}c+\textrightarrow{}\ensuremath{\Sigma}+\ensuremath{\gamma}
  and
  \ensuremath{\Xi}c0\textrightarrow{}\ensuremath{\Xi}0\ensuremath{\gamma}}},
\newblock \bibinfo{journal}{Phys. Rev. D} \bibinfo{volume}{107}
  (\bibinfo{year}{2023}) \bibinfo{pages}{032001}.
  \DOIprefix\doi{10.1103/PhysRevD.107.032001}.
  \href{http://arxiv.org/abs/2206.12517}{{\tt arXiv:2206.12517}}.
\bibitem[{Abudin\'en et~al.(2023)}]{Belle-II:2022ggx}
\bibinfo{author}{F.~Abudin\'en}, et~al. (\bibinfo{collaboration}{Belle-II}),
\newblock \bibinfo{title}{{Measurement of the \ensuremath{\Lambda}c+
  Lifetime}},
\newblock \bibinfo{journal}{Phys. Rev. Lett.} \bibinfo{volume}{130}
  (\bibinfo{year}{2023}) \bibinfo{pages}{071802}.
  \DOIprefix\doi{10.1103/PhysRevLett.130.071802}.
  \href{http://arxiv.org/abs/2206.15227}{{\tt arXiv:2206.15227}}.
\bibitem[{Li et~al.(2023{\natexlab{a}})}]{Belle:2022uod}
\bibinfo{author}{L.~K. Li}, et~al. (\bibinfo{collaboration}{Belle}),
\newblock \bibinfo{title}{{Search for CP violation and measurement of branching
  fractions and decay asymmetry parameters for
  \ensuremath{\Lambda}c+\textrightarrow{}\ensuremath{\Lambda}h+ and
  \ensuremath{\Lambda}c+\textrightarrow{}\ensuremath{\Sigma}0h+
  (h=K,\ensuremath{\pi})}},
\newblock \bibinfo{journal}{Sci. Bull.} \bibinfo{volume}{68}
  (\bibinfo{year}{2023}{\natexlab{a}}) \bibinfo{pages}{583--592}.
  \DOIprefix\doi{10.1016/j.scib.2023.02.017}.
  \href{http://arxiv.org/abs/2208.08695}{{\tt arXiv:2208.08695}}.
\bibitem[{Li et~al.(2023{\natexlab{b}})}]{Belle:2022bsi}
\bibinfo{author}{S.~X. Li}, et~al. (\bibinfo{collaboration}{Belle}),
\newblock \bibinfo{title}{{Measurements of branching fractions of $\Lambda_c^+
  \to \Sigma^+ \eta$ and $\Lambda_c^+ \to \Sigma^+ \eta'$ and asymmetry
  parameters of $\Lambda_c^+ \to \Sigma^+ \pi^0$, $\Lambda_c^+ \to \Sigma^+
  \eta$, and $\Lambda_c^+ \to \Sigma^+ \eta'$}},
\newblock \bibinfo{journal}{Phys. Rev. D} \bibinfo{volume}{107}
  (\bibinfo{year}{2023}{\natexlab{b}}) \bibinfo{pages}{032003}.
  \DOIprefix\doi{10.1103/PhysRevD.107.032003}.
  \href{http://arxiv.org/abs/2208.10825}{{\tt arXiv:2208.10825}}.
\bibitem[{Yang et~al.(2023)}]{Belle:2022cbs}
\bibinfo{author}{S.~B. Yang}, et~al. (\bibinfo{collaboration}{Belle}),
\newblock \bibinfo{title}{{Observation of a threshold cusp at the
  \ensuremath{\Lambda}\ensuremath{\eta} threshold in the pK- mass spectrum with
  \ensuremath{\Lambda}c+\textrightarrow{}pK-\ensuremath{\pi}+ decays}},
\newblock \bibinfo{journal}{Phys. Rev. D} \bibinfo{volume}{108}
  (\bibinfo{year}{2023}) \bibinfo{pages}{L031104}.
  \DOIprefix\doi{10.1103/PhysRevD.108.L031104}.
  \href{http://arxiv.org/abs/2209.00050}{{\tt arXiv:2209.00050}}.
\bibitem[{Han et~al.(2023)}]{Belle:2022yaq}
\bibinfo{author}{X.~Han}, et~al. (\bibinfo{collaboration}{Belle}),
\newblock \bibinfo{title}{{Evidence for the singly Cabibbo-suppressed decay $
  {\Omega}_c^0\to {\Xi}^{-}{\pi}^{+} $ and search for $ {\Omega}_c^0\to
  {\Xi}^{-}{K}^{+}$ and $\ensuremath{\Omega}^{-}{K}^{+}$ decays at Belle}},
\newblock \bibinfo{journal}{JHEP} \bibinfo{volume}{01} (\bibinfo{year}{2023})
  \bibinfo{pages}{055}. \DOIprefix\doi{10.1007/JHEP01(2023)055}.
  \href{http://arxiv.org/abs/2209.08583}{{\tt arXiv:2209.08583}}.
\bibitem[{Li et~al.(2023)}]{Belle:2022pwd}
\bibinfo{author}{L.~K. Li}, et~al. (\bibinfo{collaboration}{Belle}),
\newblock \bibinfo{title}{{Measurement of branching fractions of
  $\Lambda_c^+\to{}pK_S^0K_S^0$ and $\Lambda_c^+\to{}pK_S^0\eta$ at Belle}},
\newblock \bibinfo{journal}{Phys. Rev. D} \bibinfo{volume}{107}
  (\bibinfo{year}{2023}) \bibinfo{pages}{032004}.
  \DOIprefix\doi{10.1103/PhysRevD.107.032004}.
  \href{http://arxiv.org/abs/2210.01995}{{\tt arXiv:2210.01995}}.
\bibitem[{Ma et~al.(2023)}]{Belle:2022ywa}
\bibinfo{author}{Y.~Ma}, et~al. (\bibinfo{collaboration}{Belle}),
\newblock \bibinfo{title}{{First Observation of
  \ensuremath{\Lambda}\ensuremath{\pi}+ and
  \ensuremath{\Lambda}\ensuremath{\pi}- Signals near the
  K\textasciimacron{}N(I=1) Mass Threshold in
  \ensuremath{\Lambda}c+\textrightarrow{}\ensuremath{\Lambda}\ensuremath{\pi}+\ensuremath{\pi}+\ensuremath{\pi}-
  Decay}},
\newblock \bibinfo{journal}{Phys. Rev. Lett.} \bibinfo{volume}{130}
  (\bibinfo{year}{2023}) \bibinfo{pages}{151903}.
  \DOIprefix\doi{10.1103/PhysRevLett.130.151903}.
  \href{http://arxiv.org/abs/2211.11151}{{\tt arXiv:2211.11151}}.
\bibitem[{Cui et~al.(2024)}]{Belle:2023ngs}
\bibinfo{author}{J.~X. Cui}, et~al. (\bibinfo{collaboration}{Belle}),
\newblock \bibinfo{title}{{Search for the semileptonic decays
  \ensuremath{\Xi}c0\textrightarrow{}\ensuremath{\Xi}0\ensuremath{\ell}+\ensuremath{\ell}-
  at Belle}},
\newblock \bibinfo{journal}{Phys. Rev. D} \bibinfo{volume}{109}
  (\bibinfo{year}{2024}) \bibinfo{pages}{052003}.
  \DOIprefix\doi{10.1103/PhysRevD.109.052003}.
  \href{http://arxiv.org/abs/2312.02580}{{\tt arXiv:2312.02580}}.
\bibitem[{Adachi et~al.(2024)}]{Belle-II:2024jql}
\bibinfo{author}{I.~Adachi}, et~al. (\bibinfo{collaboration}{Belle-II, Belle}),
\newblock \bibinfo{title}{{Measurements of the branching fractions of
  $\Xi_{c}^{0}\to\Xi^{0}\pi^{0}$, $\Xi_{c}^{0}\to\Xi^{0}\eta$, and
  $\Xi_{c}^{0}\to\Xi^{0}\eta^{\prime}$ and asymmetry parameter of
  $\Xi_{c}^{0}\to\Xi^{0}\pi^{0}$}}  (\bibinfo{year}{2024}).
  \href{http://arxiv.org/abs/2406.04642}{{\tt arXiv:2406.04642}}.
\bibitem[{Ablikim et~al.(2015)}]{BESIII:2015ysy}
\bibinfo{author}{M.~Ablikim}, et~al. (\bibinfo{collaboration}{BESIII}),
\newblock \bibinfo{title}{{Measurement of the absolute branching fraction for
  $\Lambda^+_{c}\to \Lambda e^+\nu_e$}},
\newblock \bibinfo{journal}{Phys. Rev. Lett.} \bibinfo{volume}{115}
  (\bibinfo{year}{2015}) \bibinfo{pages}{221805}.
  \DOIprefix\doi{10.1103/PhysRevLett.115.221805}.
  \href{http://arxiv.org/abs/1510.02610}{{\tt arXiv:1510.02610}}.
\bibitem[{Ablikim et~al.(2016{\natexlab{a}})}]{BESIII:2015bjk}
\bibinfo{author}{M.~Ablikim}, et~al. (\bibinfo{collaboration}{BESIII}),
\newblock \bibinfo{title}{{Measurements of absolute hadronic branching
  fractions of $\Lambda_{c}^{+}$ baryon}},
\newblock \bibinfo{journal}{Phys. Rev. Lett.} \bibinfo{volume}{116}
  (\bibinfo{year}{2016}{\natexlab{a}}) \bibinfo{pages}{052001}.
  \DOIprefix\doi{10.1103/PhysRevLett.116.052001}.
  \href{http://arxiv.org/abs/1511.08380}{{\tt arXiv:1511.08380}}.
\bibitem[{Ablikim et~al.(2016{\natexlab{b}})}]{BESIII:2016ozn}
\bibinfo{author}{M.~Ablikim}, et~al. (\bibinfo{collaboration}{BESIII}),
\newblock \bibinfo{title}{{Measurement of Singly Cabibbo Suppressed Decays
  $\Lambda_c^{+}\to p\pi^{+}\pi^{-}$ and $\Lambda_c^{+}\to pK^{+}K^{-}$}},
\newblock \bibinfo{journal}{Phys. Rev. Lett.} \bibinfo{volume}{117}
  (\bibinfo{year}{2016}{\natexlab{b}}) \bibinfo{pages}{232002}.
  \DOIprefix\doi{10.1103/PhysRevLett.117.232002}.
  \href{http://arxiv.org/abs/1608.00407}{{\tt arXiv:1608.00407}},
  \bibinfo{note}{[Addendum: Phys.Rev.Lett. 120, 029903 (2018)]}.
\bibitem[{Ablikim et~al.(2017{\natexlab{a}})}]{BESIII:2016yrc}
\bibinfo{author}{M.~Ablikim}, et~al. (\bibinfo{collaboration}{BESIII}),
\newblock \bibinfo{title}{{Observation of $\Lambda^+_{c}\to nK^0_S\pi^+$}},
\newblock \bibinfo{journal}{Phys. Rev. Lett.} \bibinfo{volume}{118}
  (\bibinfo{year}{2017}{\natexlab{a}}) \bibinfo{pages}{112001}.
  \DOIprefix\doi{10.1103/PhysRevLett.118.112001}.
  \href{http://arxiv.org/abs/1611.02797}{{\tt arXiv:1611.02797}}.
\bibitem[{Ablikim et~al.(2017{\natexlab{b}})}]{BESIII:2017fim}
\bibinfo{author}{M.~Ablikim}, et~al. (\bibinfo{collaboration}{BESIII}),
\newblock \bibinfo{title}{{Evidence for the singly-Cabibbo-suppressed decay
  $\Lambda_{c}^{+} \to p\eta$ and search for $\Lambda_{c}^{+} \to p\pi^{0}$}},
\newblock \bibinfo{journal}{Phys. Rev. D} \bibinfo{volume}{95}
  (\bibinfo{year}{2017}{\natexlab{b}}) \bibinfo{pages}{111102}.
  \DOIprefix\doi{10.1103/PhysRevD.95.111102}.
  \href{http://arxiv.org/abs/1702.05279}{{\tt arXiv:1702.05279}}.
\bibitem[{Ablikim et~al.(2017{\natexlab{c}})}]{BESIII:2017rfd}
\bibinfo{author}{M.~Ablikim}, et~al. (\bibinfo{collaboration}{BESIII}),
\newblock \bibinfo{title}{{Observation of the decay $\Lambda_c^+\rightarrow
  \Sigma^- \pi^+\pi^+\pi^0$}},
\newblock \bibinfo{journal}{Phys. Lett. B} \bibinfo{volume}{772}
  (\bibinfo{year}{2017}{\natexlab{c}}) \bibinfo{pages}{388--393}.
  \DOIprefix\doi{10.1016/j.physletb.2017.06.065}.
  \href{http://arxiv.org/abs/1705.11109}{{\tt arXiv:1705.11109}}.
\bibitem[{Ablikim et~al.(2018{\natexlab{a}})}]{BESIII:2018cvs}
\bibinfo{author}{M.~Ablikim}, et~al. (\bibinfo{collaboration}{BESIII}),
\newblock \bibinfo{title}{{Measurements of absolute branching fractions for
  $\Lambda^+_c\to\Xi^0K^+$ and $\Xi(1530)^0K^+$}},
\newblock \bibinfo{journal}{Phys. Lett. B} \bibinfo{volume}{783}
  (\bibinfo{year}{2018}{\natexlab{a}}) \bibinfo{pages}{200--206}.
  \DOIprefix\doi{10.1016/j.physletb.2018.06.046}.
  \href{http://arxiv.org/abs/1803.04299}{{\tt arXiv:1803.04299}}.
\bibitem[{Ablikim et~al.(2018{\natexlab{b}})}]{BESIII:2018ciw}
\bibinfo{author}{M.~Ablikim}, et~al. (\bibinfo{collaboration}{BESIII}),
\newblock \bibinfo{title}{{Measurement of absolute branching fraction of the
  inclusive decay $\Lambda_{c}^{+} \to \Lambda + X$}},
\newblock \bibinfo{journal}{Phys. Rev. Lett.} \bibinfo{volume}{121}
  (\bibinfo{year}{2018}{\natexlab{b}}) \bibinfo{pages}{062003}.
  \DOIprefix\doi{10.1103/PhysRevLett.121.062003}.
  \href{http://arxiv.org/abs/1803.05706}{{\tt arXiv:1803.05706}}.
\bibitem[{Ablikim et~al.(2018{\natexlab{c}})}]{BESIII:2018mug}
\bibinfo{author}{M.~Ablikim}, et~al. (\bibinfo{collaboration}{BESIII}),
\newblock \bibinfo{title}{{Measurement of the absolute branching fraction of
  the inclusive semileptonic $\Lambda_c^+$ decay}},
\newblock \bibinfo{journal}{Phys. Rev. Lett.} \bibinfo{volume}{121}
  (\bibinfo{year}{2018}{\natexlab{c}}) \bibinfo{pages}{251801}.
  \DOIprefix\doi{10.1103/PhysRevLett.121.251801}.
  \href{http://arxiv.org/abs/1805.09060}{{\tt arXiv:1805.09060}}.
\bibitem[{Ablikim et~al.(2019{\natexlab{a}})}]{BESIII:2018cdl}
\bibinfo{author}{M.~Ablikim}, et~al. (\bibinfo{collaboration}{BESIII}),
\newblock \bibinfo{title}{{Evidence for the decays of
  $\Lambda^+_{c}\to\Sigma^+\eta$ and $\Sigma^+\eta^\prime$}},
\newblock \bibinfo{journal}{Chin. Phys. C} \bibinfo{volume}{43}
  (\bibinfo{year}{2019}{\natexlab{a}}) \bibinfo{pages}{083002}.
  \DOIprefix\doi{10.1088/1674-1137/43/8/083002}.
  \href{http://arxiv.org/abs/1811.08028}{{\tt arXiv:1811.08028}}.
\bibitem[{Ablikim et~al.(2019{\natexlab{b}})}]{BESIII:2019odb}
\bibinfo{author}{M.~Ablikim}, et~al. (\bibinfo{collaboration}{BESIII}),
\newblock \bibinfo{title}{{Measurements of Weak Decay Asymmetries of
  $\Lambda_c^+\to pK_S^0$, $\Lambda\pi^+$, $\Sigma^+\pi^0$, and
  $\Sigma^0\pi^+$}},
\newblock \bibinfo{journal}{Phys. Rev. D} \bibinfo{volume}{100}
  (\bibinfo{year}{2019}{\natexlab{b}}) \bibinfo{pages}{072004}.
  \DOIprefix\doi{10.1103/PhysRevD.100.072004}.
  \href{http://arxiv.org/abs/1905.04707}{{\tt arXiv:1905.04707}}.
\bibitem[{Ablikim et~al.(2020)}]{BESIII:2020cpu}
\bibinfo{author}{M.~Ablikim}, et~al. (\bibinfo{collaboration}{BESIII}),
\newblock \bibinfo{title}{{Measurement of the absolute branching fraction of
  the inclusive decay $\Lambda_c^+ \to K_S^0X$}},
\newblock \bibinfo{journal}{Eur. Phys. J. C} \bibinfo{volume}{80}
  (\bibinfo{year}{2020}) \bibinfo{pages}{935}.
  \DOIprefix\doi{10.1140/epjc/s10052-020-08447-0}.
  \href{http://arxiv.org/abs/2005.11211}{{\tt arXiv:2005.11211}}.
\bibitem[{Ablikim et~al.(2021)}]{BESIII:2020kzc}
\bibinfo{author}{M.~Ablikim}, et~al. (\bibinfo{collaboration}{BESIII}),
\newblock \bibinfo{title}{{Measurement of the absolute branching fraction of
  $\Lambda_c^+\to p K^0_{\mathrm{S}}\eta$ decays}},
\newblock \bibinfo{journal}{Phys. Lett. B} \bibinfo{volume}{817}
  (\bibinfo{year}{2021}) \bibinfo{pages}{136327}.
  \DOIprefix\doi{10.1016/j.physletb.2021.136327}.
  \href{http://arxiv.org/abs/2012.11106}{{\tt arXiv:2012.11106}}.
\bibitem[{Ablikim et~al.(2022{\natexlab{a}})}]{BESIII:2022bkj}
\bibinfo{author}{M.~Ablikim}, et~al. (\bibinfo{collaboration}{BESIII}),
\newblock \bibinfo{title}{{Observation of the Singly Cabibbo Suppressed Decay
  $\Lambda^+_c \to n \pi^+$}},
\newblock \bibinfo{journal}{Phys. Rev. Lett.} \bibinfo{volume}{128}
  (\bibinfo{year}{2022}{\natexlab{a}}) \bibinfo{pages}{142001}.
  \DOIprefix\doi{10.1103/PhysRevLett.128.142001}.
  \href{http://arxiv.org/abs/2201.02056}{{\tt arXiv:2201.02056}}.
\bibitem[{Ablikim et~al.(2022{\natexlab{b}})}]{BESIII:2022wxj}
\bibinfo{author}{M.~Ablikim}, et~al. (\bibinfo{collaboration}{BESIII}),
\newblock \bibinfo{title}{{Measurement of Branching Fractions of Singly
  Cabibbo-suppressed Decays $\Lambda_c^+ \rightarrow \Sigma^{0} K^+$ and
  $\Sigma^{+} K_{S}^0$}},
\newblock \bibinfo{journal}{Phys. Rev. D} \bibinfo{volume}{106}
  (\bibinfo{year}{2022}{\natexlab{b}}) \bibinfo{pages}{052003}.
  \DOIprefix\doi{10.1103/PhysRevD.106.052003}.
  \href{http://arxiv.org/abs/2207.10906}{{\tt arXiv:2207.10906}}.
\bibitem[{Ablikim et~al.(2022{\natexlab{c}})}]{BESIII:2022qaf}
\bibinfo{author}{M.~Ablikim}, et~al. (\bibinfo{collaboration}{BESIII}),
\newblock \bibinfo{title}{{First observation of the semileptonic decay
  \ensuremath{\Lambda}c+\textrightarrow{}pK-e+\ensuremath{\nu}e}},
\newblock \bibinfo{journal}{Phys. Rev. D} \bibinfo{volume}{106}
  (\bibinfo{year}{2022}{\natexlab{c}}) \bibinfo{pages}{112010}.
  \DOIprefix\doi{10.1103/PhysRevD.106.112010}.
  \href{http://arxiv.org/abs/2207.11483}{{\tt arXiv:2207.11483}}.
\bibitem[{Ablikim et~al.(2022{\natexlab{d}})}]{BESIII:2022izy}
\bibinfo{author}{M.~Ablikim}, et~al. (\bibinfo{collaboration}{BESIII}),
\newblock \bibinfo{title}{{Measurement of the absolute branching fraction of
  the singly Cabibbo suppressed decay $\Lambda^{+}_{c}\to p\eta^{\prime}$}},
\newblock \bibinfo{journal}{Phys. Rev. D} \bibinfo{volume}{106}
  (\bibinfo{year}{2022}{\natexlab{d}}) \bibinfo{pages}{072002}.
  \DOIprefix\doi{10.1103/PhysRevD.106.072002}.
  \href{http://arxiv.org/abs/2207.14461}{{\tt arXiv:2207.14461}}.
\bibitem[{Ablikim et~al.(2022{\natexlab{e}})}]{BESIII:2022ysa}
\bibinfo{author}{M.~Ablikim}, et~al. (\bibinfo{collaboration}{BESIII}),
\newblock \bibinfo{title}{{Study of the Semileptonic Decay
  \ensuremath{\Lambda}c+\textrightarrow{}\ensuremath{\Lambda}e+\ensuremath{\nu}e}},
\newblock \bibinfo{journal}{Phys. Rev. Lett.} \bibinfo{volume}{129}
  (\bibinfo{year}{2022}{\natexlab{e}}) \bibinfo{pages}{231803}.
  \DOIprefix\doi{10.1103/PhysRevLett.129.231803}.
  \href{http://arxiv.org/abs/2207.14149}{{\tt arXiv:2207.14149}}.
\bibitem[{Ablikim et~al.(2022{\natexlab{f}})}]{BESIII:2022vrr}
\bibinfo{author}{M.~Ablikim}, et~al. (\bibinfo{collaboration}{BESIII}),
\newblock \bibinfo{title}{{Search for a massless dark photon in
  \ensuremath{\Lambda}c+\textrightarrow{}p\ensuremath{\gamma}' decay}},
\newblock \bibinfo{journal}{Phys. Rev. D} \bibinfo{volume}{106}
  (\bibinfo{year}{2022}{\natexlab{f}}) \bibinfo{pages}{072008}.
  \DOIprefix\doi{10.1103/PhysRevD.106.072008}.
  \href{http://arxiv.org/abs/2208.04496}{{\tt arXiv:2208.04496}}.
\bibitem[{Ablikim et~al.(2022{\natexlab{g}})}]{BESIII:2022tnm}
\bibinfo{author}{M.~Ablikim}, et~al. (\bibinfo{collaboration}{BESIII}),
\newblock \bibinfo{title}{{Measurement of the branching fraction of the singly
  Cabibbo-suppressed decay
  \ensuremath{\Lambda}c+\textrightarrow{}\ensuremath{\Lambda}K+}},
\newblock \bibinfo{journal}{Phys. Rev. D} \bibinfo{volume}{106}
  (\bibinfo{year}{2022}{\natexlab{g}}) \bibinfo{pages}{L111101}.
  \DOIprefix\doi{10.1103/PhysRevD.106.L111101}.
  \href{http://arxiv.org/abs/2208.04001}{{\tt arXiv:2208.04001}}.
\bibitem[{Ablikim et~al.(2022{\natexlab{h}})}]{BESIII:2022udq}
\bibinfo{author}{M.~Ablikim}, et~al. (\bibinfo{collaboration}{BESIII}),
\newblock \bibinfo{title}{{Partial wave analysis of the charmed baryon hadronic
  decay $ {\Lambda}_c^{+} $\textrightarrow{}
  \ensuremath{\Lambda}\ensuremath{\pi}$^{+}$\ensuremath{\pi}$^{0}$}},
\newblock \bibinfo{journal}{JHEP} \bibinfo{volume}{12}
  (\bibinfo{year}{2022}{\natexlab{h}}) \bibinfo{pages}{033}.
  \DOIprefix\doi{10.1007/JHEP12(2022)033}.
  \href{http://arxiv.org/abs/2209.08464}{{\tt arXiv:2209.08464}}.
\bibitem[{Ablikim et~al.(2023{\natexlab{a}})}]{BESIII:2022xne}
\bibinfo{author}{M.~Ablikim}, et~al. (\bibinfo{collaboration}{(BESIII),,
  BESIII}),
\newblock \bibinfo{title}{{Observations of the Cabibbo-Suppressed decays
  $\Lambda_{c}^{+}\to n\pi^{+}\pi^{0}$, $n\pi^{+}\pi^{-}\pi^{+}$ and the
  Cabibbo-Favored decay $\Lambda_{c}^{+}\to nK^{-}\pi^{+}\pi^{+}$}},
\newblock \bibinfo{journal}{Chin. Phys. C} \bibinfo{volume}{47}
  (\bibinfo{year}{2023}{\natexlab{a}}) \bibinfo{pages}{023001}.
  \DOIprefix\doi{10.1088/1674-1137/ac9d29}.
  \href{http://arxiv.org/abs/2210.03375}{{\tt arXiv:2210.03375}}.
\bibitem[{Ablikim et~al.(2023{\natexlab{b}})}]{BESIII:2022cmg}
\bibinfo{author}{M.~Ablikim}, et~al. (\bibinfo{collaboration}{BESIII}),
\newblock \bibinfo{title}{{Improved measurement of the absolute branching
  fraction of inclusive semileptonic \ensuremath{\Lambda}c+ decay}},
\newblock \bibinfo{journal}{Phys. Rev. D} \bibinfo{volume}{107}
  (\bibinfo{year}{2023}{\natexlab{b}}) \bibinfo{pages}{052005}.
  \DOIprefix\doi{10.1103/PhysRevD.107.052005}.
  \href{http://arxiv.org/abs/2212.03753}{{\tt arXiv:2212.03753}}.
\bibitem[{Ablikim et~al.(2023{\natexlab{c}})}]{BESIII:2022aok}
\bibinfo{author}{M.~Ablikim}, et~al. (\bibinfo{collaboration}{BESIII}),
\newblock \bibinfo{title}{{Search for the weak radiative decay
  \ensuremath{\Lambda}c+\textrightarrow{}\ensuremath{\Sigma}+\ensuremath{\gamma}
  at BESIII}},
\newblock \bibinfo{journal}{Phys. Rev. D} \bibinfo{volume}{107}
  (\bibinfo{year}{2023}{\natexlab{c}}) \bibinfo{pages}{052002}.
  \DOIprefix\doi{10.1103/PhysRevD.107.052002}.
  \href{http://arxiv.org/abs/2212.07214}{{\tt arXiv:2212.07214}}.
\bibitem[{Ablikim et~al.(2023{\natexlab{d}})}]{BESIII:2023jem}
\bibinfo{author}{M.~Ablikim}, et~al. (\bibinfo{collaboration}{BESIII}),
\newblock \bibinfo{title}{{Search for the semi-leptonic decays
  \ensuremath{\Lambda}c+\textrightarrow{}\ensuremath{\Lambda}\ensuremath{\pi}+\ensuremath{\pi}\ensuremath{-}e+\ensuremath{\nu}e
  and
  \ensuremath{\Lambda}c+\textrightarrow{}pKS0\ensuremath{\pi}\ensuremath{-}e+\ensuremath{\nu}e}},
\newblock \bibinfo{journal}{Phys. Lett. B} \bibinfo{volume}{843}
  (\bibinfo{year}{2023}{\natexlab{d}}) \bibinfo{pages}{137993}.
  \DOIprefix\doi{10.1016/j.physletb.2023.137993}.
  \href{http://arxiv.org/abs/2302.07529}{{\tt arXiv:2302.07529}}.
\bibitem[{Ablikim et~al.(2023{\natexlab{e}})}]{BESIII:2023rky}
\bibinfo{author}{M.~Ablikim}, et~al. (\bibinfo{collaboration}{BESIII}),
\newblock \bibinfo{title}{{Measurement of branching fractions of
  ${\Lambda}_{\textrm{c}}^{+}$ decays to
  $\ensuremath{\Sigma}^{+}{K}^{+}{K}^{-}$,
  $\ensuremath{\Sigma}^{+}\ensuremath{\phi}$ and
  $\ensuremath{\Sigma}^{+}{K}^{+}\ensuremath{\pi}^{-}(\ensuremath{\pi}^{0}$)}},
\newblock \bibinfo{journal}{JHEP} \bibinfo{volume}{09}
  (\bibinfo{year}{2023}{\natexlab{e}}) \bibinfo{pages}{125}.
  \DOIprefix\doi{10.1007/JHEP09(2023)125}.
  \href{http://arxiv.org/abs/2304.09405}{{\tt arXiv:2304.09405}}.
\bibitem[{Ablikim et~al.(2023{\natexlab{f}})}]{BESIII:2023vfi}
\bibinfo{author}{M.~Ablikim}, et~al. (\bibinfo{collaboration}{BESIII}),
\newblock \bibinfo{title}{{Study of $\Lambda_c^+ \rightarrow \Lambda
  \mu^+\nu_{\mu}$ and test of lepton flavor universality with $\Lambda_c^+
  \rightarrow \Lambda \ell^+\nu_{\ell}$ decays}},
\newblock \bibinfo{journal}{Phys. Rev. D} \bibinfo{volume}{108}
  (\bibinfo{year}{2023}{\natexlab{f}}) \bibinfo{pages}{L031105}.
  \DOIprefix\doi{10.1103/PhysRevD.108.L031105}.
  \href{http://arxiv.org/abs/2306.02624}{{\tt arXiv:2306.02624}}.
\bibitem[{Ablikim et~al.(2023{\natexlab{g}})}]{BESIII:2023ooh}
\bibinfo{author}{M.~Ablikim}, et~al. (\bibinfo{collaboration}{BESIII}),
\newblock \bibinfo{title}{{Measurement of the branching fractions of the singly
  Cabibbo-suppressed decays $ {\Lambda}_{\textrm{c}}^{+}\to \textrm{p}\eta $
  and $ {\Lambda}_{\textrm{c}}^{+}\to \textrm{p}\omega $}},
\newblock \bibinfo{journal}{JHEP} \bibinfo{volume}{11}
  (\bibinfo{year}{2023}{\natexlab{g}}) \bibinfo{pages}{137}.
  \DOIprefix\doi{10.1007/JHEP11(2023)137}.
  \href{http://arxiv.org/abs/2307.09266}{{\tt arXiv:2307.09266}}.
\bibitem[{Ablikim et~al.(2024{\natexlab{a}})}]{BESIII:2023wrw}
\bibinfo{author}{M.~Ablikim}, et~al. (\bibinfo{collaboration}{BESIII}),
\newblock \bibinfo{title}{{First Measurement of the Decay Asymmetry in the Pure
  W-Boson-Exchange Decay
  \ensuremath{\Lambda}c+\textrightarrow{}\ensuremath{\Xi}0K+}},
\newblock \bibinfo{journal}{Phys. Rev. Lett.} \bibinfo{volume}{132}
  (\bibinfo{year}{2024}{\natexlab{a}}) \bibinfo{pages}{031801}.
  \DOIprefix\doi{10.1103/PhysRevLett.132.031801}.
  \href{http://arxiv.org/abs/2309.02774}{{\tt arXiv:2309.02774}}.
\bibitem[{Ablikim et~al.(2024{\natexlab{b}})}]{BESIII:2023iwu}
\bibinfo{author}{M.~Ablikim}, et~al. (\bibinfo{collaboration}{BESIII}),
\newblock \bibinfo{title}{{Observation of the singly Cabibbo-suppressed decay
  \ensuremath{\Lambda}c+\textrightarrow{}\ensuremath{\Sigma}-K+\ensuremath{\pi}+}},
\newblock \bibinfo{journal}{Phys. Rev. D} \bibinfo{volume}{109}
  (\bibinfo{year}{2024}{\natexlab{b}}) \bibinfo{pages}{L071103}.
  \DOIprefix\doi{10.1103/PhysRevD.109.L071103}.
  \href{http://arxiv.org/abs/2309.05484}{{\tt arXiv:2309.05484}}.
\bibitem[{Ablikim et~al.(2024{\natexlab{c}})}]{BESIII:2023dvx}
\bibinfo{author}{M.~Ablikim}, et~al. (\bibinfo{collaboration}{BESIII}),
\newblock \bibinfo{title}{{Measurement of the absolute branching fraction of
  the three-body decay
  \ensuremath{\Lambda}c+\textrightarrow{}\ensuremath{\Xi}0K+\ensuremath{\pi}0
  and search for \ensuremath{\Lambda}c+\textrightarrow{}nK+\ensuremath{\pi}0,
  \ensuremath{\Sigma}0K+\ensuremath{\pi}0, and
  \ensuremath{\Lambda}K+\ensuremath{\pi}0}},
\newblock \bibinfo{journal}{Phys. Rev. D} \bibinfo{volume}{109}
  (\bibinfo{year}{2024}{\natexlab{c}}) \bibinfo{pages}{052001}.
  \DOIprefix\doi{10.1103/PhysRevD.109.052001}.
  \href{http://arxiv.org/abs/2311.02347}{{\tt arXiv:2311.02347}}.
\bibitem[{Ablikim et~al.(2024{\natexlab{d}})}]{BESIII:2023uvs}
\bibinfo{author}{M.~Ablikim}, et~al. (\bibinfo{collaboration}{BESIII}),
\newblock \bibinfo{title}{{Evidence of the singly Cabibbo suppressed decay
  \ensuremath{\Lambda}c+\textrightarrow{}p\ensuremath{\pi}0}},
\newblock \bibinfo{journal}{Phys. Rev. D} \bibinfo{volume}{109}
  (\bibinfo{year}{2024}{\natexlab{d}}) \bibinfo{pages}{L091101}.
  \DOIprefix\doi{10.1103/PhysRevD.109.L091101}.
  \href{http://arxiv.org/abs/2311.06883}{{\tt arXiv:2311.06883}}.
\bibitem[{Ablikim et~al.(2024{\natexlab{e}})}]{BESIII:2023sdr}
\bibinfo{author}{M.~Ablikim}, et~al. (\bibinfo{collaboration}{BESIII}),
\newblock \bibinfo{title}{{First observation of
  \ensuremath{\Lambda}c+\textrightarrow{}\ensuremath{\Lambda}K+\ensuremath{\pi}0
  and evidence of
  \ensuremath{\Lambda}c+\textrightarrow{}\ensuremath{\Lambda}K+\ensuremath{\pi}+\ensuremath{\pi}-}},
\newblock \bibinfo{journal}{Phys. Rev. D} \bibinfo{volume}{109}
  (\bibinfo{year}{2024}{\natexlab{e}}) \bibinfo{pages}{032003}.
  \DOIprefix\doi{10.1103/PhysRevD.109.032003}.
  \href{http://arxiv.org/abs/2311.12903}{{\tt arXiv:2311.12903}}.
\bibitem[{Ablikim et~al.(2024{\natexlab{f}})}]{BESIII:2023pia}
\bibinfo{author}{M.~Ablikim}, et~al. (\bibinfo{collaboration}{BESIII}),
\newblock \bibinfo{title}{{Measurement of branching fractions for
  \ensuremath{\Lambda}c+\textrightarrow{}nKS0\ensuremath{\pi}+ and
  \ensuremath{\Lambda}c+\textrightarrow{}nKS0K+}},
\newblock \bibinfo{journal}{Phys. Rev. D} \bibinfo{volume}{109}
  (\bibinfo{year}{2024}{\natexlab{f}}) \bibinfo{pages}{072010}.
  \DOIprefix\doi{10.1103/PhysRevD.109.072010}.
  \href{http://arxiv.org/abs/2311.17131}{{\tt arXiv:2311.17131}}.
\bibitem[{Ablikim et~al.(2024{\natexlab{g}})}]{BESIII:2024xgl}
\bibinfo{author}{M.~Ablikim}, et~al. (\bibinfo{collaboration}{BESIII}),
\newblock \bibinfo{title}{{First observation of the decay
  \ensuremath{\Lambda}c+\textrightarrow{}nKS0\ensuremath{\pi}+\ensuremath{\pi}0}},
\newblock \bibinfo{journal}{Phys. Rev. D} \bibinfo{volume}{109}
  (\bibinfo{year}{2024}{\natexlab{g}}) \bibinfo{pages}{053005}.
  \DOIprefix\doi{10.1103/PhysRevD.109.053005}.
  \href{http://arxiv.org/abs/2401.06813}{{\tt arXiv:2401.06813}}.
\bibitem[{Ablikim et~al.(2024{\natexlab{h}})}]{BESIII:2024sfz}
\bibinfo{author}{M.~Ablikim}, et~al. (\bibinfo{collaboration}{BESIII}),
\newblock \bibinfo{title}{{Measurements of $K_S^0$-$K_L^0$ asymmetries in the
  decays $\Lambda_c^+ \to pK_{L,S}^0$, $pK_{L,S}^0\pi^+\pi^-$ and
  $pK_{L,S}^0\pi^0$}}  (\bibinfo{year}{2024}{\natexlab{h}}).
  \href{http://arxiv.org/abs/2406.18083}{{\tt arXiv:2406.18083}}.
\bibitem[{Ablikim et~al.(2024{\natexlab{i}})}]{BESIII:2024mbf}
\bibinfo{author}{M.~Ablikim}, et~al. (\bibinfo{collaboration}{BESIII}),
\newblock \bibinfo{title}{{Observation of $\Lambda_c^+ \to \Lambda a_0(980)^+$
  and Evidence for $\Sigma(1380)^+$ in $\Lambda_c^+ \to \Lambda \pi^+ \eta$}}
  (\bibinfo{year}{2024}{\natexlab{i}}).
  \href{http://arxiv.org/abs/2407.12270}{{\tt arXiv:2407.12270}}.
\bibitem[{Savage(1991)}]{Savage:1991wu}
\bibinfo{author}{M.~J. Savage},
\newblock \bibinfo{title}{{SU(3) violations in the nonleptonic decay of charmed
  hadrons}},
\newblock \bibinfo{journal}{Phys. Lett. B} \bibinfo{volume}{257}
  (\bibinfo{year}{1991}) \bibinfo{pages}{414--418}.
  \DOIprefix\doi{10.1016/0370-2693(91)91917-K}.
\bibitem[{Sheikholeslami et~al.(1991)Sheikholeslami, Khanna, and
  Verma}]{Sheikholeslami:1991ab}
\bibinfo{author}{S.~M. Sheikholeslami}, \bibinfo{author}{M.~P. Khanna},
  \bibinfo{author}{R.~C. Verma},
\newblock \bibinfo{title}{{Cabibbo enhanced weak decays of charmed baryons in
  the SU(4) semidynamical scheme}},
\newblock \bibinfo{journal}{Phys. Rev. D} \bibinfo{volume}{43}
  (\bibinfo{year}{1991}) \bibinfo{pages}{170--178}.
  \DOIprefix\doi{10.1103/PhysRevD.43.170}.
\bibitem[{Verma and Khanna(1996)}]{Verma:1995dk}
\bibinfo{author}{R.~C. Verma}, \bibinfo{author}{M.~P. Khanna},
\newblock \bibinfo{title}{{Cabibbo favored hadronic decays of charmed baryons
  in flavor SU(3)}},
\newblock \bibinfo{journal}{Phys. Rev. D} \bibinfo{volume}{53}
  (\bibinfo{year}{1996}) \bibinfo{pages}{3723--3730}.
  \DOIprefix\doi{10.1103/PhysRevD.53.3723}.
  \href{http://arxiv.org/abs/hep-ph/9506394}{{\tt arXiv:hep-ph/9506394}}.
\bibitem[{Sharma and Verma(1997)}]{Sharma:1996sc}
\bibinfo{author}{K.~K. Sharma}, \bibinfo{author}{R.~C. Verma},
\newblock \bibinfo{title}{{SU(3) flavor analysis of two-body weak decays of
  charmed baryons}},
\newblock \bibinfo{journal}{Phys. Rev. D} \bibinfo{volume}{55}
  (\bibinfo{year}{1997}) \bibinfo{pages}{7067--7074}.
  \DOIprefix\doi{10.1103/PhysRevD.55.7067}.
  \href{http://arxiv.org/abs/hep-ph/9704391}{{\tt arXiv:hep-ph/9704391}}.
\bibitem[{Chen et~al.(2003)Chen, Guo, Li, and Wang}]{Chen:2002jr}
\bibinfo{author}{S.-L. Chen}, \bibinfo{author}{X.-H. Guo},
  \bibinfo{author}{X.-Q. Li}, \bibinfo{author}{G.-L. Wang},
\newblock \bibinfo{title}{{Cabibbo suppressed nonleptonic decays of Lambda(c)
  and final state interaction}},
\newblock \bibinfo{journal}{Commun. Theor. Phys.} \bibinfo{volume}{40}
  (\bibinfo{year}{2003}) \bibinfo{pages}{563--572}.
  \DOIprefix\doi{10.1088/0253-6102/40/5/563}.
  \href{http://arxiv.org/abs/hep-ph/0208006}{{\tt arXiv:hep-ph/0208006}}.
\bibitem[{L\"u et~al.(2016)L\"u, Wang, and Yu}]{Lu:2016ogy}
\bibinfo{author}{C.-D. L\"u}, \bibinfo{author}{W.~Wang}, \bibinfo{author}{F.-S.
  Yu},
\newblock \bibinfo{title}{{Test flavor SU(3) symmetry in exclusive $\Lambda_c$
  decays}},
\newblock \bibinfo{journal}{Phys. Rev. D} \bibinfo{volume}{93}
  (\bibinfo{year}{2016}) \bibinfo{pages}{056008}.
  \DOIprefix\doi{10.1103/PhysRevD.93.056008}.
  \href{http://arxiv.org/abs/1601.04241}{{\tt arXiv:1601.04241}}.
\bibitem[{Wang et~al.(2018)Wang, Guo, Long, and Yu}]{Wang:2017gxe}
\bibinfo{author}{D.~Wang}, \bibinfo{author}{P.-F. Guo}, \bibinfo{author}{W.-H.
  Long}, \bibinfo{author}{F.-S. Yu},
\newblock \bibinfo{title}{{K$_{S}^{0}$ \ensuremath{-} K$_{L}^{0}$ asymmetries
  and CP violation in charmed baryon decays into neutral kaons}},
\newblock \bibinfo{journal}{JHEP} \bibinfo{volume}{03} (\bibinfo{year}{2018})
  \bibinfo{pages}{066}. \DOIprefix\doi{10.1007/JHEP03(2018)066}.
  \href{http://arxiv.org/abs/1709.09873}{{\tt arXiv:1709.09873}}.
\bibitem[{Geng et~al.(2018)Geng, Hsiao, Lin, and Liu}]{Geng:2017esc}
\bibinfo{author}{C.~Q. Geng}, \bibinfo{author}{Y.~K. Hsiao},
  \bibinfo{author}{Y.-H. Lin}, \bibinfo{author}{L.-L. Liu},
\newblock \bibinfo{title}{{Non-leptonic two-body weak decays of
  $\Lambda_c(2286)$}},
\newblock \bibinfo{journal}{Phys. Lett. B} \bibinfo{volume}{776}
  (\bibinfo{year}{2018}) \bibinfo{pages}{265--269}.
  \DOIprefix\doi{10.1016/j.physletb.2017.11.062}.
  \href{http://arxiv.org/abs/1708.02460}{{\tt arXiv:1708.02460}}.
\bibitem[{Geng et~al.(2017)Geng, Hsiao, Liu, and Tsai}]{Geng:2017mxn}
\bibinfo{author}{C.~Q. Geng}, \bibinfo{author}{Y.~K. Hsiao},
  \bibinfo{author}{C.-W. Liu}, \bibinfo{author}{T.-H. Tsai},
\newblock \bibinfo{title}{{Charmed Baryon Weak Decays with SU(3) Flavor
  Symmetry}},
\newblock \bibinfo{journal}{JHEP} \bibinfo{volume}{11} (\bibinfo{year}{2017})
  \bibinfo{pages}{147}. \DOIprefix\doi{10.1007/JHEP11(2017)147}.
  \href{http://arxiv.org/abs/1709.00808}{{\tt arXiv:1709.00808}}.
\bibitem[{Cheng et~al.(2018)Cheng, Kang, and Xu}]{Cheng:2018hwl}
\bibinfo{author}{H.-Y. Cheng}, \bibinfo{author}{X.-W. Kang},
  \bibinfo{author}{F.~Xu},
\newblock \bibinfo{title}{{Singly Cabibbo-suppressed hadronic decays of
  $\Lambda_c^+$}},
\newblock \bibinfo{journal}{Phys. Rev. D} \bibinfo{volume}{97}
  (\bibinfo{year}{2018}) \bibinfo{pages}{074028}.
  \DOIprefix\doi{10.1103/PhysRevD.97.074028}.
  \href{http://arxiv.org/abs/1801.08625}{{\tt arXiv:1801.08625}}.
\bibitem[{Zou et~al.(2020)Zou, Xu, Meng, and Cheng}]{Zou:2019kzq}
\bibinfo{author}{J.~Zou}, \bibinfo{author}{F.~Xu}, \bibinfo{author}{G.~Meng},
  \bibinfo{author}{H.-Y. Cheng},
\newblock \bibinfo{title}{{Two-body hadronic weak decays of antitriplet charmed
  baryons}},
\newblock \bibinfo{journal}{Phys. Rev. D} \bibinfo{volume}{101}
  (\bibinfo{year}{2020}) \bibinfo{pages}{014011}.
  \DOIprefix\doi{10.1103/PhysRevD.101.014011}.
  \href{http://arxiv.org/abs/1910.13626}{{\tt arXiv:1910.13626}}.
\bibitem[{Jia et~al.(2020)Jia, Wang, and Yu}]{Jia:2019zxi}
\bibinfo{author}{C.-P. Jia}, \bibinfo{author}{D.~Wang}, \bibinfo{author}{F.-S.
  Yu},
\newblock \bibinfo{title}{{Charmed baryon decays in $SU(3)_F$ symmetry}},
\newblock \bibinfo{journal}{Nucl. Phys. B} \bibinfo{volume}{956}
  (\bibinfo{year}{2020}) \bibinfo{pages}{115048}.
  \DOIprefix\doi{10.1016/j.nuclphysb.2020.115048}.
  \href{http://arxiv.org/abs/1910.00876}{{\tt arXiv:1910.00876}}.
\bibitem[{Hsiao et~al.(2019)Hsiao, Yao, and Zhao}]{Hsiao:2019yur}
\bibinfo{author}{Y.~K. Hsiao}, \bibinfo{author}{Y.~Yao}, \bibinfo{author}{H.~J.
  Zhao},
\newblock \bibinfo{title}{{Two-body charmed baryon decays involving vector
  meson with $SU(3)$ flavor symmetry}},
\newblock \bibinfo{journal}{Phys. Lett. B} \bibinfo{volume}{792}
  (\bibinfo{year}{2019}) \bibinfo{pages}{35--39}.
  \DOIprefix\doi{10.1016/j.physletb.2019.03.031}.
  \href{http://arxiv.org/abs/1902.08783}{{\tt arXiv:1902.08783}}.
\bibitem[{Geng et~al.(2019)Geng, Liu, and Tsai}]{Geng:2019xbo}
\bibinfo{author}{C.~Q. Geng}, \bibinfo{author}{C.-W. Liu},
  \bibinfo{author}{T.-H. Tsai},
\newblock \bibinfo{title}{{Asymmetries of anti-triplet charmed baryon decays}},
\newblock \bibinfo{journal}{Phys. Lett. B} \bibinfo{volume}{794}
  (\bibinfo{year}{2019}) \bibinfo{pages}{19--28}.
  \DOIprefix\doi{10.1016/j.physletb.2019.05.024}.
  \href{http://arxiv.org/abs/1902.06189}{{\tt arXiv:1902.06189}}.
\bibitem[{Cen et~al.(2019)Cen, Geng, Liu, and Tsai}]{Cen:2019ims}
\bibinfo{author}{J.-Y. Cen}, \bibinfo{author}{C.-Q. Geng},
  \bibinfo{author}{C.-W. Liu}, \bibinfo{author}{T.-H. Tsai},
\newblock \bibinfo{title}{{Up-down asymmetries of charmed baryon three-body
  decays}},
\newblock \bibinfo{journal}{Eur. Phys. J. C} \bibinfo{volume}{79}
  (\bibinfo{year}{2019}) \bibinfo{pages}{946}.
  \DOIprefix\doi{10.1140/epjc/s10052-019-7467-z}.
  \href{http://arxiv.org/abs/1906.01848}{{\tt arXiv:1906.01848}}.
\bibitem[{Meng et~al.(2020)Meng, Wong, and Xu}]{Meng:2020euv}
\bibinfo{author}{G.~Meng}, \bibinfo{author}{S.~M.-Y. Wong},
  \bibinfo{author}{F.~Xu},
\newblock \bibinfo{title}{{Doubly Cabibbo-suppressed decays of antitriplet
  charmed baryons}},
\newblock \bibinfo{journal}{JHEP} \bibinfo{volume}{11} (\bibinfo{year}{2020})
  \bibinfo{pages}{126}. \DOIprefix\doi{10.1007/JHEP11(2020)126}.
  \href{http://arxiv.org/abs/2005.12111}{{\tt arXiv:2005.12111}}.
\bibitem[{Geng et~al.(2020{\natexlab{a}})Geng, Liu, and Tsai}]{Geng:2020zgr}
\bibinfo{author}{C.~Q. Geng}, \bibinfo{author}{C.-W. Liu},
  \bibinfo{author}{T.-H. Tsai},
\newblock \bibinfo{title}{{Charmed Baryon Weak Decays with Vector Mesons}},
\newblock \bibinfo{journal}{Phys. Rev. D} \bibinfo{volume}{101}
  (\bibinfo{year}{2020}{\natexlab{a}}) \bibinfo{pages}{053002}.
  \DOIprefix\doi{10.1103/PhysRevD.101.053002}.
  \href{http://arxiv.org/abs/2001.05079}{{\tt arXiv:2001.05079}}.
\bibitem[{Geng et~al.(2020{\natexlab{b}})Geng, Liu, and Tsai}]{Geng:2020tlx}
\bibinfo{author}{C.~Q. Geng}, \bibinfo{author}{C.-W. Liu},
  \bibinfo{author}{T.-H. Tsai},
\newblock \bibinfo{title}{{Mixing effects of $\Sigma^0-\Lambda^0$ in
  $\Lambda_c^+$ decays}},
\newblock \bibinfo{journal}{Phys. Rev. D} \bibinfo{volume}{101}
  (\bibinfo{year}{2020}{\natexlab{b}}) \bibinfo{pages}{054005}.
  \DOIprefix\doi{10.1103/PhysRevD.101.054005}.
  \href{http://arxiv.org/abs/2002.09583}{{\tt arXiv:2002.09583}}.
\bibitem[{Ke and Li(2020)}]{Ke:2020uks}
\bibinfo{author}{H.-W. Ke}, \bibinfo{author}{X.-Q. Li},
\newblock \bibinfo{title}{{A natural interpretation on the data of
  $\Lambda_c\to\Sigma\pi$}},
\newblock \bibinfo{journal}{Phys. Rev. D} \bibinfo{volume}{102}
  (\bibinfo{year}{2020}) \bibinfo{pages}{113013}.
  \DOIprefix\doi{10.1103/PhysRevD.102.113013}.
  \href{http://arxiv.org/abs/2008.12163}{{\tt arXiv:2008.12163}}.
\bibitem[{Li et~al.(2021)Li, Liu, and Yu}]{Li:2021qod}
\bibinfo{author}{Y.-S. Li}, \bibinfo{author}{X.~Liu}, \bibinfo{author}{F.-S.
  Yu},
\newblock \bibinfo{title}{{Revisiting semileptonic decays of $\Lambda_{b(c)}$
  supported by baryon spectroscopy}},
\newblock \bibinfo{journal}{Phys. Rev. D} \bibinfo{volume}{104}
  (\bibinfo{year}{2021}) \bibinfo{pages}{013005}.
  \DOIprefix\doi{10.1103/PhysRevD.104.013005}.
  \href{http://arxiv.org/abs/2104.04962}{{\tt arXiv:2104.04962}}.
\bibitem[{Xu et~al.(2021)Xu, Wen, and Zhong}]{Xu:2021mkg}
\bibinfo{author}{F.~Xu}, \bibinfo{author}{Q.~Wen}, \bibinfo{author}{H.~Zhong},
\newblock \bibinfo{title}{{$K_S^0-K_L^0$ Asymmetries in Weak Decays of Charmed
  Baryons}},
\newblock \bibinfo{journal}{LHEP} \bibinfo{volume}{2021} (\bibinfo{year}{2021})
  \bibinfo{pages}{218}. \DOIprefix\doi{10.31526/lhep.2021.218}.
\bibitem[{Hsiao et~al.(2022)Hsiao, Wang, and Zhao}]{Hsiao:2021nsc}
\bibinfo{author}{Y.~K. Hsiao}, \bibinfo{author}{Y.~L. Wang},
  \bibinfo{author}{H.~J. Zhao},
\newblock \bibinfo{title}{{Equivalent SU(3)$_{f}$ approaches for two-body
  anti-triplet charmed baryon decays}},
\newblock \bibinfo{journal}{JHEP} \bibinfo{volume}{09} (\bibinfo{year}{2022})
  \bibinfo{pages}{035}. \DOIprefix\doi{10.1007/JHEP09(2022)035}.
  \href{http://arxiv.org/abs/2111.04124}{{\tt arXiv:2111.04124}}.
\bibitem[{Wang and Yu(2024)}]{Wang:2022tcm}
\bibinfo{author}{J.-P. Wang}, \bibinfo{author}{F.-S. Yu},
\newblock \bibinfo{title}{{Probing hyperon CP violation with charmed baryon
  decays}},
\newblock \bibinfo{journal}{Phys. Lett. B} \bibinfo{volume}{849}
  (\bibinfo{year}{2024}) \bibinfo{pages}{138460}.
  \DOIprefix\doi{10.1016/j.physletb.2024.138460}.
  \href{http://arxiv.org/abs/2208.01589}{{\tt arXiv:2208.01589}}.
\bibitem[{Cheng and Xu(2022)}]{Cheng:2022kea}
\bibinfo{author}{H.-Y. Cheng}, \bibinfo{author}{F.~Xu},
\newblock \bibinfo{title}{{Heavy-flavor-conserving hadronic weak decays of
  charmed and bottom baryons}},
\newblock \bibinfo{journal}{Phys. Rev. D} \bibinfo{volume}{105}
  (\bibinfo{year}{2022}) \bibinfo{pages}{094011}.
  \DOIprefix\doi{10.1103/PhysRevD.105.094011}.
  \href{http://arxiv.org/abs/2204.03149}{{\tt arXiv:2204.03149}}.
\bibitem[{Zhong et~al.(2023)Zhong, Xu, Wen, and Gu}]{Zhong:2022exp}
\bibinfo{author}{H.~Zhong}, \bibinfo{author}{F.~Xu}, \bibinfo{author}{Q.~Wen},
  \bibinfo{author}{Y.~Gu},
\newblock \bibinfo{title}{{Weak decays of antitriplet charmed baryons from the
  perspective of flavor symmetry}},
\newblock \bibinfo{journal}{JHEP} \bibinfo{volume}{02} (\bibinfo{year}{2023})
  \bibinfo{pages}{235}. \DOIprefix\doi{10.1007/JHEP02(2023)235}.
  \href{http://arxiv.org/abs/2210.12728}{{\tt arXiv:2210.12728}}.
\bibitem[{Liu and Yang(2023)}]{Liu:2023pyk}
\bibinfo{author}{H.~Liu}, \bibinfo{author}{C.~Yang},
\newblock \bibinfo{title}{{Two-body hadronic decays of \ensuremath{\Xi}c0 in
  light front approach}},
\newblock \bibinfo{journal}{Phys. Rev. D} \bibinfo{volume}{108}
  (\bibinfo{year}{2023}) \bibinfo{pages}{093011}.
  \DOIprefix\doi{10.1103/PhysRevD.108.093011}.
  \href{http://arxiv.org/abs/2304.12128}{{\tt arXiv:2304.12128}}.
\bibitem[{Xing et~al.(2023)Xing, He, Huang, and Yang}]{Xing:2023bzh}
\bibinfo{author}{Z.-P. Xing}, \bibinfo{author}{X.-G. He},
  \bibinfo{author}{F.~Huang}, \bibinfo{author}{C.~Yang},
\newblock \bibinfo{title}{{Global analysis of measured and unmeasured hadronic
  two-body weak decays of antitriplet charmed baryons}},
\newblock \bibinfo{journal}{Phys. Rev. D} \bibinfo{volume}{108}
  (\bibinfo{year}{2023}) \bibinfo{pages}{053004}.
  \DOIprefix\doi{10.1103/PhysRevD.108.053004}.
  \href{http://arxiv.org/abs/2305.14854}{{\tt arXiv:2305.14854}}.
\bibitem[{Liu(2024)}]{Liu:2023dvg}
\bibinfo{author}{C.-W. Liu},
\newblock \bibinfo{title}{{Nonleptonic two-body weak decays of charmed
  baryons}},
\newblock \bibinfo{journal}{Phys. Rev. D} \bibinfo{volume}{109}
  (\bibinfo{year}{2024}) \bibinfo{pages}{033004}.
  \DOIprefix\doi{10.1103/PhysRevD.109.033004}.
  \href{http://arxiv.org/abs/2308.07754}{{\tt arXiv:2308.07754}}.
\bibitem[{Geng et~al.(2024)Geng, He, Jin, Liu, and Yang}]{Geng:2023pkr}
\bibinfo{author}{C.-Q. Geng}, \bibinfo{author}{X.-G. He},
  \bibinfo{author}{X.-N. Jin}, \bibinfo{author}{C.-W. Liu},
  \bibinfo{author}{C.~Yang},
\newblock \bibinfo{title}{{Complete determination of SU(3)F amplitudes and
  strong phase in \ensuremath{\Lambda}c+\textrightarrow{}\ensuremath{\Xi}0K+}},
\newblock \bibinfo{journal}{Phys. Rev. D} \bibinfo{volume}{109}
  (\bibinfo{year}{2024}) \bibinfo{pages}{L071302}.
  \DOIprefix\doi{10.1103/PhysRevD.109.L071302}.
  \href{http://arxiv.org/abs/2310.05491}{{\tt arXiv:2310.05491}}.
\bibitem[{Shi and Zhao(2024)}]{Shi:2024plf}
\bibinfo{author}{Y.-J. Shi}, \bibinfo{author}{Z.-X. Zhao},
\newblock \bibinfo{title}{{Light-cone sum rules study on the purely
  non-factorizable $\Lambda_{c}^{+}\to\Xi^{0}K^{+}$ decay}}
  (\bibinfo{year}{2024}). \href{http://arxiv.org/abs/2407.07431}{{\tt
  arXiv:2407.07431}}.
\bibitem[{Zhong et~al.(2024{\natexlab{a}})Zhong, Xu, and Cheng}]{Zhong:2024zme}
\bibinfo{author}{H.~Zhong}, \bibinfo{author}{F.~Xu}, \bibinfo{author}{H.-Y.
  Cheng},
\newblock \bibinfo{title}{{Topological Diagrams and Hadronic Weak Decays of
  Charmed Baryons}}  (\bibinfo{year}{2024}{\natexlab{a}}).
  \href{http://arxiv.org/abs/2401.15926}{{\tt arXiv:2401.15926}}.
\bibitem[{Zhong et~al.(2024{\natexlab{b}})Zhong, Xu, and Cheng}]{Zhong:2024qqs}
\bibinfo{author}{H.~Zhong}, \bibinfo{author}{F.~Xu}, \bibinfo{author}{H.-Y.
  Cheng},
\newblock \bibinfo{title}{{Analysis of hadronic weak decays of charmed baryons
  in the topological diagrammatic approach}},
\newblock \bibinfo{journal}{Phys. Rev. D} \bibinfo{volume}{109}
  (\bibinfo{year}{2024}{\natexlab{b}}) \bibinfo{pages}{114027}.
  \DOIprefix\doi{10.1103/PhysRevD.109.114027}.
  \href{http://arxiv.org/abs/2404.01350}{{\tt arXiv:2404.01350}}.
\bibitem[{Geng et~al.(2024)Geng, Liu, and Liu}]{Geng:2024sgq}
\bibinfo{author}{C.-Q. Geng}, \bibinfo{author}{C.-W. Liu},
  \bibinfo{author}{S.-L. Liu},
\newblock \bibinfo{title}{{Nonleptonic three-body charmed baryon weak decays
  with H(15)}},
\newblock \bibinfo{journal}{Phys. Rev. D} \bibinfo{volume}{109}
  (\bibinfo{year}{2024}) \bibinfo{pages}{093002}.
  \DOIprefix\doi{10.1103/PhysRevD.109.093002}.
  \href{http://arxiv.org/abs/2403.06469}{{\tt arXiv:2403.06469}}.
\bibitem[{Wang and Luo(2024)}]{Wang:2024ztg}
\bibinfo{author}{D.~Wang}, \bibinfo{author}{J.-F. Luo},
\newblock \bibinfo{title}{{Topological amplitudes of charmed baryon decays in
  the $SU(3)_F$ limit}}  (\bibinfo{year}{2024}).
  \href{http://arxiv.org/abs/2406.14061}{{\tt arXiv:2406.14061}}.
\bibitem[{Wang(2024)}]{Wang:2024nxb}
\bibinfo{author}{D.~Wang},
\newblock \bibinfo{title}{{Topological diagram analysis of
  $\mathcal{B}_{c\overline 3}\to \mathcal{B}_{10}M$ decays in the $SU(3)_F$
  limit and beyond}}  (\bibinfo{year}{2024}).
  \href{http://arxiv.org/abs/2408.02015}{{\tt arXiv:2408.02015}}.
\bibitem[{Grossman and Schacht(2019)}]{Grossman:2018ptn}
\bibinfo{author}{Y.~Grossman}, \bibinfo{author}{S.~Schacht},
\newblock \bibinfo{title}{{U-Spin Sum Rules for CP Asymmetries of Three-Body
  Charmed Baryon Decays}},
\newblock \bibinfo{journal}{Phys. Rev. D} \bibinfo{volume}{99}
  (\bibinfo{year}{2019}) \bibinfo{pages}{033005}.
  \DOIprefix\doi{10.1103/PhysRevD.99.033005}.
  \href{http://arxiv.org/abs/1811.11188}{{\tt arXiv:1811.11188}}.
\bibitem[{Wang(2019)}]{Wang:2019dls}
\bibinfo{author}{D.~Wang},
\newblock \bibinfo{title}{{Sum rules for $CP$ asymmetries of charmed baryon
  decays in the $SU(3)_F$ limit}},
\newblock \bibinfo{journal}{Eur. Phys. J. C} \bibinfo{volume}{79}
  (\bibinfo{year}{2019}) \bibinfo{pages}{429}.
  \DOIprefix\doi{10.1140/epjc/s10052-019-6925-y}.
  \href{http://arxiv.org/abs/1901.01776}{{\tt arXiv:1901.01776}}.
\bibitem[{He and Liu(2024)}]{He:2024pxh}
\bibinfo{author}{X.-G. He}, \bibinfo{author}{C.-W. Liu},
\newblock \bibinfo{title}{{Large CP violation in charmed baryon decays}}
  (\bibinfo{year}{2024}). \href{http://arxiv.org/abs/2404.19166}{{\tt
  arXiv:2404.19166}}.
\bibitem[{Sun et~al.(2024)Sun, Zhu, and Xing}]{Sun:2024mmk}
\bibinfo{author}{J.~Sun}, \bibinfo{author}{R.~Zhu}, \bibinfo{author}{Z.-P.
  Xing},
\newblock \bibinfo{title}{{Observable CP-violation in charmed baryons decays
  with SU(3) symmetry analysis}}  (\bibinfo{year}{2024}).
  \href{http://arxiv.org/abs/2407.00426}{{\tt arXiv:2407.00426}}.
\bibitem[{Xing et~al.(2024)Xing, Shi, Sun, and Xing}]{Xing:2024nvg}
\bibinfo{author}{Z.-P. Xing}, \bibinfo{author}{Y.-J. Shi},
  \bibinfo{author}{J.~Sun}, \bibinfo{author}{Y.~Xing},
\newblock \bibinfo{title}{{SU(3) symmetry analysis in charmed baryon two body
  decays with penguin diagram contribution}}  (\bibinfo{year}{2024}).
  \href{http://arxiv.org/abs/2407.09234}{{\tt arXiv:2407.09234}}.
\bibitem[{K\"orner et~al.(1978)K\"orner, Kramer, and Willrodt}]{Korner:1978ec}
\bibinfo{author}{J.~G. K\"orner}, \bibinfo{author}{G.~Kramer},
  \bibinfo{author}{J.~Willrodt},
\newblock \bibinfo{title}{{Weak Decays of the Charmed Baryon C$_0^+$ and the
  Inclusive Yield of $\Lambda$ and $p$}},
\newblock \bibinfo{journal}{Phys. Lett. B} \bibinfo{volume}{78}
  (\bibinfo{year}{1978}) \bibinfo{pages}{492}.
  \DOIprefix\doi{10.1016/0370-2693(78)90495-1}, \bibinfo{note}{[Erratum:
  Phys.Lett.B 81, 419--419 (1979)]}.
\bibitem[{Kohara(1991)}]{Kohara:1991ug}
\bibinfo{author}{Y.~Kohara},
\newblock \bibinfo{title}{{Quark diagram analysis of charmed baryon decays}},
\newblock \bibinfo{journal}{Phys. Rev. D} \bibinfo{volume}{44}
  (\bibinfo{year}{1991}) \bibinfo{pages}{2799--2802}.
  \DOIprefix\doi{10.1103/PhysRevD.44.2799}.
\bibitem[{Cheng and Tseng(1992)}]{Cheng:1991sn}
\bibinfo{author}{H.-Y. Cheng}, \bibinfo{author}{B.~Tseng},
\newblock \bibinfo{title}{{Nonleptonic weak decays of charmed baryons}},
\newblock \bibinfo{journal}{Phys. Rev. D} \bibinfo{volume}{46}
  (\bibinfo{year}{1992}) \bibinfo{pages}{1042}.
  \DOIprefix\doi{10.1103/PhysRevD.46.1042}, \bibinfo{note}{[Erratum: Phys.Rev.D
  55, 1697 (1997)]}.
\bibitem[{Korner and Kramer(1992)}]{Korner:1992wi}
\bibinfo{author}{J.~G. Korner}, \bibinfo{author}{M.~Kramer},
\newblock \bibinfo{title}{{Exclusive nonleptonic charm baryon decays}},
\newblock \bibinfo{journal}{Z. Phys. C} \bibinfo{volume}{55}
  (\bibinfo{year}{1992}) \bibinfo{pages}{659--670}.
  \DOIprefix\doi{10.1007/BF01561305}.
\bibitem[{Xu and Kamal(1992{\natexlab{a}})}]{Xu:1992vc}
\bibinfo{author}{Q.~P. Xu}, \bibinfo{author}{A.~N. Kamal},
\newblock \bibinfo{title}{{Cabibbo favored nonleptonic decays of charmed
  baryons}},
\newblock \bibinfo{journal}{Phys. Rev. D} \bibinfo{volume}{46}
  (\bibinfo{year}{1992}{\natexlab{a}}) \bibinfo{pages}{270--278}.
  \DOIprefix\doi{10.1103/PhysRevD.46.270}.
\bibitem[{Xu and Kamal(1992{\natexlab{b}})}]{Xu:1992sw}
\bibinfo{author}{Q.~P. Xu}, \bibinfo{author}{A.~N. Kamal},
\newblock \bibinfo{title}{{The Nonleptonic charmed baryon decays: B(c)
  ---\ensuremath{>} B (3/2+, decuplet) + P(0-) or V(1-)}},
\newblock \bibinfo{journal}{Phys. Rev. D} \bibinfo{volume}{46}
  (\bibinfo{year}{1992}{\natexlab{b}}) \bibinfo{pages}{3836--3844}.
  \DOIprefix\doi{10.1103/PhysRevD.46.3836}.
\bibitem[{Zenczykowski(1994)}]{Zenczykowski:1993jm}
\bibinfo{author}{P.~Zenczykowski},
\newblock \bibinfo{title}{{Nonleptonic charmed baryon decays: Symmetry
  properties of parity violating amplitudes}},
\newblock \bibinfo{journal}{Phys. Rev. D} \bibinfo{volume}{50}
  (\bibinfo{year}{1994}) \bibinfo{pages}{5787--5792}.
  \DOIprefix\doi{10.1103/PhysRevD.50.5787}.
\bibitem[{Cheng and Tseng(1993)}]{Cheng:1993gf}
\bibinfo{author}{H.-Y. Cheng}, \bibinfo{author}{B.~Tseng},
\newblock \bibinfo{title}{{Cabibbo allowed nonleptonic weak decays of charmed
  baryons}},
\newblock \bibinfo{journal}{Phys. Rev. D} \bibinfo{volume}{48}
  (\bibinfo{year}{1993}) \bibinfo{pages}{4188--4202}.
  \DOIprefix\doi{10.1103/PhysRevD.48.4188}.
  \href{http://arxiv.org/abs/hep-ph/9304286}{{\tt arXiv:hep-ph/9304286}}.
\bibitem[{Uppal et~al.(1994)Uppal, Verma, and Khanna}]{Uppal:1994pt}
\bibinfo{author}{T.~Uppal}, \bibinfo{author}{R.~C. Verma},
  \bibinfo{author}{M.~P. Khanna},
\newblock \bibinfo{title}{{Constituent quark model analysis of weak mesonic
  decays of charm baryons}},
\newblock \bibinfo{journal}{Phys. Rev. D} \bibinfo{volume}{49}
  (\bibinfo{year}{1994}) \bibinfo{pages}{3417--3425}.
  \DOIprefix\doi{10.1103/PhysRevD.49.3417}.
\bibitem[{Chau et~al.(1996)Chau, Cheng, and Tseng}]{Chau:1995gk}
\bibinfo{author}{L.-L. Chau}, \bibinfo{author}{H.-Y. Cheng},
  \bibinfo{author}{B.~Tseng},
\newblock \bibinfo{title}{{Analysis of two-body decays of charmed baryons using
  the quark diagram scheme}},
\newblock \bibinfo{journal}{Phys. Rev. D} \bibinfo{volume}{54}
  (\bibinfo{year}{1996}) \bibinfo{pages}{2132--2160}.
  \DOIprefix\doi{10.1103/PhysRevD.54.2132}.
  \href{http://arxiv.org/abs/hep-ph/9508382}{{\tt arXiv:hep-ph/9508382}}.
\bibitem[{Ivanov et~al.(1998)Ivanov, Korner, Lyubovitskij, and
  Rusetsky}]{Ivanov:1997ra}
\bibinfo{author}{M.~A. Ivanov}, \bibinfo{author}{J.~G. Korner},
  \bibinfo{author}{V.~E. Lyubovitskij}, \bibinfo{author}{A.~G. Rusetsky},
\newblock \bibinfo{title}{{Exclusive nonleptonic decays of bottom and charm
  baryons in a relativistic three quark model: Evaluation of nonfactorizing
  diagrams}},
\newblock \bibinfo{journal}{Phys. Rev. D} \bibinfo{volume}{57}
  (\bibinfo{year}{1998}) \bibinfo{pages}{5632--5652}.
  \DOIprefix\doi{10.1103/PhysRevD.57.5632}.
  \href{http://arxiv.org/abs/hep-ph/9709372}{{\tt arXiv:hep-ph/9709372}}.
\bibitem[{Sharma and Verma(1999)}]{Sharma:1998rd}
\bibinfo{author}{K.~K. Sharma}, \bibinfo{author}{R.~C. Verma},
\newblock \bibinfo{title}{{A Study of weak mesonic decays of Lambda(c) and
  Xi(c) baryons on the basis of HQET results}},
\newblock \bibinfo{journal}{Eur. Phys. J. C} \bibinfo{volume}{7}
  (\bibinfo{year}{1999}) \bibinfo{pages}{217--224}.
  \DOIprefix\doi{10.1007/s100529801008}.
  \href{http://arxiv.org/abs/hep-ph/9803302}{{\tt arXiv:hep-ph/9803302}}.
\bibitem[{Gutsche et~al.(2018)Gutsche, Ivanov, K\"orner, and
  Lyubovitskij}]{Gutsche:2018utw}
\bibinfo{author}{T.~Gutsche}, \bibinfo{author}{M.~A. Ivanov},
  \bibinfo{author}{J.~G. K\"orner}, \bibinfo{author}{V.~E. Lyubovitskij},
\newblock \bibinfo{title}{{Nonleptonic two-body decays of single heavy baryons
  $\Lambda_Q$, $\Xi_Q$, and $\Omega_Q$ $(Q=b,c)$ induced by $W$ emission in the
  covariant confined quark model}},
\newblock \bibinfo{journal}{Phys. Rev. D} \bibinfo{volume}{98}
  (\bibinfo{year}{2018}) \bibinfo{pages}{074011}.
  \DOIprefix\doi{10.1103/PhysRevD.98.074011}.
  \href{http://arxiv.org/abs/1806.11549}{{\tt arXiv:1806.11549}}.
\bibitem[{Niu et~al.(2020)Niu, Richard, Wang, and Zhao}]{Niu:2020gjw}
\bibinfo{author}{P.-Y. Niu}, \bibinfo{author}{J.-M. Richard},
  \bibinfo{author}{Q.~Wang}, \bibinfo{author}{Q.~Zhao},
\newblock \bibinfo{title}{{Hadronic weak decays of $\Lambda_c$ in the quark
  model}},
\newblock \bibinfo{journal}{Phys. Rev. D} \bibinfo{volume}{102}
  (\bibinfo{year}{2020}) \bibinfo{pages}{073005}.
  \DOIprefix\doi{10.1103/PhysRevD.102.073005}.
  \href{http://arxiv.org/abs/2003.09323}{{\tt arXiv:2003.09323}}.
\bibitem[{Li et~al.(2024)Li, Liu, Wang, Li, Geng, and Xie}]{Li:2024rqb}
\bibinfo{author}{Y.~Li}, \bibinfo{author}{S.-W. Liu},
  \bibinfo{author}{E.~Wang}, \bibinfo{author}{D.-M. Li}, \bibinfo{author}{L.-S.
  Geng}, \bibinfo{author}{J.-J. Xie},
\newblock \bibinfo{title}{{Theoretical study of $N(1535)$ and $\Sigma^*(1/2^-)$
  in the Cabibbo-favored process $\Lambda_c^+ \to p \bar{K}^0\eta$}}
  (\bibinfo{year}{2024}). \href{http://arxiv.org/abs/2406.01209}{{\tt
  arXiv:2406.01209}}.
\bibitem[{Zhang et~al.(2024)Zhang, Duan, Lyu, Wang, Zhu, and
  Wang}]{Zhang:2024jby}
\bibinfo{author}{S.-C. Zhang}, \bibinfo{author}{M.-Y. Duan},
  \bibinfo{author}{W.-T. Lyu}, \bibinfo{author}{G.-Y. Wang},
  \bibinfo{author}{J.-Y. Zhu}, \bibinfo{author}{E.~Wang},
\newblock \bibinfo{title}{{Explore the properties of $\Lambda(1670)$ in the
  Cabibbo-favored process $\Lambda^+_c \to p K^- \pi^+$ decay}}
  (\bibinfo{year}{2024}). \href{http://arxiv.org/abs/2405.14235}{{\tt
  arXiv:2405.14235}}.
\bibitem[{Lyu et~al.(2024)Lyu, Zhang, Wang, Wu, Wang, Geng, and
  Xie}]{Lyu:2024qgc}
\bibinfo{author}{W.-T. Lyu}, \bibinfo{author}{S.-C. Zhang},
  \bibinfo{author}{G.-Y. Wang}, \bibinfo{author}{J.-J. Wu},
  \bibinfo{author}{E.~Wang}, \bibinfo{author}{L.-S. Geng},
  \bibinfo{author}{J.-J. Xie},
\newblock \bibinfo{title}{{Evidence of the low-lying baryon $\Sigma^*(1/2^-)$
  in the process $\Lambda_c^+\to \eta\pi^+\Lambda$}}  (\bibinfo{year}{2024}).
  \href{http://arxiv.org/abs/2405.09226}{{\tt arXiv:2405.09226}}.
\bibitem[{Feng et~al.(2023)Feng, Wei, Duan, Wang, and Li}]{Feng:2020jvp}
\bibinfo{author}{X.-C. Feng}, \bibinfo{author}{L.-L. Wei},
  \bibinfo{author}{M.-Y. Duan}, \bibinfo{author}{E.~Wang},
  \bibinfo{author}{D.-M. Li},
\newblock \bibinfo{title}{{The a0(980) in the single Cabibbo-suppressed process
  \ensuremath{\Lambda}c \textrightarrow{}
  \ensuremath{\pi}0\ensuremath{\eta}p}},
\newblock \bibinfo{journal}{Phys. Lett. B} \bibinfo{volume}{846}
  (\bibinfo{year}{2023}) \bibinfo{pages}{138185}.
  \DOIprefix\doi{10.1016/j.physletb.2023.138185}.
  \href{http://arxiv.org/abs/2009.08600}{{\tt arXiv:2009.08600}}.
\bibitem[{Aaij et~al.(2019)}]{LHCb:2019hro}
\bibinfo{author}{R.~Aaij}, et~al. (\bibinfo{collaboration}{LHCb}),
\newblock \bibinfo{title}{{Observation of CP Violation in Charm Decays}},
\newblock \bibinfo{journal}{Phys. Rev. Lett.} \bibinfo{volume}{122}
  (\bibinfo{year}{2019}) \bibinfo{pages}{211803}.
  \DOIprefix\doi{10.1103/PhysRevLett.122.211803}.
  \href{http://arxiv.org/abs/1903.08726}{{\tt arXiv:1903.08726}}.
\bibitem[{Li et~al.(2012)Li, Lu, and Yu}]{Li:2012cfa}
\bibinfo{author}{H.-n. Li}, \bibinfo{author}{C.-D. Lu}, \bibinfo{author}{F.-S.
  Yu},
\newblock \bibinfo{title}{{Branching ratios and direct CP asymmetries in $D\to
  PP$ decays}},
\newblock \bibinfo{journal}{Phys. Rev. D} \bibinfo{volume}{86}
  (\bibinfo{year}{2012}) \bibinfo{pages}{036012}.
  \DOIprefix\doi{10.1103/PhysRevD.86.036012}.
  \href{http://arxiv.org/abs/1203.3120}{{\tt arXiv:1203.3120}}.
\bibitem[{Li et~al.(2019)Li, L\"u, and Yu}]{Li:2019hho}
\bibinfo{author}{H.-N. Li}, \bibinfo{author}{C.-D. L\"u},
  \bibinfo{author}{F.-S. Yu},
\newblock \bibinfo{title}{{Implications on the first observation of charm CPV
  at LHCb}}  (\bibinfo{year}{2019}).
  \href{http://arxiv.org/abs/1903.10638}{{\tt arXiv:1903.10638}}.
\bibitem[{Bediaga et~al.(2023)Bediaga, Frederico, and
  Magalh\~aes}]{Bediaga:2022sxw}
\bibinfo{author}{I.~Bediaga}, \bibinfo{author}{T.~Frederico},
  \bibinfo{author}{P.~C. Magalh\~aes},
\newblock \bibinfo{title}{{Enhanced Charm CP Asymmetries from Final State
  Interactions}},
\newblock \bibinfo{journal}{Phys. Rev. Lett.} \bibinfo{volume}{131}
  (\bibinfo{year}{2023}) \bibinfo{pages}{051802}.
  \DOIprefix\doi{10.1103/PhysRevLett.131.051802}.
  \href{http://arxiv.org/abs/2203.04056}{{\tt arXiv:2203.04056}}.
\bibitem[{Pich et~al.(2023)Pich, Solomonidi, and Vale~Silva}]{Pich:2023kim}
\bibinfo{author}{A.~Pich}, \bibinfo{author}{E.~Solomonidi},
  \bibinfo{author}{L.~Vale~Silva},
\newblock \bibinfo{title}{{Final-state interactions in the CP asymmetries of
  charm-meson two-body decays}},
\newblock \bibinfo{journal}{Phys. Rev. D} \bibinfo{volume}{108}
  (\bibinfo{year}{2023}) \bibinfo{pages}{036026}.
  \DOIprefix\doi{10.1103/PhysRevD.108.036026}.
  \href{http://arxiv.org/abs/2305.11951}{{\tt arXiv:2305.11951}}.
\bibitem[{Cheng et~al.(2005)Cheng, Chua, and Soni}]{Cheng:2004ru}
\bibinfo{author}{H.-Y. Cheng}, \bibinfo{author}{C.-K. Chua},
  \bibinfo{author}{A.~Soni},
\newblock \bibinfo{title}{{Final state interactions in hadronic B decays}},
\newblock \bibinfo{journal}{Phys. Rev. D} \bibinfo{volume}{71}
  (\bibinfo{year}{2005}) \bibinfo{pages}{014030}.
  \DOIprefix\doi{10.1103/PhysRevD.71.014030}.
  \href{http://arxiv.org/abs/hep-ph/0409317}{{\tt arXiv:hep-ph/0409317}}.
\bibitem[{Yu et~al.(2018)Yu, Jiang, Li, L\"u, Wang, and Zhao}]{Yu:2017zst}
\bibinfo{author}{F.-S. Yu}, \bibinfo{author}{H.-Y. Jiang},
  \bibinfo{author}{R.-H. Li}, \bibinfo{author}{C.-D. L\"u},
  \bibinfo{author}{W.~Wang}, \bibinfo{author}{Z.-X. Zhao},
\newblock \bibinfo{title}{{Discovery Potentials of Doubly Charmed Baryons}},
\newblock \bibinfo{journal}{Chin. Phys. C} \bibinfo{volume}{42}
  (\bibinfo{year}{2018}) \bibinfo{pages}{051001}.
  \DOIprefix\doi{10.1088/1674-1137/42/5/051001}.
  \href{http://arxiv.org/abs/1703.09086}{{\tt arXiv:1703.09086}}.
\bibitem[{Chau(1983)}]{Chau:1982da}
\bibinfo{author}{L.-L. Chau},
\newblock \bibinfo{title}{{Quark Mixing in Weak Interactions}},
\newblock \bibinfo{journal}{Phys. Rept.} \bibinfo{volume}{95}
  (\bibinfo{year}{1983}) \bibinfo{pages}{1--94}.
  \DOIprefix\doi{10.1016/0370-1573(83)90043-1}.
\bibitem[{Chau and Cheng(1986)}]{Chau:1986jb}
\bibinfo{author}{L.~L. Chau}, \bibinfo{author}{H.~Y. Cheng},
\newblock \bibinfo{title}{{Quark Diagram Analysis of Two-body Charm Decays}},
\newblock \bibinfo{journal}{Phys. Rev. Lett.} \bibinfo{volume}{56}
  (\bibinfo{year}{1986}) \bibinfo{pages}{1655--1658}.
  \DOIprefix\doi{10.1103/PhysRevLett.56.1655}.
\bibitem[{Biyajima et~al.(1987)Biyajima, Shirane, and
  Terazawa}]{Biyajima:1987qm}
\bibinfo{author}{M.~Biyajima}, \bibinfo{author}{K.~Shirane},
  \bibinfo{author}{O.~Terazawa},
\newblock \bibinfo{title}{{CALCULATIONS OF STANDARD HIGGS BOSON PRODUCTION
  CROSS-SECTIONS IN E+ E- COLLISIONS BY MEANS OF A REASONABLE SET OF
  PARAMETERS. (REVISED VERSION)}},
\newblock \bibinfo{journal}{Phys. Rev. D} \bibinfo{volume}{36}
  (\bibinfo{year}{1987}) \bibinfo{pages}{2161--2164}.
  \DOIprefix\doi{10.1103/PhysRevD.36.2161}.
\bibitem[{Chau and Cheng(1989)}]{Chau:1988kb}
\bibinfo{author}{L.-L. Chau}, \bibinfo{author}{H.-Y. Cheng},
\newblock \bibinfo{title}{{Comments on {QCD} Sum Rule Calculations of Exclusive
  Two-body Decays of Charmed Mesons}},
\newblock \bibinfo{journal}{Mod. Phys. Lett. A} \bibinfo{volume}{4}
  (\bibinfo{year}{1989}) \bibinfo{pages}{877}.
  \DOIprefix\doi{10.1142/S0217732389001039}.
\bibitem[{Qin et~al.(2023)Qin, Qiu, and Yu}]{Qin:2022nof}
\bibinfo{author}{Q.~Qin}, \bibinfo{author}{J.-L. Qiu}, \bibinfo{author}{F.-S.
  Yu},
\newblock \bibinfo{title}{{Diagrammatic analysis of hidden- and open-charm
  tetraquark production in B decays}},
\newblock \bibinfo{journal}{Eur. Phys. J. C} \bibinfo{volume}{83}
  (\bibinfo{year}{2023}) \bibinfo{pages}{227}.
  \DOIprefix\doi{10.1140/epjc/s10052-023-11375-4}.
  \href{http://arxiv.org/abs/2212.03590}{{\tt arXiv:2212.03590}}.
\bibitem[{Fu-Sheng et~al.(2011)Fu-Sheng, Wang, and Lu}]{Fu-Sheng:2011fji}
\bibinfo{author}{Y.~Fu-Sheng}, \bibinfo{author}{X.-X. Wang},
  \bibinfo{author}{C.-D. Lu},
\newblock \bibinfo{title}{{Nonleptonic Two Body Decays of Charmed Mesons}},
\newblock \bibinfo{journal}{Phys. Rev. D} \bibinfo{volume}{84}
  (\bibinfo{year}{2011}) \bibinfo{pages}{074019}.
  \DOIprefix\doi{10.1103/PhysRevD.84.074019}.
  \href{http://arxiv.org/abs/1101.4714}{{\tt arXiv:1101.4714}}.
\bibitem[{Qin et~al.(2014)Qin, Li, L\"u, and Yu}]{Qin:2013tje}
\bibinfo{author}{Q.~Qin}, \bibinfo{author}{H.-n. Li}, \bibinfo{author}{C.-D.
  L\"u}, \bibinfo{author}{F.-S. Yu},
\newblock \bibinfo{title}{{Branching ratios and direct CP asymmetries in $D\to
  PV$ decays}},
\newblock \bibinfo{journal}{Phys. Rev. D} \bibinfo{volume}{89}
  (\bibinfo{year}{2014}) \bibinfo{pages}{054006}.
  \DOIprefix\doi{10.1103/PhysRevD.89.054006}.
  \href{http://arxiv.org/abs/1305.7021}{{\tt arXiv:1305.7021}}.
\bibitem[{Han et~al.(2021{\natexlab{a}})Han, Jiang, Liu, Xiao, and
  Yu}]{Han:2021azw}
\bibinfo{author}{J.-J. Han}, \bibinfo{author}{H.-Y. Jiang},
  \bibinfo{author}{W.~Liu}, \bibinfo{author}{Z.-J. Xiao},
  \bibinfo{author}{F.-S. Yu},
\newblock \bibinfo{title}{{Rescattering mechanism of weak decays of
  double-charm baryons}},
\newblock \bibinfo{journal}{Chin. Phys. C} \bibinfo{volume}{45}
  (\bibinfo{year}{2021}{\natexlab{a}}) \bibinfo{pages}{053105}.
  \DOIprefix\doi{10.1088/1674-1137/abec68}.
  \href{http://arxiv.org/abs/2101.12019}{{\tt arXiv:2101.12019}}.
\bibitem[{Han et~al.(2021{\natexlab{b}})Han, Zhang, Jiang, Xiao, and
  Yu}]{Han:2021gkl}
\bibinfo{author}{J.-J. Han}, \bibinfo{author}{R.-X. Zhang},
  \bibinfo{author}{H.-Y. Jiang}, \bibinfo{author}{Z.-J. Xiao},
  \bibinfo{author}{F.-S. Yu},
\newblock \bibinfo{title}{{Weak decays of bottom-charm baryons: $\mathcal
  {B}_{bc}\rightarrow \mathcal {B}_bP$}},
\newblock \bibinfo{journal}{Eur. Phys. J. C} \bibinfo{volume}{81}
  (\bibinfo{year}{2021}{\natexlab{b}}) \bibinfo{pages}{539}.
  \DOIprefix\doi{10.1140/epjc/s10052-021-09239-w}.
  \href{http://arxiv.org/abs/2102.00961}{{\tt arXiv:2102.00961}}.
\bibitem[{Leibovich et~al.(2004)Leibovich, Ligeti, Stewart, and
  Wise}]{Leibovich:2003tw}
\bibinfo{author}{A.~K. Leibovich}, \bibinfo{author}{Z.~Ligeti},
  \bibinfo{author}{I.~W. Stewart}, \bibinfo{author}{M.~B. Wise},
\newblock \bibinfo{title}{{Predictions for nonleptonic Lambda(b) and Theta(b)
  decays}},
\newblock \bibinfo{journal}{Phys. Lett. B} \bibinfo{volume}{586}
  (\bibinfo{year}{2004}) \bibinfo{pages}{337--344}.
  \DOIprefix\doi{10.1016/j.physletb.2004.02.033}.
  \href{http://arxiv.org/abs/hep-ph/0312319}{{\tt arXiv:hep-ph/0312319}}.
\bibitem[{Mantry et~al.(2003)Mantry, Pirjol, and Stewart}]{Mantry:2003uz}
\bibinfo{author}{S.~Mantry}, \bibinfo{author}{D.~Pirjol},
  \bibinfo{author}{I.~W. Stewart},
\newblock \bibinfo{title}{{Strong phases and factorization for color suppressed
  decays}},
\newblock \bibinfo{journal}{Phys. Rev. D} \bibinfo{volume}{68}
  (\bibinfo{year}{2003}) \bibinfo{pages}{114009}.
  \DOIprefix\doi{10.1103/PhysRevD.68.114009}.
  \href{http://arxiv.org/abs/hep-ph/0306254}{{\tt arXiv:hep-ph/0306254}}.
\bibitem[{Cheng(1997)}]{Cheng:1996cs}
\bibinfo{author}{H.-Y. Cheng},
\newblock \bibinfo{title}{{Nonleptonic weak decays of bottom baryons}},
\newblock \bibinfo{journal}{Phys. Rev. D} \bibinfo{volume}{56}
  (\bibinfo{year}{1997}) \bibinfo{pages}{2799--2811}.
  \DOIprefix\doi{10.1103/PhysRevD.56.2799}.
  \href{http://arxiv.org/abs/hep-ph/9612223}{{\tt arXiv:hep-ph/9612223}},
  \bibinfo{note}{[Erratum: Phys.Rev.D 99, 079901 (2019)]}.
\bibitem[{Wolfenstein(1991)}]{Wolfenstein:1990ks}
\bibinfo{author}{L.~Wolfenstein},
\newblock \bibinfo{title}{{Final state interactions and CP violation in weak
  decays}},
\newblock \bibinfo{journal}{Phys. Rev. D} \bibinfo{volume}{43}
  (\bibinfo{year}{1991}) \bibinfo{pages}{151--156}.
  \DOIprefix\doi{10.1103/PhysRevD.43.151}.
\bibitem[{Suzuki and Wolfenstein(1999)}]{Suzuki:1999uc}
\bibinfo{author}{M.~Suzuki}, \bibinfo{author}{L.~Wolfenstein},
\newblock \bibinfo{title}{{Final state interaction phase in B decays}},
\newblock \bibinfo{journal}{Phys. Rev. D} \bibinfo{volume}{60}
  (\bibinfo{year}{1999}) \bibinfo{pages}{074019}.
  \DOIprefix\doi{10.1103/PhysRevD.60.074019}.
  \href{http://arxiv.org/abs/hep-ph/9903477}{{\tt arXiv:hep-ph/9903477}}.
\bibitem[{Cheng and Chua(2020)}]{Cheng:2020ipp}
\bibinfo{author}{H.-Y. Cheng}, \bibinfo{author}{C.-K. Chua},
\newblock \bibinfo{title}{{Branching fractions and $CP$ violation in $B^-\to
  K^+K^-\pi^-$ and $B^-\to \pi^+\pi^-\pi^-$ decays}},
\newblock \bibinfo{journal}{Phys. Rev. D} \bibinfo{volume}{102}
  (\bibinfo{year}{2020}) \bibinfo{pages}{053006}.
  \DOIprefix\doi{10.1103/PhysRevD.102.053006}.
  \href{http://arxiv.org/abs/2007.02558}{{\tt arXiv:2007.02558}}.
\bibitem[{Chua(2008)}]{Chua:2007cm}
\bibinfo{author}{C.-K. Chua},
\newblock \bibinfo{title}{{Rescattering effects in charmless anti-B(u,d,s)
  ---\ensuremath{>} P P decays}},
\newblock \bibinfo{journal}{Phys. Rev. D} \bibinfo{volume}{78}
  (\bibinfo{year}{2008}) \bibinfo{pages}{076002}.
  \DOIprefix\doi{10.1103/PhysRevD.78.076002}.
  \href{http://arxiv.org/abs/0712.4187}{{\tt arXiv:0712.4187}}.
\bibitem[{Dedonder et~al.(2011)Dedonder, Furman, Kaminski, Lesniak, and
  Loiseau}]{Dedonder:2010fg}
\bibinfo{author}{J.~P. Dedonder}, \bibinfo{author}{A.~Furman},
  \bibinfo{author}{R.~Kaminski}, \bibinfo{author}{L.~Lesniak},
  \bibinfo{author}{B.~Loiseau},
\newblock \bibinfo{title}{{S-, P- and D-wave final state interactions and CP
  violation in B+- --\ensuremath{>} pi+- pi-+ pi+- decays}},
\newblock \bibinfo{journal}{Acta Phys. Polon. B} \bibinfo{volume}{42}
  (\bibinfo{year}{2011}) \bibinfo{pages}{2013}.
  \DOIprefix\doi{10.5506/APhysPolB.42.2013}.
  \href{http://arxiv.org/abs/1011.0960}{{\tt arXiv:1011.0960}}.
\bibitem[{Bediaga et~al.(2014)Bediaga, Frederico, and
  Louren\c{c}o}]{Bediaga:2013ela}
\bibinfo{author}{I.~Bediaga}, \bibinfo{author}{T.~Frederico},
  \bibinfo{author}{O.~Louren\c{c}o},
\newblock \bibinfo{title}{{CP violation and CPT invariance in $B^\pm$ decays
  with final state interactions}},
\newblock \bibinfo{journal}{Phys. Rev. D} \bibinfo{volume}{89}
  (\bibinfo{year}{2014}) \bibinfo{pages}{094013}.
  \DOIprefix\doi{10.1103/PhysRevD.89.094013}.
  \href{http://arxiv.org/abs/1307.8164}{{\tt arXiv:1307.8164}}.
\bibitem[{Cheng and Chua(2013)}]{Cheng:2013dua}
\bibinfo{author}{H.-Y. Cheng}, \bibinfo{author}{C.-K. Chua},
\newblock \bibinfo{title}{{Branching Fractions and Direct CP Violation in
  Charmless Three-body Decays of B Mesons}},
\newblock \bibinfo{journal}{Phys. Rev. D} \bibinfo{volume}{88}
  (\bibinfo{year}{2013}) \bibinfo{pages}{114014}.
  \DOIprefix\doi{10.1103/PhysRevD.88.114014}.
  \href{http://arxiv.org/abs/1308.5139}{{\tt arXiv:1308.5139}}.
\bibitem[{Buccella et~al.(2019)Buccella, Paul, and
  Santorelli}]{Buccella:2019kpn}
\bibinfo{author}{F.~Buccella}, \bibinfo{author}{A.~Paul},
  \bibinfo{author}{P.~Santorelli},
\newblock \bibinfo{title}{{$SU(3)_F$ breaking through final state interactions
  and $CP$ asymmetries in $D \to PP$ decays}},
\newblock \bibinfo{journal}{Phys. Rev. D} \bibinfo{volume}{99}
  (\bibinfo{year}{2019}) \bibinfo{pages}{113001}.
  \DOIprefix\doi{10.1103/PhysRevD.99.113001}.
  \href{http://arxiv.org/abs/1902.05564}{{\tt arXiv:1902.05564}}.
\bibitem[{Cheng and Chiang(2010)}]{Cheng:2010ry}
\bibinfo{author}{H.-Y. Cheng}, \bibinfo{author}{C.-W. Chiang},
\newblock \bibinfo{title}{{Two-body hadronic charmed meson decays}},
\newblock \bibinfo{journal}{Phys. Rev. D} \bibinfo{volume}{81}
  (\bibinfo{year}{2010}) \bibinfo{pages}{074021}.
  \DOIprefix\doi{10.1103/PhysRevD.81.074021}.
  \href{http://arxiv.org/abs/1001.0987}{{\tt arXiv:1001.0987}}.
\bibitem[{Magalhaes et~al.(2011)Magalhaes, Robilotta, Guimaraes, Frederico,
  de~Paula, Bediaga, Reis, Maekawa, and Zarnauskas}]{Magalhaes:2011sh}
\bibinfo{author}{P.~C. Magalhaes}, \bibinfo{author}{M.~R. Robilotta},
  \bibinfo{author}{K.~S. F.~F. Guimaraes}, \bibinfo{author}{T.~Frederico},
  \bibinfo{author}{W.~de~Paula}, \bibinfo{author}{I.~Bediaga},
  \bibinfo{author}{A.~C.~d. Reis}, \bibinfo{author}{C.~M. Maekawa},
  \bibinfo{author}{G.~R.~S. Zarnauskas},
\newblock \bibinfo{title}{{Towards three-body unitarity in $D^+ \to K^- \pi^+
  \pi^+$}},
\newblock \bibinfo{journal}{Phys. Rev. D} \bibinfo{volume}{84}
  (\bibinfo{year}{2011}) \bibinfo{pages}{094001}.
  \DOIprefix\doi{10.1103/PhysRevD.84.094001}.
  \href{http://arxiv.org/abs/1105.5120}{{\tt arXiv:1105.5120}}.
\bibitem[{Garrote et~al.(2023)Garrote, Cuervo, Magalh\~aes, and
  Pel\'aez}]{Garrote:2022uub}
\bibinfo{author}{R.~A. Garrote}, \bibinfo{author}{J.~Cuervo},
  \bibinfo{author}{P.~C. Magalh\~aes}, \bibinfo{author}{J.~R. Pel\'aez},
\newblock \bibinfo{title}{{Dispersive
  \ensuremath{\pi}\ensuremath{\pi}\textrightarrow{}KK\textasciimacron{}
  Amplitude and Giant CP Violation in B to Three Light-Meson Decays at LHCb}},
\newblock \bibinfo{journal}{Phys. Rev. Lett.} \bibinfo{volume}{130}
  (\bibinfo{year}{2023}) \bibinfo{pages}{201901}.
  \DOIprefix\doi{10.1103/PhysRevLett.130.201901}.
  \href{http://arxiv.org/abs/2210.08354}{{\tt arXiv:2210.08354}}.
\bibitem[{Zhou(2004)}]{Zhou:2004gm}
\bibinfo{author}{Y.~Zhou},
\newblock \bibinfo{title}{{Imaginary part of Feynman amplitude, cutting rules
  and optical theorem}}  (\bibinfo{year}{2004}).
  \href{http://arxiv.org/abs/hep-ph/0412204}{{\tt arXiv:hep-ph/0412204}}.
\bibitem[{Aste(York)}]{Andreas:1995ap}
\bibinfo{author}{A.~Aste},
\newblock \bibinfo{title}{{Finite Quantum Electrodynamics, the Causal
  Approach}},
\newblock \bibinfo{journal}{Annals of Physics} \bibinfo{volume}{257}
  (\bibinfo{year}{1997,2nd ed., Springer, Berlin/Heidelberg/New York})
  \bibinfo{pages}{158}.
\bibitem[{Gortchakov et~al.(1996)Gortchakov, Locher, Markushin, and von
  Rotz}]{Gortchakov:1995im}
\bibinfo{author}{O.~Gortchakov}, \bibinfo{author}{M.~P. Locher},
  \bibinfo{author}{V.~E. Markushin}, \bibinfo{author}{S.~von Rotz},
\newblock \bibinfo{title}{{Two meson doorway calculation for anti-p p
  ---\ensuremath{>} phi pi including off-shell effects and the OZI rule}},
\newblock \bibinfo{journal}{Z. Phys. A} \bibinfo{volume}{353}
  (\bibinfo{year}{1996}) \bibinfo{pages}{447--453}.
  \DOIprefix\doi{10.1007/BF01285155}.
\bibitem[{Workman et~al.(2022)}]{ParticleDataGroup:2022pth}
\bibinfo{author}{R.~L. Workman}, et~al. (\bibinfo{collaboration}{Particle Data
  Group}),
\newblock \bibinfo{title}{{Review of Particle Physics}},
\newblock \bibinfo{journal}{PTEP} \bibinfo{volume}{2022} (\bibinfo{year}{2022})
  \bibinfo{pages}{083C01}. \DOIprefix\doi{10.1093/ptep/ptac097}.
\bibitem[{Ball et~al.(2007)Ball, Jones, and Zwicky}]{Ball:2006eu}
\bibinfo{author}{P.~Ball}, \bibinfo{author}{G.~W. Jones},
  \bibinfo{author}{R.~Zwicky},
\newblock \bibinfo{title}{{$B \to V \gamma$ beyond QCD factorisation}},
\newblock \bibinfo{journal}{Phys. Rev. D} \bibinfo{volume}{75}
  (\bibinfo{year}{2007}) \bibinfo{pages}{054004}.
  \DOIprefix\doi{10.1103/PhysRevD.75.054004}.
  \href{http://arxiv.org/abs/hep-ph/0612081}{{\tt arXiv:hep-ph/0612081}}.
\bibitem[{Detmold and Meinel(2016)}]{Detmold:2016pkz}
\bibinfo{author}{W.~Detmold}, \bibinfo{author}{S.~Meinel},
\newblock \bibinfo{title}{{$\Lambda_b \to \Lambda \ell^+ \ell^-$ form factors,
  differential branching fraction, and angular observables from lattice QCD
  with relativistic $b$ quarks}},
\newblock \bibinfo{journal}{Phys. Rev. D} \bibinfo{volume}{93}
  (\bibinfo{year}{2016}) \bibinfo{pages}{074501}.
  \DOIprefix\doi{10.1103/PhysRevD.93.074501}.
  \href{http://arxiv.org/abs/1602.01399}{{\tt arXiv:1602.01399}}.
\bibitem[{Meinel(2017)}]{Meinel:2016dqj}
\bibinfo{author}{S.~Meinel},
\newblock \bibinfo{title}{{$\Lambda_c \to \Lambda l^+ \nu_l$ form factors and
  decay rates from lattice QCD with physical quark masses}},
\newblock \bibinfo{journal}{Phys. Rev. Lett.} \bibinfo{volume}{118}
  (\bibinfo{year}{2017}) \bibinfo{pages}{082001}.
  \DOIprefix\doi{10.1103/PhysRevLett.118.082001}.
  \href{http://arxiv.org/abs/1611.09696}{{\tt arXiv:1611.09696}}.
\bibitem[{Meinel(2018)}]{Meinel:2017ggx}
\bibinfo{author}{S.~Meinel},
\newblock \bibinfo{title}{{$\Lambda_c \to N$ form factors from lattice QCD and
  phenomenology of $\Lambda_c \to n \ell^+ \nu_\ell$ and $\Lambda_c \to p \mu^+
  \mu^-$ decays}},
\newblock \bibinfo{journal}{Phys. Rev. D} \bibinfo{volume}{97}
  (\bibinfo{year}{2018}) \bibinfo{pages}{034511}.
  \DOIprefix\doi{10.1103/PhysRevD.97.034511}.
  \href{http://arxiv.org/abs/1712.05783}{{\tt arXiv:1712.05783}}.
\bibitem[{Bahtiyar(2021)}]{Bahtiyar:2021voz}
\bibinfo{author}{H.~Bahtiyar},
\newblock \bibinfo{title}{{$\Lambda_c \to \Lambda$ Form Factors in Lattice
  QCD}},
\newblock \bibinfo{journal}{Turk. J. Phys.} \bibinfo{volume}{45}
  (\bibinfo{year}{2021}) \bibinfo{pages}{4}.
  \DOIprefix\doi{10.3906/fiz-2104-28}.
  \href{http://arxiv.org/abs/2107.13909}{{\tt arXiv:2107.13909}}.
\bibitem[{Zhao(2018)}]{Zhao:2018zcb}
\bibinfo{author}{Z.-X. Zhao},
\newblock \bibinfo{title}{{Weak decays of heavy baryons in the light-front
  approach}},
\newblock \bibinfo{journal}{Chin. Phys. C} \bibinfo{volume}{42}
  (\bibinfo{year}{2018}) \bibinfo{pages}{093101}.
  \DOIprefix\doi{10.1088/1674-1137/42/9/093101}.
  \href{http://arxiv.org/abs/1803.02292}{{\tt arXiv:1803.02292}}.
\bibitem[{Liu et~al.(2009)Liu, Huang, and Wang}]{Liu:2009sn}
\bibinfo{author}{Y.-L. Liu}, \bibinfo{author}{M.-Q. Huang},
  \bibinfo{author}{D.-W. Wang},
\newblock \bibinfo{title}{{Improved analysis on the semi-leptonic decay
  Lambda(c) ---\ensuremath{>} Lambda l+ nu from QCD light-cone sum rules}},
\newblock \bibinfo{journal}{Phys. Rev. D} \bibinfo{volume}{80}
  (\bibinfo{year}{2009}) \bibinfo{pages}{074011}.
  \DOIprefix\doi{10.1103/PhysRevD.80.074011}.
  \href{http://arxiv.org/abs/0910.1160}{{\tt arXiv:0910.1160}}.
\bibitem[{Gutsche et~al.(2016)Gutsche, Ivanov, Korner, Lyubovitskij, and
  Santorelli}]{Gutsche:2015rrt}
\bibinfo{author}{T.~Gutsche}, \bibinfo{author}{M.~A. Ivanov},
  \bibinfo{author}{J.~G. Korner}, \bibinfo{author}{V.~E. Lyubovitskij},
  \bibinfo{author}{P.~Santorelli},
\newblock \bibinfo{title}{{Semileptonic decays $\Lambda_c^+ \to \Lambda \ell^+
  \nu_\ell\,\,(\ell=e,\mu)$ in the covariant quark model and comparison with
  the new absolute branching fraction measurements of Belle and BESIII}},
\newblock \bibinfo{journal}{Phys. Rev. D} \bibinfo{volume}{93}
  (\bibinfo{year}{2016}) \bibinfo{pages}{034008}.
  \DOIprefix\doi{10.1103/PhysRevD.93.034008}.
  \href{http://arxiv.org/abs/1512.02168}{{\tt arXiv:1512.02168}}.
\bibitem[{Faustov and Galkin(2018)}]{Faustov:2018dkn}
\bibinfo{author}{R.~N. Faustov}, \bibinfo{author}{V.~O. Galkin},
\newblock \bibinfo{title}{{Rare $\Lambda _c\rightarrow p \ell ^+\ell ^-$ decay
  in the relativistic quark model}},
\newblock \bibinfo{journal}{Eur. Phys. J. C} \bibinfo{volume}{78}
  (\bibinfo{year}{2018}) \bibinfo{pages}{527}.
  \DOIprefix\doi{10.1140/epjc/s10052-018-6010-y}.
  \href{http://arxiv.org/abs/1805.02516}{{\tt arXiv:1805.02516}}.
\bibitem[{Aliev et~al.(2006)Aliev, Ozpineci, Yakovlev, and
  Zamiralov}]{Aliev:2006xr}
\bibinfo{author}{T.~M. Aliev}, \bibinfo{author}{A.~Ozpineci},
  \bibinfo{author}{S.~B. Yakovlev}, \bibinfo{author}{V.~Zamiralov},
\newblock \bibinfo{title}{{Meson-octet-baryon couplings using light cone QCD
  sum rules}},
\newblock \bibinfo{journal}{Phys. Rev. D} \bibinfo{volume}{74}
  (\bibinfo{year}{2006}) \bibinfo{pages}{116001}.
  \DOIprefix\doi{10.1103/PhysRevD.74.116001}.
\bibitem[{Aliev et~al.(2009)Aliev, Ozpineci, Savci, and
  Zamiralov}]{Aliev:2009ei}
\bibinfo{author}{T.~M. Aliev}, \bibinfo{author}{A.~Ozpineci},
  \bibinfo{author}{M.~Savci}, \bibinfo{author}{V.~S. Zamiralov},
\newblock \bibinfo{title}{{Vector meson-baryon strong coupling contants in
  light cone QCD sum rules}},
\newblock \bibinfo{journal}{Phys. Rev. D} \bibinfo{volume}{80}
  (\bibinfo{year}{2009}) \bibinfo{pages}{016010}.
  \DOIprefix\doi{10.1103/PhysRevD.80.016010}.
  \href{http://arxiv.org/abs/0905.4664}{{\tt arXiv:0905.4664}}.
\bibitem[{Janssen et~al.(1996)Janssen, Holinde, and Speth}]{Janssen:1996kx}
\bibinfo{author}{G.~Janssen}, \bibinfo{author}{K.~Holinde},
  \bibinfo{author}{J.~Speth},
\newblock \bibinfo{title}{{pi rho correlations in the N N potential}},
\newblock \bibinfo{journal}{Phys. Rev. C} \bibinfo{volume}{54}
  (\bibinfo{year}{1996}) \bibinfo{pages}{2218--2234}.
  \DOIprefix\doi{10.1103/PhysRevC.54.2218}.
\bibitem[{Oset and Ramos(2010)}]{Oset:2010tof}
\bibinfo{author}{E.~Oset}, \bibinfo{author}{A.~Ramos},
\newblock \bibinfo{title}{{Dynamically generated resonances from the vector
  octet-baryon octet interaction}},
\newblock \bibinfo{journal}{Eur. Phys. J. A} \bibinfo{volume}{44}
  (\bibinfo{year}{2010}) \bibinfo{pages}{445--454}.
  \DOIprefix\doi{10.1140/epja/i2010-10957-3}.
  \href{http://arxiv.org/abs/0905.0973}{{\tt arXiv:0905.0973}}.
\bibitem[{Bramon et~al.(1995)Bramon, Grau, and Pancheri}]{Bramon:1994pq}
\bibinfo{author}{A.~Bramon}, \bibinfo{author}{A.~Grau},
  \bibinfo{author}{G.~Pancheri},
\newblock \bibinfo{title}{{Radiative vector meson decays in SU(3) broken
  effective chiral Lagrangians}},
\newblock \bibinfo{journal}{Phys. Lett. B} \bibinfo{volume}{344}
  (\bibinfo{year}{1995}) \bibinfo{pages}{240--244}.
  \DOIprefix\doi{10.1016/0370-2693(94)01543-L}.
\bibitem[{Aaij et~al.(2024)}]{LHCb:2024djr}
\bibinfo{author}{R.~Aaij}, et~al. (\bibinfo{collaboration}{LHCb}),
\newblock \bibinfo{title}{{Search for the rare $\Lambda_c^+ \to p \mu^+ \mu^-$
  decay}}  (\bibinfo{year}{2024}). \href{http://arxiv.org/abs/2407.11474}{{\tt
  arXiv:2407.11474}}.
\bibitem[{Hong et~al.(2023)Hong, Yan, Ping, and Luo}]{Hong:2022prk}
\bibinfo{author}{P.-C. Hong}, \bibinfo{author}{F.~Yan}, \bibinfo{author}{R.-G.
  Ping}, \bibinfo{author}{T.~Luo},
\newblock \bibinfo{title}{{Study of parity violation in and decays*}},
\newblock \bibinfo{journal}{Chin. Phys. C} \bibinfo{volume}{47}
  (\bibinfo{year}{2023}) \bibinfo{pages}{053101}.
  \DOIprefix\doi{10.1088/1674-1137/acb7ce}.
  \href{http://arxiv.org/abs/2211.16014}{{\tt arXiv:2211.16014}}.
\bibitem[{Fajfer et~al.(2003)Fajfer, Prapotnik, Singer, and
  Zupan}]{Fajfer:2003ag}
\bibinfo{author}{S.~Fajfer}, \bibinfo{author}{A.~Prapotnik},
  \bibinfo{author}{P.~Singer}, \bibinfo{author}{J.~Zupan},
\newblock \bibinfo{title}{{Final state interactions in the D+(s)
  ---\ensuremath{>} omega pi+ and D+(s) ---\ensuremath{>} rho0 pi+ decays}},
\newblock \bibinfo{journal}{Phys. Rev. D} \bibinfo{volume}{68}
  (\bibinfo{year}{2003}) \bibinfo{pages}{094012}.
  \DOIprefix\doi{10.1103/PhysRevD.68.094012}.
  \href{http://arxiv.org/abs/hep-ph/0308100}{{\tt arXiv:hep-ph/0308100}}.
\bibitem[{Ronchen et~al.(2013)Ronchen, Doring, Huang, Haberzettl, Haidenbauer,
  Hanhart, Krewald, Meissner, and Nakayama}]{Ronchen:2012eg}
\bibinfo{author}{D.~Ronchen}, \bibinfo{author}{M.~Doring},
  \bibinfo{author}{F.~Huang}, \bibinfo{author}{H.~Haberzettl},
  \bibinfo{author}{J.~Haidenbauer}, \bibinfo{author}{C.~Hanhart},
  \bibinfo{author}{S.~Krewald}, \bibinfo{author}{U.~G. Meissner},
  \bibinfo{author}{K.~Nakayama},
\newblock \bibinfo{title}{{Coupled-channel dynamics in the reactions piN
  --\ensuremath{>} piN, etaN, KLambda, KSigma}},
\newblock \bibinfo{journal}{Eur. Phys. J. A} \bibinfo{volume}{49}
  (\bibinfo{year}{2013}) \bibinfo{pages}{44}.
  \DOIprefix\doi{10.1140/epja/i2013-13044-5}.
  \href{http://arxiv.org/abs/1211.6998}{{\tt arXiv:1211.6998}}.

\end{thebibliography}

\bibliographystyle{elsarticle-num-names}

\end{document}